\documentclass[11pt,a4paper]{article}

\usepackage[left=30mm,right=30mm]{geometry}

\usepackage{hyperref}
\usepackage{comment}

\usepackage{cite}

\pdfoutput=1 
\usepackage{mathtools}
\usepackage{environ}
\usepackage{xcolor}
\usepackage{pgfplots}
\pgfplotsset{compat=1.15}
\usepackage{tikz}
\usepackage[tikz]{bclogo}
\usepackage{tkz-euclide}
\usetikzlibrary{calc,arrows,decorations}
\usepackage{amssymb,amsmath}
\usepackage{todonotes}
\usepackage{dsfont}
\usepackage{fancyvrb}
\usepackage{fontawesome}
\usepackage{bbding}
\numberwithin{equation}{section}

\def\ket#1{|{#1}\rangle}
\def\bra#1{\langle{#1}|}
\def\Tr{\text{Tr}}

\newcommand{\be}[0]{\begin{equation}}
\newcommand{\ee}[0]{\end{equation}}
\newcommand{\bea}{\begin{eqnarray}}
\newcommand{\eea}{\end{eqnarray}}

\begin{document}


\vspace*{15pt}
\begin{center}
{\Large {\bf Correlation functions and quantum measures\\of descendant states}}

\vspace{35pt}
{\bf Enrico M.~Brehm\textsuperscript{\SnowflakeChevron}, Matteo Broccoli\textsuperscript{\faTree}}

\vspace{30pt}

{\it
Max-Planck-Institut für Gravitationsphysik,\\
Albert-Einstein-Institut, \\
Potsdam-Golm, D-14476, 
Germany.
}

\vspace{15pt}
{\textsuperscript{\SnowflakeChevron}\href{mailto:enrico.brehm@aei.mpg.de}{enrico.brehm@aei.mpg.de}, \textsuperscript{\faTree}\href{mailto:matteo.broccoli@aei.mpg.de}{matteo.broccoli@aei.mpg.de} }

\vspace{55pt}

{ABSTRACT}
\end{center}

\noindent
We discuss a computer implementation of a recursive formula to calculate correlation functions of descendant states in two-dimensional CFT.
This allows us to obtain any $N$-point function of vacuum descendants, or to express the correlator as a differential operator acting on the respective primary correlator in case of non-vacuum descendants.
With this tool at hand, we then study some entanglement and distinguishability measures between descendant states, namely the R\'enyi entropy, trace square distance and sandwiched R\'enyi divergence.
Our results provide a test of the conjectured R\'enyi QNEC and 
new tools to analyse the holographic description of descendant states at large $c$.
\noindent 


\thispagestyle{empty}

\pagebreak

\setcounter{tocdepth}{2}
\tableofcontents
\noindent\rule{\textwidth}{1pt}
\vspace{10pt}

\newpage
\section{Introduction}

The space of states lies at the heart of the kinematic information about a quantum system. Even in the finite dimensional case we are far from fully understanding its mathematical structures and their connections to the physics of the system. More so in infinite dimensions, i.e. in the case of quantum field theories. 

One  essential feature of quantum states is entanglement. It plays a crucial role in quantum information theory and beyond that provides ways to characterise quantum fluctuations. For example, the entanglement of the ground state alone can help classifying quantum phases and tell us about possible topological structure \cite{Kitaev:2005dm,Li:2008aa,Jiang:2012aa} or whether a system is close to criticality \cite{Amico:2007ag}. Therefore measures of entanglement of quantum states play a crucial role in describing the structure of state spaces.  

Another standard way to understand these structures is the development of methods to compare different states. Quickly one comes to realize that even if the microscopic realization of two states is quite different their meso- or macroscopic features might be very similar. An immediate example are different energy eigenstates. One can also go the opposite way. Imagine two states with macroscopically very similar features, they e.g. share the same energy. How deep do we have to dig to see the difference in these states, or in other words how distinguishable are they? 

Mathematical measures of distinguishability can attach a lot of structure to the space of states. Ideally this structure has physical significance, i.e. it helps to explain physical phenomena. 
For instance, distinguishability measures help to put the Eigenstate Thermalization Hypothesis \cite{srednicki1996thermal,Deutsch:1991,rigol2008thermalization} on a more quantitative footing, and, as another example, they should govern the `indistinguishability' of black hole microstates in AdS \cite{Strominger:1996sh,Strominger:1997eq}.

We here want to investigate some entanglement and distinguisability measures in the context of two dimensional conformal field theory. The latter are among the best understood and most studied quantum field theories, play a crucial role in the perturbative description of string theory and appear as fixed points of renormalization group flow such that they describe the dynamics of statistical and condensed matter systems at criticality. In some cases they can even be solved exactly \cite{Belavin:1984vu} and under certain conditions -- the case of rational theories with a finite number of primary operators -- all possible CFTs have been classified \cite{Cappelli:1986hf}.
Their huge amount of symmetry allows to explicitly compute partition and correlation functions as well as their conformal transformation rules. It is not a coincidence that all the measures we will use can be computed by particularly transformed correlation functions. 

We put our focus on so-called descendant states -- states excited by Virasoro generators -- on a circle of length $L$. Then we consider subsystems of size $l < L$ onto which we reduce the pure states of the full system. How to compute entanglement for these kind of construction was shown in \cite{Palmai:2014jqa,Taddia:2016dbm}. We will use similar methods to also compute distinguishability measure for these reduced density matrices. 

As will become clear when we introduce the methods to compute the entanglement and distinguishability measures, it is in principle possible to compute algebraic expressions for any descendant, in particular for descendants of the vacuum. In practice, the algebraic expressions become cumbersome and are easier to tackle by computer algebra programs. We use Mathematica for our computations and explicitly display important parts of our code in the appendices. The notebooks with the remaining code are openly accessible. The heart of the code is a function that implements a recursive algorithm to compute generic correlators of descendants. In case of vacuum descendants it results in an analytic expression of the insertion points and the central charge of the theory. In case of descendants of arbitrary primary states the function returns a differential operator acting on the respective primary correlator. 

With this tool at hand, we are able to compute, for instance, the Sandwiched R\'enyi Divergence (SRD) and the Trace Squared Distance (TSD) which have not been computed for descendant states before. In case of the R\'enyi entropy we can expand on existing results. The outcomes for the SRD for example allow us to test a generalisation of the quantum null energy condition suggested in \cite{Lashkari:2018nsl}. Results that we compute for vacuum descendants are universal and, in particular, can be studied at large central charge, i.e. the regime where two dimensional conformal field theories may have a semi-classical gravitational dual in $AdS_3$.
We will show results for vacuum descendant states in this limit.

We will organise the paper as follows. In section \ref{sec:CFTtec} we review all the CFT techniques that we need later. In the following section \ref{sec:qmeasures} we discuss the quantum measures that we want to compute, namely the R\'enyi entanglement entropy as a measure of entanglement, and the sandwiched R\'enyi divergence and the trace square distance as measures of distinguishability between states reduced to a subsystem. In section \ref{sec:universal} we focus on results for descendants of the vacuum. These will apply to all theories with a unique vacuum and, hence, we call them universal. In particular these results can be computed explicitly up to rather high excitation. In the following section \ref{sec:nonuniversal} we show the results for descendants of generic primary states. These results depend on the primary correlators that are theory dependent and, hence, are non-universal. Therefore we compute results in two explicit models, namely the critical Ising model the three-state Potts model.

\section{Review of some CFT techniques}\label{sec:CFTtec}

\subsection{Notation and definitions} 

We want to introduce a notation for the states and fields appearing in our expressions. Consider the Virasoro representation $R_p$, whose primary state has conformal dimension $\Delta = h+ \bar{h}$, with the chiral and anti-chiral conformal weights $h,\bar{h}$, and is denoted by $\ket{\Delta}$. Chiral descendant states are written as $\ket{\Delta,\{(m_i,n_i)\}} = \prod_i L_{-m_i}^{n_i}\ket{\Delta}$, with the chiral copy of the Virasoro generators $L_m$. For anti-chiral descendants one simply uses the anti-chiral copy of the Virasoro algebra. Any state in $R_p$ can be written as a linear combination of the latter states. 

In two-dimensional CFT the operator-state correspondence holds, where the operators are local quantum fields on the space-time of the theory. For any state $\ket{s}$ we denote the respective field as $f_{\ket{s}}$. The primary field that corresponds to the primary state $\ket{\Delta}$ is then $f_{\ket{\Delta}}$. Descendant fields are given by
\begin{equation}
    f_{ \ket{\Delta,\{(m_i,n_i)\}}} = \prod_i \hat{L}_{-m_i}^{n_i} f_{\ket{\Delta}}\,,
\end{equation}
where 
\begin{equation}\label{eq:desfield}
    \hat{L}_{-m} g(w) := \oint_{\gamma_w} \frac{dz}{2\pi i} \frac{1}{(z-w)^{m-1}} T(z) g(w) 
\end{equation}
for any field $g$; $\gamma_w$ is a closed path surrounding $w$. $\hat{L}_{-m} g(w)$ is the $m$th `expansion coefficient' in the OPE of the energy momentum tensor $T$ with the field $g$. 

A field's dual is the field that corresponds to the dual vector. We denote the field dual to $f_{\ket{s}}(z,\bar{z})$ by
\begin{equation}
    f_{\bra{s}}(\bar{z},z) := \left(f_{\ket{s}}(z,\bar{z})\right)^\dagger\,.
\end{equation}

\noindent
Note that it is most naturally defined on the complex plane.

The duality structure of the Hilbert space is fixed by the definitions $L_{-n}^\dagger = L_{n}$ and $\bra{\Delta}\Delta'\rangle = \delta_{\Delta,\Delta'}$. This structure needs to be recovered from the two point function of the respective fields when the two points coincide, i.e 
\begin{equation}\label{eq:contraint0}
    \bra{s}s'\rangle \equiv \lim_{z\to w}\left\langle  f_{\bra{s}}(\bar{z},z) f_{\ket{s'}}(w,\bar{w}) \right\rangle\,.
\end{equation}

\noindent
To achieve this one chooses radial quantization around the second insertion point $w$ and defines the dual field $f_{\bra{s}}(\bar{z},z)$ as the outcome of the transformation $G(z)=\frac{1}{z-w}+w$ of the field $f_{\ket{s}}(z,\bar{z})$ at the unit circle surrounding $w$.
With the help of the transformation rules that we define in the following section \ref{sec:trafo} we can therefore write
\begin{equation}\label{eq:DualFeld}
    f_{\bra{s}}(\bar{z},z) = f_{\Gamma_{G} \ket{s}}\left(\frac1{z-w}+w,\frac1{\bar{z}-\bar{w}}+\bar{w}\right)\,,
\end{equation}
where the action $\Gamma_G$ on the local Hilber space takes the simple form 
\begin{equation}
    \Gamma_G = \left(-\frac{1}{(z-w)^2}\right)^{L_0}\left(-\frac{1}{(\bar{z}-\bar{w})^2}\right)^{\bar{L}_0} \exp\left(\frac{L_1}{w-z}+\frac{\bar{L}_1}{\bar{w}-\bar{z}}\right)\,.
\end{equation}

\noindent
In what follows we will use radial quantization around the origin of the complex plane, i.e. we will choose $w=0$. Note, that \eqref{eq:DualFeld} gives \eqref{eq:contraint0} up to a phase factor $(-1)^{S_p}$, where $S_p$ is the conformal spin of the primary state $\ket{s}$ is built from.

\subsection{Transformation of states and fields}
\label{sec:trafo}

The transformation rule for arbitrary chiral fields was first presented in \cite{Gaberdiel:1994fs}. We will, however, use the (equivalent) method introduced in \cite{frenkel2004vertex} (section 6.3). 

There is a natural action $M(G)$ of a conformal transformation $G$ on any Virasoro module and, hence, on the full space of states. For a field $f_{\ket{s}}(w)$ we need to know how the transformation acts locally around $w$ and transform the field accordingly. It works as follows:

Consider a conformal transformation $G$ and choose local coordinates around the insertion point $w$ and the point $G(w)$. The induced local coordinate change can be written as $\mathfrak{g}(z) = \sum_{k=1}^\infty a_k z^k$, where $z$ are the local coordinates around $w$ that are mapped to the local coordinates $\mathfrak{g}(z)$ around $G(w)$. Now solve the equation 
\begin{equation}
    v_0 \exp\left(\sum_{j=1}^\infty v_j t^{j+1}\partial_t\right)t = \mathfrak{g}(t)
\end{equation}
for the coefficients $v_j$ order by order in $t$. The local action of $G$ on the module is then given by $M(G) := \exp\left(-\sum_{j=1}^\infty v_j L_j\right) v_0^{-L_0}$. The inverse, that we will rather use, is then given by
\begin{equation}
    \Gamma := M(g)^{-1} = v_0^{L_0} \exp\left(\sum_{j=1}^\infty v_j L_j\right)\,,
\end{equation}
such that we can write
\begin{equation}
    f_{\ket{s}}(G(w)) = f_{\ket{s'} = \Gamma \ket{s}}(w)\,.
\end{equation}

\noindent
Note that for a descendant at level $k$ we only need the coefficients $v_j$ up to $j=k$. A Mathematica code to obtain the relation between the coefficients $v_j$ and $a_k$ is given in appendix~\ref{app:matv}.

\subsection{Computing correlation functions of descendant fields on the plane}

We will be interested in computing correlation functions 
\begin{equation}
    \langle \prod_{i=1}^N f_{\ket{s_i}}(z_i) \rangle \, ,
\end{equation}
where $\ket{s_i}$ are some descendant states. 

To get a handle on them we use Ward identities in a particular way. Therefore, consider a meromorphic function $\rho(z)$ that has singularities at most at $z\in \left\{z_i\right\}\cup \{0,\infty\} $, i.e. at the insertion points and at the singular points of the energy momentum tensor. Let us make the particular choice
\begin{equation}
    \rho(z) = \prod_{i=1}^N (z-z_i)^{a_i}
\end{equation}
for $a_i\in\mathbb{Z}$, which is in particular regular at $0$. Now, consider the integral identity
\begin{equation}
    \sum_{i=1}^N \oint_{\gamma_{z_i}} \frac{dz}{2\pi i} \rho(z) \left\langle T(z) g_i(z_i)  \prod_{j\neq i} g_j(z_j) \right\rangle 
    = - \oint_{\gamma_\infty} \frac{dz}{2\pi i} \rho(z) \left\langle T(z) \prod_{j=1}^N g_j(z_j) \right\rangle\,,  
\end{equation}
where $g_j$ are arbitrary fields, e.g. descendant fields. The latter identity simply follows from deforming the integral contour accordingly. The r.h.s. vanishes for $\sum_{i=1}^N a_i \le2$. Next, we consider the functions
\begin{equation}
    \rho_i(z) := \prod_{j\neq i} (z-z_j)^{a_j} = \frac{\rho(z)}{(z-z_i)^{a_i}}
\end{equation}
for which we need the expansion around $z_i$,
\begin{equation}
    \rho_i(z) \equiv \sum_{n=0}^\infty \rho_i^{(n)} \, (z-z_i)^n\,.
\end{equation}

\noindent
Note, that the expansion coefficients $\rho_i^{(n)}$ are some rational expressions that depend on all $z_j\neq z_i$ and $a_j$.

Now, using the definition of $\hat{L}_m$, \eqref{eq:desfield}, and the latter expansion we obtain
\begin{equation}
    \sum_{i=1}^N \sum_{n=0}^\infty \rho_i^{(n)} \left\langle\left(\hat{L}_{a_i+n-1}g_i(z_i)\right)   \prod_{j\neq i} g_j(z_j)\right\rangle = 0\,\label{eq:WardIdNpt}
\end{equation}
for $\sum a_i\leq 2$. Note that, even if not written explicitly, the sums over $n$ do always terminate for descendant fields $g_i$. Note further that these relations among correlation functions depend on the choice of $a_i$ but the correlators that can be computed from these relations are unique.

\subsubsection{Example for particular choices and explicit recursive formula}

One very immediate choice is $a_i = 1-m$ and $a_{j\neq i}=0$ which gives the relation
\begin{align}
    \left\langle \left(\hat{L}_{-m}g_i(z_i)\right) \prod_{j\neq i}g_j(z_j)\right\rangle = - \sum_{j\neq i} \sum_{n=0}^{\text{lvl}(g_j)+1} \rho_j^{(n)} \left\langle \left(\hat{L}_{n-1} g_j(z_j)\right) \prod_{k\neq j}g_k(z_k) \right\rangle \label{eq:rec1}
\end{align}
with 
\begin{equation}
    \rho_j^{(n)} = (-1)^{n}\binom{n+m-2}{n} (z_j-z_i)^{1-m-n}\,.
\end{equation}

\noindent
For $m>1$ we see that the total level of each correlator on the r.h.s., i.e. the sum over all levels of fields appearing in the correlation functions, is lower than the one on the l.h.s. We, hence, can express correlation functions of higher total level by correlators of lower total level. One way of computing correlation functions of descendants is using the above formula recursively until there are only $L_{-1}$ left. These simply act as derivative operators on the respective primary. 

The Mathematica code that uses above equation recursively and computes arbitrary correlation functions of vacuum descendants is given in appendix \ref{app:VacDesCorr}. It produces an algebraic expression of the insertion points and the central charge $c$. The Mathematica code to compute correlation function for descendants of generic primary fields is given in appendix \ref{app:PrimDesCorr}. It produces a derivative operator that acts on the respective primary correlator, which in general is theory dependent. 

\section{Review of some quantum measures in CFT}\label{sec:qmeasures}

We want to consider an isolated quantum system living on a circle of length $L$ whose (low-energy) physics is governed by a (1+1)-dimensional effective field theory. At some critical value of its couplings the theory becomes conformal. This is what we want to assume. Then, the system is in some pure state of a (1+1)d CFT, associated with a density matrix $\rho = \ket{s}\bra{s}$. 

Let us further consider a spatial bipartition into a region $A$ of size $l<L$ and its complement $\overline{A}$. Assume a situation where one has no access to the complement, i.e. all measurements are restricted to the subregion $A$. Our ignorance of the complement means that the state in the region we have access to can be  reduced to the density matrix
\begin{equation}
    \rho_A = \Tr_{\overline{A}}\rho\,,
\end{equation}
where $\Tr_{\overline{A}}$ is the partial trace over the degrees of freedom of the complement. In fact, a physically realistic CFT observer can only access a restricted amount of information by measurements which in the present case is modeled by restricting the measurement to a spatial region $A$. 

Our focus of interest lies in reduced density matrices that originate from descendant states of the full system. We, in particular, want to study their entanglement and measures of distinguishability between them. 

\subsection{Entanglement measure: R\'enyi entropy} \label{sec:Renyi}

The $n$th R\'enyi entropy \cite{renyi2012probability,nielsen_chuang_2010} is defined as
\begin{equation}
    S_n(A) = \frac{1}{1-n} \log \Tr_A \rho_A^n\,.
\end{equation}

\noindent 
For $n\to 1$ it converges to the (von Neumann) entanglement entropy $S(A) = -\Tr \rho_A \log\rho_A$ which is the most common entanglement measure \cite{nielsen_chuang_2010}. However, in particular in field theories, there exist alluring analytical tools that make it much easier to compute R\'enyi entropies for $n>1$ than the entanglement entropy. Additionally, many key properties of the entanglement entropy, such as the proportionality of ground state entanglement to the central charge in critical systems and the area law of gapped states, hold for R\'enyi entropies too. In principle, the knowledge of the  R\'enyi entropy for all $n\in \mathbb{N}$ allows to determine all eigenvalues of the reduced density matrix $\rho_A$. 

In the present case, the full system can be described by a CFT on the Euclidean space-time manifold of an infinite cylinder for which we choose complex coordinates $u = x + i \tau$ with $\tau\in\mathbb{R}$ and $x + L \equiv x \in \left(-\frac L2 ,\frac L2\right]$. The variable $\tau$ is regarded as the time coordinate and $x$ is the spatial coordinate. As subsystem $A$ we choose the spatial interval $\left(-\frac{l}2,\frac{l}{2}\right)$\,. In 2d CFT, the trace over the $n$th power of the reduced density matrix $\rho_A = \Tr_{\overline{A}}\ket{s}\bra{s}$ is equivalent to a $2n$-point function on the so-called \textit{replica manifold} which is given by $n$ copies of the cylinder glued together cyclically across branch cuts along the subsystem $A$ at $\tau =0$ \cite{Holzhey:1994we,Calabrese:2009qy}. The exponential map $z(u) = \exp\left(2\pi i u/L\right)$ maps the latter manifold to the $n$-sheeted plane $\Sigma_n$, where the branch cut now extends between $\exp\left(\pm i \pi \frac{l}{L}\right)$\,. The $2n$ fields are those that correspond to the state $\ket{s}$ and its dual $\bra{s}$, where one of each is inserted at the origin of each sheet:
\begin{align}
    \Tr_A \rho_A^n  
                    &= \mathcal{N}_n \left\langle \prod_{k=1}^n f_{\bra{s}}(0_k)f_{\ket{s}}(0_k)\right\rangle_{\Sigma_n}\\
                    &= \mathcal{N}_n \left\langle \prod_{k=1}^n f_{\Gamma_{-1/z}\ket{s}}(\infty_k)f_{\ket{s}}(0_k)\right\rangle_{\Sigma_n}\,.
\end{align}

\noindent
The constant $\mathcal{N}_n =  Z(\Sigma_n)/Z(\mathbb{C})^n = \left(\frac{L}{\pi a} \sin\left(\frac{\pi l}{L}\right)\right)^{\frac{c}{3} \left(n-\frac{1}{n}\right)}$, $Z$ being the partition function on the respective manifold, ensures the normalization $\Tr_A \rho_A =1$, with some UV regulator $a$ (for example some lattice spacing). In the second line we use the definition of the dual state. 

One way to compute the above correlation function is to use a uniformization map from $\Sigma_n$ to the complex plane. It is given by composing a Möbius transformation with the $n$th root,
\begin{equation}\label{eq:uniformization}
     w(z) = \left(\frac{z e^{ -i\pi \frac{l}{L}} - 1}{z -  e^{-i\pi\frac{l}{L}} }\right)^{\frac1n}\,.
\end{equation}

\noindent 
The $2n$ fields are mapped to the insertion points 
\begin{align}
        w(0_k) &= \exp\left(\frac{i \pi l}{ n L}+\frac{2\pi i(k-1)}{n} \right)\label{eq:InsPoints}\\
        w(\infty_k) &= \exp\left(-\frac{i \pi l}{ n L}+\frac{2\pi i (k-1)}{n}\right)\nonumber
\end{align}
on the unite circle, and the fields have to transform as described in section \ref{sec:trafo}. The change of local coordinates is given in \ref{app:uniformization}. The local action is denoted by $\Gamma_{w(z)} \equiv \Gamma_{k,l}$ and for the dual fields we get $\Gamma_{w(1/z)} = \Gamma_{w(z)} \Gamma_{1/z} \equiv \Gamma_{k,-l}$.

Putting all together we see that computing the $n$th R\'enyi entropy is basically equivalent to computing a $2n$ point function of particularly transformed fields: 

\begin{align}\label{eq:RFE}
    e^{(1-n)S_n(A)} = \Tr_A \rho_A^n \equiv \mathcal{N}_n \left\langle \prod_{k=1}^n f_{\Gamma_{k,l} \ket{s}}\left(w(0_k)\right)f_{\Gamma_{k,-l} \ket{s}}\left(w(\infty_k)\right) \right\rangle_{\mathbb{C}}=: \mathcal{N}_n F_{\ket{s}}^{(n)} \,. 
\end{align}

\noindent 
See also \cite{Palmai:2014jqa,Taddia:2016dbm} for derivations of the latter formula. Other computations of the entanglement entropy of excited states (not necessarily descendants) can also be found in \cite{Alcaraz:2011tn,Berganza:2011mh,Mosaffa:2012mz,Bhattacharya:2012mi,Taddia_2013,Caputa:2014vaa,Asplund:2014coa,Nozaki:2014uaa,Caputa:2014eta,Zhang:2020ouz,Zhang:2020txb}.

\subsection{Distance measures}

Distance and other similarity measures between density matrices provide quantitative methods to evaluate how distinguishable they are, where distinguishability in particular refers to the outcome of generic measurements in the different states. There is not a single best measure and not even agreement upon criteria to evaluate different distance measures. Most of them are designed such that they provide the space of (not necessarily pure) states with some additional structure that ideally allows to draw some physically relevant conclusions about the system under consideration. In case of reduced density matrices distance measures quantify how distinguishable they are by measurements confined to the subregion~$A$.  

We want to consider two of these measurements for reduced density matrices in two dimensional CFT. Let us denote the reduced density matrices as $\rho_i = \Tr_{\overline{A}} \ket{s_i}\bra{s_i}$, with $\rho_0 \equiv \Tr_{\overline{A}} \ket{0}\bra{0}$ the reduce density matrix of the vacuum.

\subsubsection{Relative entropy}

The relative entropy between two reduced density matrices $\rho_{1}$ and $\rho_{2}$ is given by 
\begin{equation}\label{eq:RelEntropy}
    S(\rho_{1},\rho_{2}) = \Tr  ( \rho_{1} \log \rho_{1}) - \Tr  ( \rho_{1} \log \rho_{2}) \,.
\end{equation}

\noindent 
It is free from UV divergencies, positive definite and one of the most commonly used distance measures in quantum information, in particular because several other important quantum information quantities are special cases of it, e.g. the quantum mutual information and quantum conditional entropy. The relative entropy also shows to be useful in high energy application when e.g. coupling theories to (semiclassical) gravity. It allows a precise formulation of the Bekenstein bound \cite{Casini_2008}, a proof of the generalized second law \cite{Wall_2010,Wall:2011hj} and the quantum Bousso bound \cite{Bousso:2014sda,Bousso:2014uxa}. It also appears in the context of holography where it can be used to formulate important bulk energy conditions (see e.g. \cite{Lin:2014hva,Lashkari:2014kda,Lashkari:2015hha}). 

However, as in the case of the entanglement entropy there exist no direct analytical tools to compute the relative entropy in generic two-dimensional conformal field theory.  There exist several R\'enyi type generalisations (see e.g. \cite{Lashkari:2014yva,Lashkari:2015dia}) that are more straight forward to compute. We here want to focus on a quite common one called the Sandwiched R\'enyi Divergence. 

\subsubsection*{Sandwiched R\'enyi divergence}\label{sec:SRD}

The Sandwiched R\'enyi Divergence (SRD) between two density matrices $\rho_1$ and $\rho_2$ is given by 
\begin{equation}
    \mathcal{S}_n(\rho_{1},\rho_{2}) = \frac{1}{n-1} \log \Tr \left(\rho_1^{\frac{1-n}{2n}} \rho_2 \rho_1^{\frac{1-n}{2n}}\right)^n\,.\label{eq:SRD}
\end{equation} 

\noindent 
It is a possible one-parameter generalization of the relative entropy \eqref{eq:RelEntropy}, with the parameter $n \in[\frac12,\infty)$ and $S(\rho_{1},\rho_{2}) \equiv \mathcal{S}_{n\to1} (\rho_{1},\rho_{2})$\,. The SRD by itself has been shown to enjoy important properties of a measure of distinguishability of quantum states. It is, in particular, positive for all states, unitarily invariant, and decreases under tracing out degrees of freedom \cite{Mueller-Lennert:2013,Wilde:2014eda,Frank_2013,Beigi_2013}.

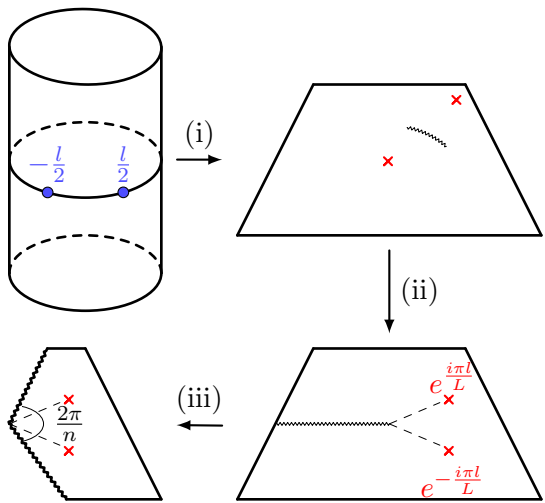
\begin{figure}[t]
\centering
\definecolor{ududff}{rgb}{0.30196078431372547,0.30196078431372547,1}
\begin{tikzpicture}[line cap=round,line join=round,>=triangle 45,x=1cm,y=1cm,scale=1]

\begin{scope}[local bounding box = cylinder]
\draw [rotate around={178.65449714894862:(1,3)},line width=1pt] (1,3) ellipse (1.0008289996875666cm and 0.4998965358776981cm);
\draw [line width=1pt] (2,3)-- (2,0);
\draw [line width=1pt] (2,0) arc(0:-180:1cm and 0.5cm);
\draw [dashed, line width=1pt] (2,0) arc(0:180:1cm and 0.5cm);
\draw [line width=1pt] (2,1.5) arc(0:-180:1cm and 0.5cm);
\draw [dashed, line width=1pt] (2,1.5) arc(0:180:1cm and 0.5cm);
\draw [line width=1pt] (0,3)-- (0,0);
\draw [fill=ududff] (0.5040500631758853,1.068897766899791) circle (2pt) node[anchor=south,color=ududff] {$-\frac{l}{2}$};
\draw [fill=ududff] (1.5,1.07) circle (2pt) node[anchor=south,color=ududff] {$\frac{l}{2}$};
\end{scope}

\begin{scope}[local bounding box = plane, shift={(5,1.5)}]
\draw[decoration = {zigzag,segment length = 0.5mm, amplitude = 0.2mm},decorate,line width=0.25pt] (0.75,0.18) arc(49.03:79.29:1.1cm);
\draw [line width=1pt] (-2,-1)-- (-1,1);
\draw [line width=1pt] (-1,1)-- (1,1);
\draw [line width=1pt] (1,1)-- (2,-1);
\draw [line width=1pt] (-2,-1)-- (2,-1);
\draw [color=red,thick] (0,0) -- ++(-2pt,-2pt) -- ++(3pt,3pt) ++(-3pt,0) -- ++(3pt,-3pt);
\draw [color=red,thick] (0.9,0.8133333333333327) -- ++(-2pt,-2pt) -- ++(3pt,3pt) ++(-3pt,0) -- ++(3pt,-3pt);
\end{scope}

\draw[shorten >=2mm,shorten <=2mm, thick,-latex] (cylinder) -- node[above]{(i)} (plane);

\begin{scope}[local bounding box = moebius, shift={(5,-2)}]
\draw [line width=1pt] (-2,-1)-- (-1,1);
\draw [line width=1pt] (-1,1)-- (1,1);
\draw [line width=1pt] (1,1)-- (2,-1);
\draw [line width=1pt] (-2,-1)-- (2,-1);
\draw[decoration = {zigzag,segment length = 0.5mm, amplitude = 0.2mm},decorate,line width=0.25pt] (0,0) -- (-1.5,0);
\node at (0.8,-0.34) (a) {};
\node at (0.8,0.34) (aa) {};
\draw [color=red,thick] (0.8,-0.34) -- ++(-2pt,-2pt) -- ++(3pt,3pt) ++(-3pt,0) -- ++(3pt,-3pt) node[anchor=north, color=red] {$e^{-\frac{i \pi l}{L}}$};
\draw[dashed, line width=0.25pt] (0,0) -- (a);
\draw [color=red,thick] (0.8,0.34) -- ++(-2pt,-2pt) -- ++(3pt,3pt) ++(-3pt,0) -- ++(3pt,-3pt) node[anchor=south, color=red] {$e^{\frac{i \pi l}{L}}$};
\draw[dashed, line width=0.25pt] (0,0) -- (aa);
\end{scope}

\draw[shorten >=2mm,shorten <=2mm, thick,-latex] (plane) --node[right]{(ii)} (moebius);

\begin{scope}[local bounding box = moebius_negpow, shift={(0,-2)}]
\draw [line width=1pt] (0.5,1)-- (1,1);
\draw [line width=1pt] (1,1)-- (2,-1);
\draw [line width=1pt] (0.75,-1)-- (2,-1);
\draw[decoration = {zigzag,segment length = 0.8mm, amplitude = 0.2mm},decorate,line width=1pt] (0,0) -- (0.5,1);
\draw[decoration = {zigzag,segment length = 0.8mm, amplitude = 0.2mm},decorate,line width=1pt] (0,0) -- (0.75,-1);
\node at (0.8,-0.34) (a) {};
\node at (0.8,0.34) (aa) {};
\draw [color=red,thick] (0.8,-0.34) -- ++(-2pt,-2pt) -- ++(3pt,3pt) ++(-3pt,0) -- ++(3pt,-3pt);
\draw[dashed, line width=0.25pt] (0,0) -- (a);
\draw [color=red,thick] (0.8,0.34) -- ++(-2pt,-2pt) -- ++(3pt,3pt) ++(-3pt,0) -- ++(3pt,-3pt);
\draw[dashed, line width=0.25pt] (0,0) -- (aa);
\begin{scope}
    \clip (0.17,-0.25) rectangle (0.5,0.25);
    \draw[line width=0.25pt] (0,0) ellipse (0.45cm and 0.25cm);
\end{scope}
\node[anchor=west] at (0.45,0) {$\frac{2\pi}{n}$}; 
\end{scope}

\draw[shorten >=2mm,shorten <=2mm, thick, -latex] (moebius) -- node[above]{(iii)} (moebius_negpow);

\end{tikzpicture}
    
\caption{Pictorial representation of the geometric setting for the SRD. (i) The reduced density matrix is represented by the sheet with respective operator insertions (red crosses) at 0 and $\infty$. (ii) A M\"obius transformation maps the insertion points to $e^{\pm \frac{i\pi l}{L}}$ and the branch cut to the negative real line. (iii) The multiplication by negative fractional powers of the reduced vacuum states is given cutting out respective parts of the sheet.}
\label{fig:SRD}
\end{figure}

In particular due to the negative fractional power of $\rho_1$, there is no general method known to compute the SRD for arbitrary states in CFT. However, if $\rho_1$ is the reduced density matrix of the theory's vacuum then there is a technique introduced in \cite{Lashkari:2018nsl} to express it in terms of correlation functions. Let us remind that the reduced density matrix for a sub-system on the cylinder is represented by a sheet of the complex plane with a brunch cut along some fraction of the unit circle with the respective operator insertions at the origin and at infinity of that sheet. In case of the vacuum the corresponding operator is the identity and, hence, we regard it as no operator insertion. Multiplication of reduced density matrices is represented by gluing them along the branch cut. Now, let us consider the M\"obius transformation
\begin{equation}\label{eq:Moebius}
   w(z) =  \frac{z e^{ -i\pi \frac{l}{L}} - 1}{z- e^{-i\pi\frac{l}{L}} }\,,
\end{equation}
which in particular maps the two insertions points $0$ and $\infty$ of a sheet to $e^{\pm \frac{i\pi l}{L} }$ and the cut to the negative real axis on every sheet. Now, the reduced density operators can be regarded as operators acting on states defined on the negative real axis by rotating them by $2\pi$ and exciting them by locally acting with the respective operators at $e^{\pm \frac{i\pi l}{L} }$. In case of the vacuum reduced density matrix this now allows to define fractional powers by rotating by a fractional angle and even negative powers by rotating by \textit{negative} angles which basically means removing a portion of the previous sheet. The latter is, however, only possible if no operator insertion is removed. In the present case, the negative power $\frac{1-n}{2n}$ corresponds to an angle $-\pi + \frac{\pi}{n}$. Hence, this construction only makes sense for $\frac{l}{L}< \frac{1}{n}$.\footnote{In \cite{Moosa:2020jwt} the interested reader can find arguments why this is not simply an artifact of the CFT construction but holds generally when one assumes that the state is prepared from a Euclidean path integral.} If this requirement holds then $\rho_0^{\frac{1-n}{2n}} \rho_2 \rho_0^{\frac{1-n}{2n}}$ can be interpreted as a part of the complex plane between angles~$\pm \frac{\pi}{n}$ with operator insertions at angles $\pm\frac{\pi l}{L}$. This procedure is pictorially presented in figure \ref{fig:SRD}. Finally, taking the cyclic trace of $n$ copies of it means gluing $n$ of these regions onto each other which results in a $2n$ point function on the complex plane:
\begin{align}
   \mathcal{F}^{(n)}_{\ket{s}} := \Tr\left(\rho_0^{\frac{1-n}{2n}} \rho_2 \rho_0^{\frac{1-n}{2n}}\right)^n = \left\langle \prod\limits_{k=0}^{n-1} f_{\Gamma_{k,l} \ket{s}}\left(e^{\frac{i\pi l}{L} + \frac{2\pi i k}{n}}\right)f_{\Gamma_{k,-l} \ket{s}}\left(e^{-\frac{i\pi l}{L} + \frac{2\pi i k}{n}}\right) \right\rangle_\mathbb{C} \label{eq:SRDcorr}
\end{align}
where, in contrast to the previous and following section, $\Gamma_{k,l}$ is the local action of the above M\"obius transformation $w(z)$ followed by a rotation $e^{\frac{2\pi i k}{n}}$ to obtain the correct gluing. As before, for the dual field one has to consider $w(1/z)$ which is done by replacing $l\to-l$\,. 

We, here, want to take the opportunity to give an explicit example of the connection between rather formal definitions of distinguishability measures and physical features of a theory. The latter is the Quantum Null Energy Condition (QNEC) which follows from the so-called Quantum Focusing Conjecture \cite{Bousso:2015mna}. The QNEC gives a lower bound on the stress-energy tensor in a relativistic quantum field theory that depends on the second variation of entanglement of a subregion. The QNEC can also be formulated solely in terms of quantum information theoretical quantities and has been shown to be equivalent to positivity of the second variation of relative entropies \cite{Leichenauer:2018obf}. After the QNEC has been proven in free and holographic theories \cite{Bousso:2015wca,Koeller:2015qmn,Malik:2019dpg} it has since been shown to hold quite generally in the context of Tomita-Takesaki modular theory \cite{Balakrishnan:2017bjg,Ceyhan:2018zfg}. Recently a generalized version of QNEC has been suggested in \cite{Lashkari:2018nsl} and later proven to be true in free theories in dimensions larger than two \cite{Moosa:2020jwt}. This generalization may be called `R\'enyi Quantum Null Energy Condition' and is formulated as the positivity of the second variation of sandwiched R\'enyi entropies. The diagonal part of the second variation is simply given by the second derivative of the SRD with respect to the subsystem size. Hence, the R\'enyi Quantum Null Energy Condition can only be true in a theory if any SRD is a convex function of the subsystem size. We will explicitly check if this is true in our results. 

\subsubsection{Trace square distance}

The Trace Square Distance (TSD) between two reduced density matrices is given by 
\begin{equation}
    T^{(2)}(\rho_{1},\rho_{2}) := \frac{\Tr|\rho_{1} -\rho_{2}|^2}{\Tr \rho_{0}^2} = \frac{\Tr \rho_1^2 + \Tr \rho_2^2 -2\Tr\rho_1\rho_2}{\Tr \rho_{0}^2}\,,
\end{equation}
where the factor $\Tr \rho_{0}^2$ in particular removes any UV divergences and allows to directly express the trace square distance in terms of four-point functions on the two-sheeted surface $\Sigma_2$ (see also \cite{Sarosi:2016oks}),
\begin{align}
     T^{(2)}(\rho_{1},\rho_{2}) \equiv\quad& \left\langle f_{\bra{1}}(0_1)f_{\ket{1}}(0_1)f_{\bra{1}}(0_2)f_{\ket{1}}(0_2)\right\rangle_{\Sigma_2} \\  
     +& \left\langle f_{\bra{2}}(0_1)f_{\ket{2}}(0_1)f_{\bra{2}}(0_2)f_{\ket{2}}(0_2)\right\rangle_{\Sigma_2} \nonumber\\
     -& 2  \left\langle f_{\bra{1}}(0_1)f_{\ket{1}}(0_1)f_{\bra{2}}(0_2)f_{\ket{2}}(0_2)\right\rangle_{\Sigma_2}\,.\nonumber
\end{align}

\noindent
Using the uniformization map \eqref{eq:uniformization} with $n=2$ we can express it in terms of four-point functions on the complex plane,
\begin{align} \label{eq:TSDcorr}
    T^{(2)}(\rho_{1},\rho_{2}) \equiv\quad& \left\langle f_{\Gamma_{1,-l}\ket{1}}\left(e^{-\frac{i\pi l}{2L}}\right)f_{\Gamma_{1,l}\ket{1}}\left(e^{\frac{i\pi l}{2L}}\right)f_{\Gamma_{2,-l}\ket{1}}\left(-e^{-\frac{i\pi l}{2L}}\right)f_{\Gamma_{2,l}\ket{1}}\left(-e^{\frac{i\pi l}{2L}}\right)\right\rangle_{\mathbb{C}} \\  
     + &\left\langle f_{\Gamma_{1,-l}\ket{2}}\left(e^{-\frac{i\pi l}{2L}}\right)f_{\Gamma_{1,l}\ket{2}}\left(e^{\frac{i\pi l}{2L}}\right)f_{\Gamma_{2,-l}\ket{2}}\left(-e^{-\frac{i\pi l}{2L}}\right)f_{\Gamma_{2,l}\ket{2}}\left(-e^{\frac{i\pi l}{2L}}\right)\right\rangle_{\mathbb{C}} \nonumber\\
     -& 2  \left\langle f_{\Gamma_{1,-l}\ket{1}}\left(e^{-\frac{i\pi l}{2L}}\right)f_{\Gamma_{1,l}\ket{1}}\left(e^{\frac{i\pi l}{2L}}\right)f_{\Gamma_{2,-l}\ket{2}}\left(-e^{-\frac{i\pi l}{2L}}\right)f_{\Gamma_{2,l}\ket{2}}\left(-e^{\frac{i\pi l}{2L}}\right)\right\rangle_{\mathbb{C}}\,.\nonumber
\end{align}

\noindent 
The trace square distance is manifestly positive and has the great advantage that we can compute it directly in terms of four-point correlators, i.e. there is no need to consider higher sheeted replica manifolds and we do not need to take any analytic continuations. 

Different trace distances between (not necessarily descendant) states in 2d CFT have e.g. be considered in \cite{Sarosi:2016oks,Zhang:2019wqo,Zhang:2019itb}. 

\section{Universal results from the vacuum representation} \label{sec:universal}

Most physically interesting conformal field theories contain a unique vacuum that naturally corresponds to the identity field. For the vacuum all the above correlation functions to compute the quantum measures become basically trivial. However, the theories also contain the whole vacuum representation which for example consists of the state $L_{-2}\ket{0}$ that corresponds to the holomorphic part of the energy momentum tensor, $T(z)$. Correlation functions of  vacuum descendant fields generically depend on the central charge of the theory and can in principle be computed explicitly using the Ward identities \eqref{eq:WardIdNpt} or \eqref{eq:rec1} recursively. 
Since all quantities discussed in section \ref{sec:qmeasures} can be expressed in terms of correlators, we can in principle compute all of them as closed form expressions, too. 
However, since we use computer algebra to perform the transformations and compute the correlation functions, computer resources are the biggest limiting factor. 
We, here, present results for all descendants up to conformal weight five and in some cases for the state $L_{-10}\ket{0}$\,. We, in particular, want to check how the measures depend on the conformal weights of the states and if states at the same conformal weight can be regarded as similar. 

\subsection{R\'enyi entanglement entropy}\label{sec:renyivac}

Only for the first few excited states in the identity tower, the expressions \eqref{eq:RFE} to compute the second R\'enyi entanglement entropy are compact enough to display them explicitly. In case of the first descendant $L_{-2}\ket{0}$, i.e. the state that corresponds to the energy momentum tensor, we get
\begin{align}\label{eq:RFE[-2]}
    F^{(2)}_{L_{-2}\ket{0}} &= \frac{c^2 \sin ^8(\pi x)}{1024}+\frac{c \sin ^4(\pi x) (\cos (2 \pi x)+7)^2}{1024}+\frac{\sin ^4(\pi x) (\cos (2 \pi x)+7)}{16
   c}\\&\quad +\frac{16200 \cos (2 \pi x)-228 \cos (4 \pi x)+120 \cos (6 \pi x)+\cos (8 \pi x)+16675}{32768}\,,\nonumber
\end{align}
where we defined $x=l/L$\,. The results for the states $L_{-n} \ket{0}$ with $n=3,4,5$ are given in \ref{app:REresultsVac}. The results here agree with those in \cite{Taddia:2016dbm} when present.

One important case is the limit of small subsystem size, i.e. when $x\ll 1$. In this limit to leading order any of the above $2n$-point functions \eqref{eq:RFE} decouple into $n$ $2$-point functions. This is because the operator product of a field and its conjugate includes the identity. Then, in the limit $x\to 0$ the respective identity block dominates and takes the form of a product of $n$ 2-point functions. Those two point functions are, however, given by the transition amplitude from the state to its dual on the $k$th sheet that decouples in the limit $x\to 0$ from all other sheets. The latter is simply given by the squared norm of the state, i.e. it gives one for normalized states. Hence, we can write
\begin{align}
    \lim_{x\to 0} F_{\ket{s}}^{(n)} &= \prod\limits_{k=1}^n \lim_{x\to 0} \langle f_{\Gamma_{k,l} \ket{s}}\left(w(0_k)\right)f_{\Gamma_{k,-l} \ket{s}}\left(w(\infty_k)\right) \rangle_\mathbb{C}\\ 
    &=\prod\limits_{k=1}^n \bra{s}s\rangle = 1\,.
\end{align}

\noindent
Hence, to order $x^0$ the descendant does not play any role at all. For the next to leading order result there are expectations from primary excitations and the change of the entanglement entropy computed from holography. E.g. in \cite{Bhattacharya:2012mi} it is shown that the change should be proportional to the excitation energy and, in particular, should be independent from $c$. Expanding the explicitly shown results \eqref{eq:RFE[-2]},\eqref{eq:RFE[-3]}, \eqref{eq:RFE[-4]}, and \eqref{eq:RFE[-5]} we obtain 
\begin{equation}
    F_{L_{-n}\ket{0}}^{(2)} = 1 - \frac{n}{2}\left(\pi x\right)^2 + O\!\left(x^4\right)\,, \quad \text{for} ~ n = 2,3,4,5\,,\label{eq:RElowx}
\end{equation}
which is in agreement with all above expectations. 

\begin{figure}[t]
    \centering
    \begin{tikzpicture}
    \node at (0,0) {\includegraphics[width=.45\textwidth]{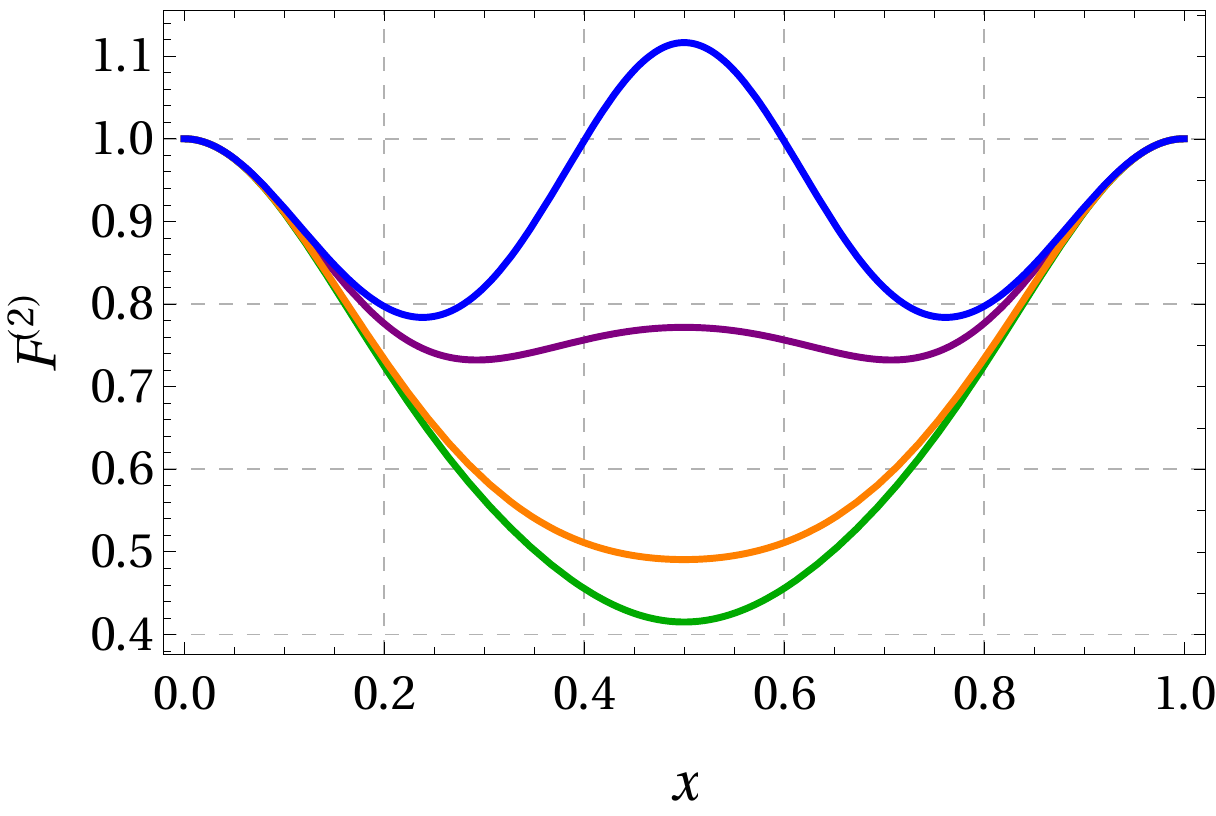}};
    \node at (.47\textwidth,0)  {\includegraphics[width=.45\textwidth]{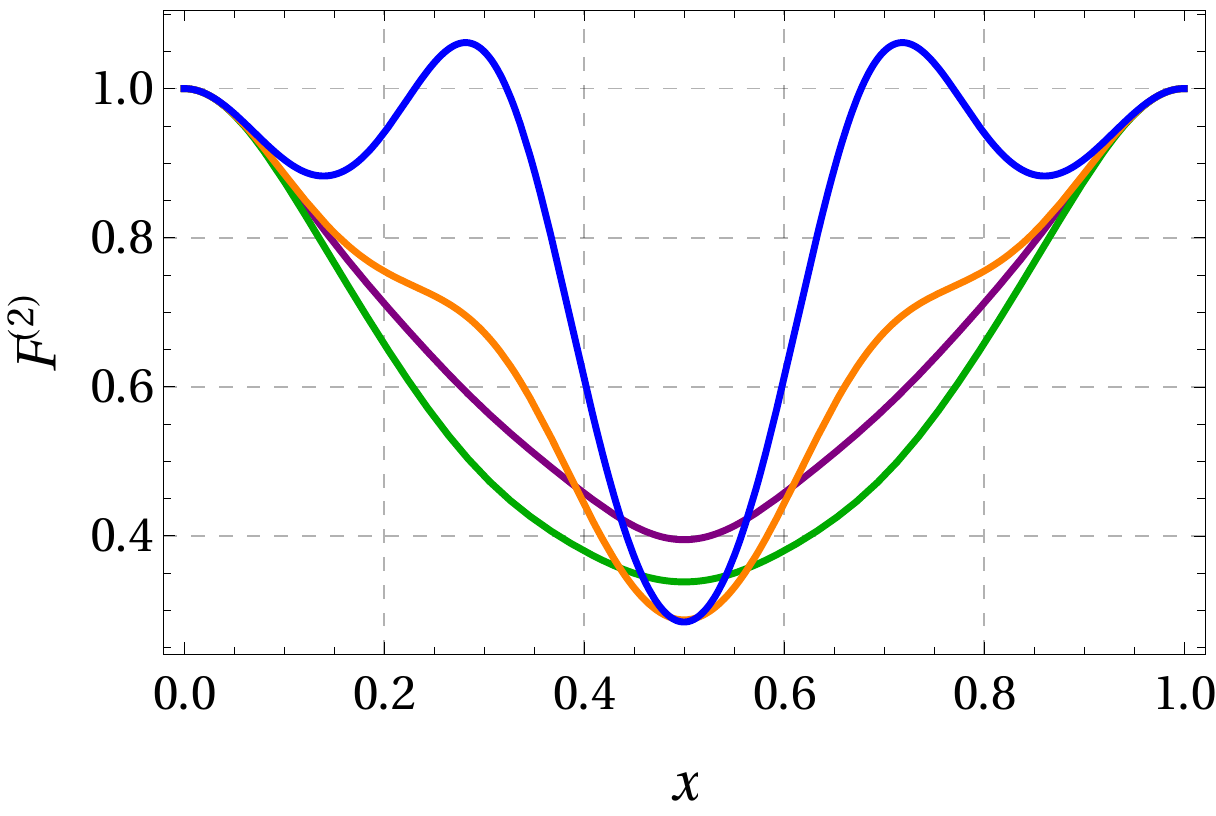}};
    \node at (0,-4.7) {\includegraphics[width=.45\textwidth]{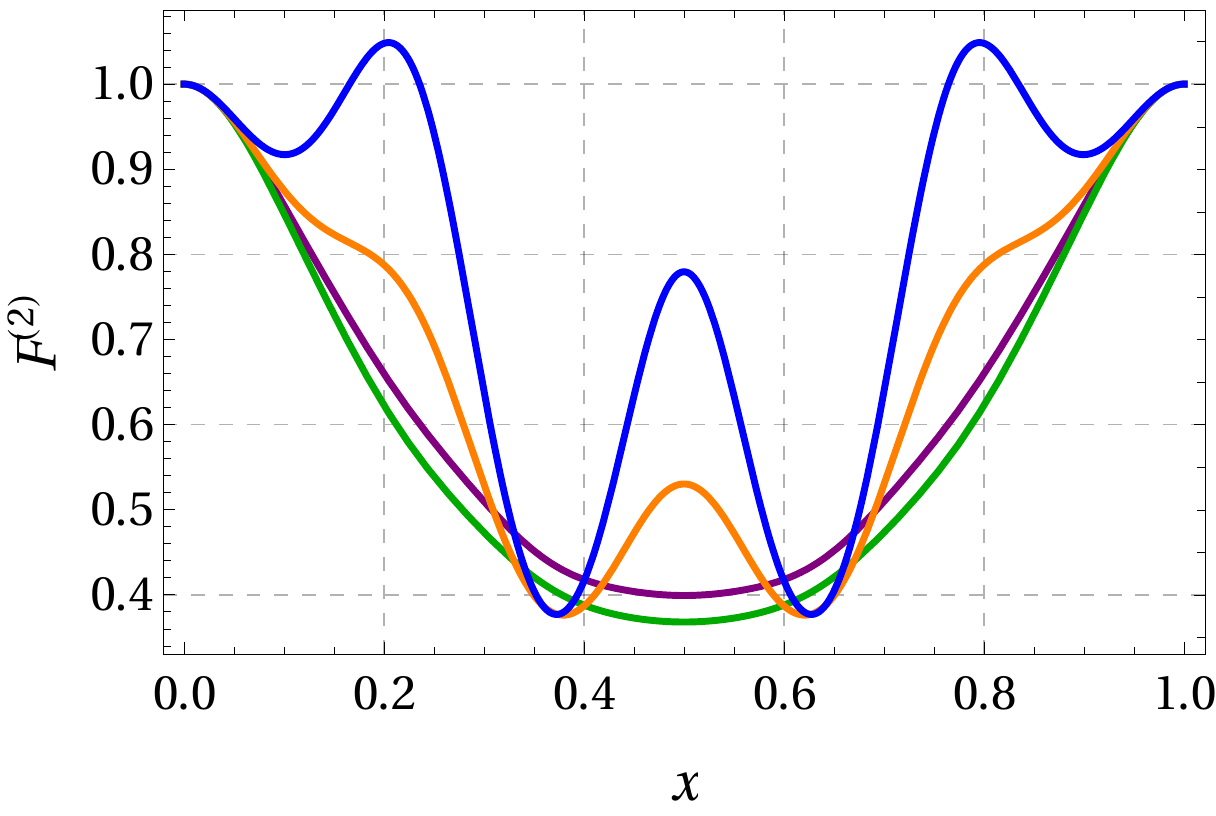}};
    \node at (.47\textwidth,-4.7)  {\includegraphics[width=.45\textwidth]{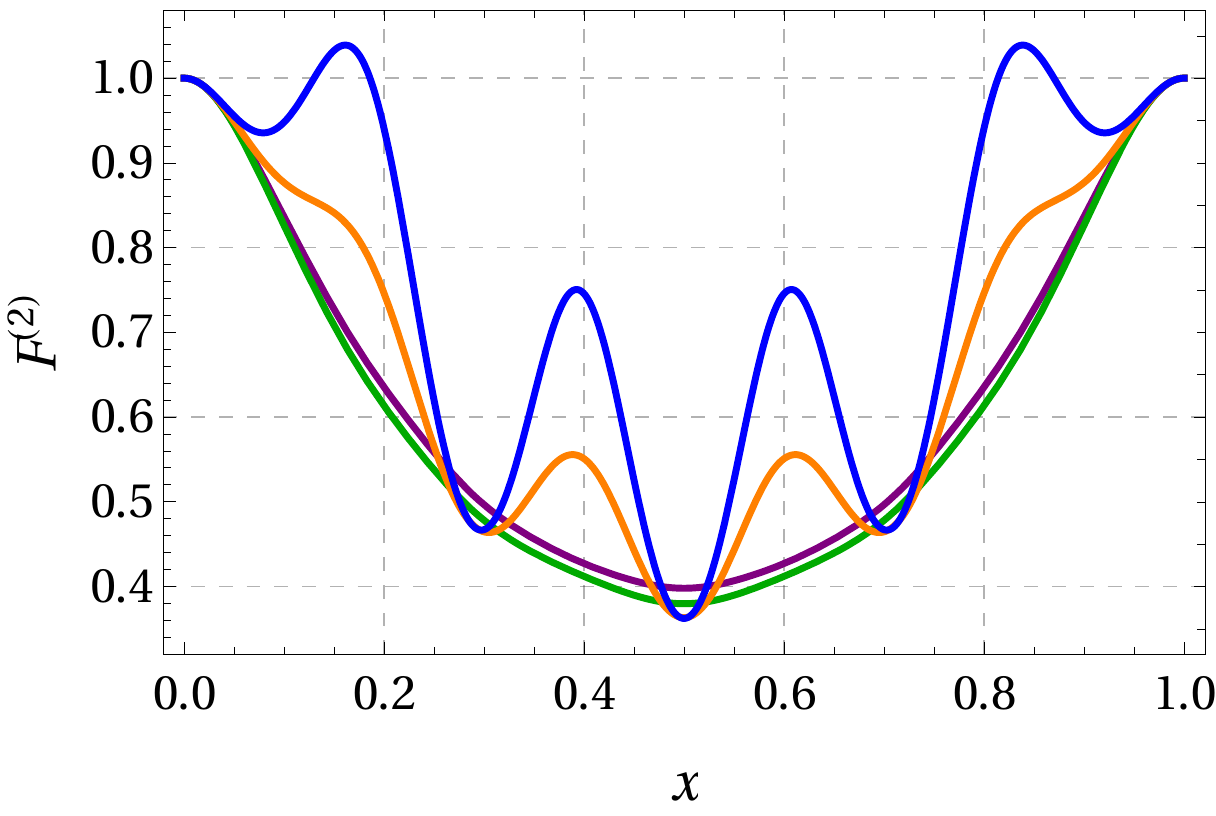}};
    \node at (.7\textwidth,-1.5) {\includegraphics[width=.1\textwidth]{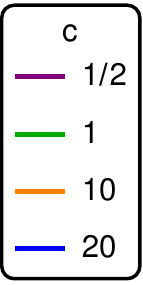}};
    \node at (.4,1.5) {(a)};
    \node at (7.5,1.5) {(b)};
    \node at (.4,-3.2) {(c)};
    \node at (7.5,-3.2) {(d)};
    \end{tikzpicture}
    \caption{The correlator $F^{(2)}_{\ket{s}}$ for (a) $\ket{s} = L_{-2}\ket{0}$, (b) $\ket{s} = L_{-3}\ket{0}$, (c) $\ket{s} = L_{-4}\ket{0}$, (d) $\ket{s} = L_{-5}\ket{0}$, for several values of the central charge.}
    \label{fig:RE2lowA}
\end{figure}

In figure \ref{fig:RE2lowA} we show the results for $F^{(2)}_{\ket{s}}$ for the states $\ket{s} = L_{-n}\ket{0}$, $n=2,3,4,5$\,. The first observation is that at large $c$ the correlator shows an oscillating behaviour with oscillation period proportional to $1/n$. In fact, we can see this also from the explicit results \eqref{eq:RFE[-2]},\eqref{eq:RFE[-3]},\eqref{eq:RFE[-5]},\eqref{eq:RFE[-5]} where at large central charge the term proportional to $c^2$ dominates. 
Note, that the correlator $F^{(n)}$ can become larger than one at large central charge and, hence, its contribution to R\'enyi entropy $S^{(n)}$ can get negative. For example, in case of $n=2$ and $\ket{s} = L_{-2}\ket{0}$ this happens at $x=1/2$ for $c\gtrsim 18.3745$.  

\begin{figure}[t]
    \centering
    \begin{tikzpicture}
    \node at (0,0) {\includegraphics[width=.45\textwidth]{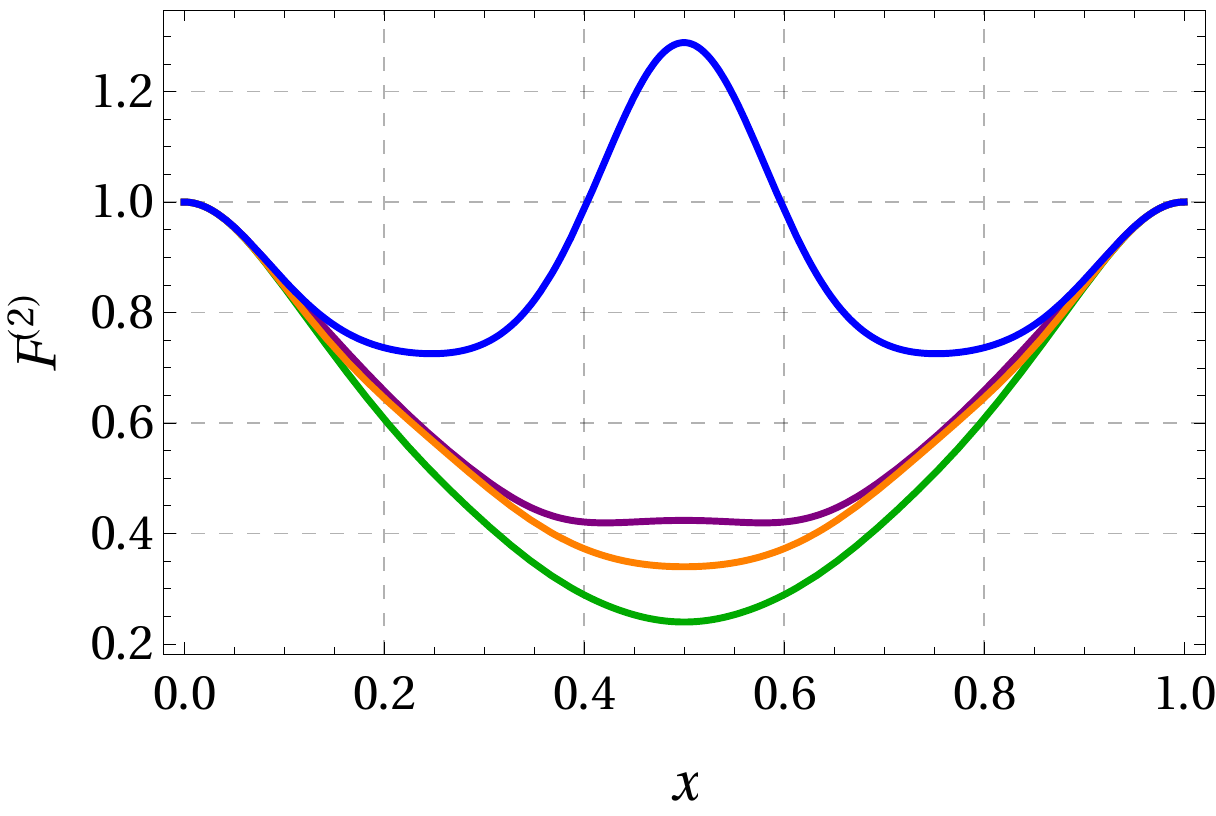}};
    \node at (.47\textwidth,0)  {\includegraphics[width=.45\textwidth]{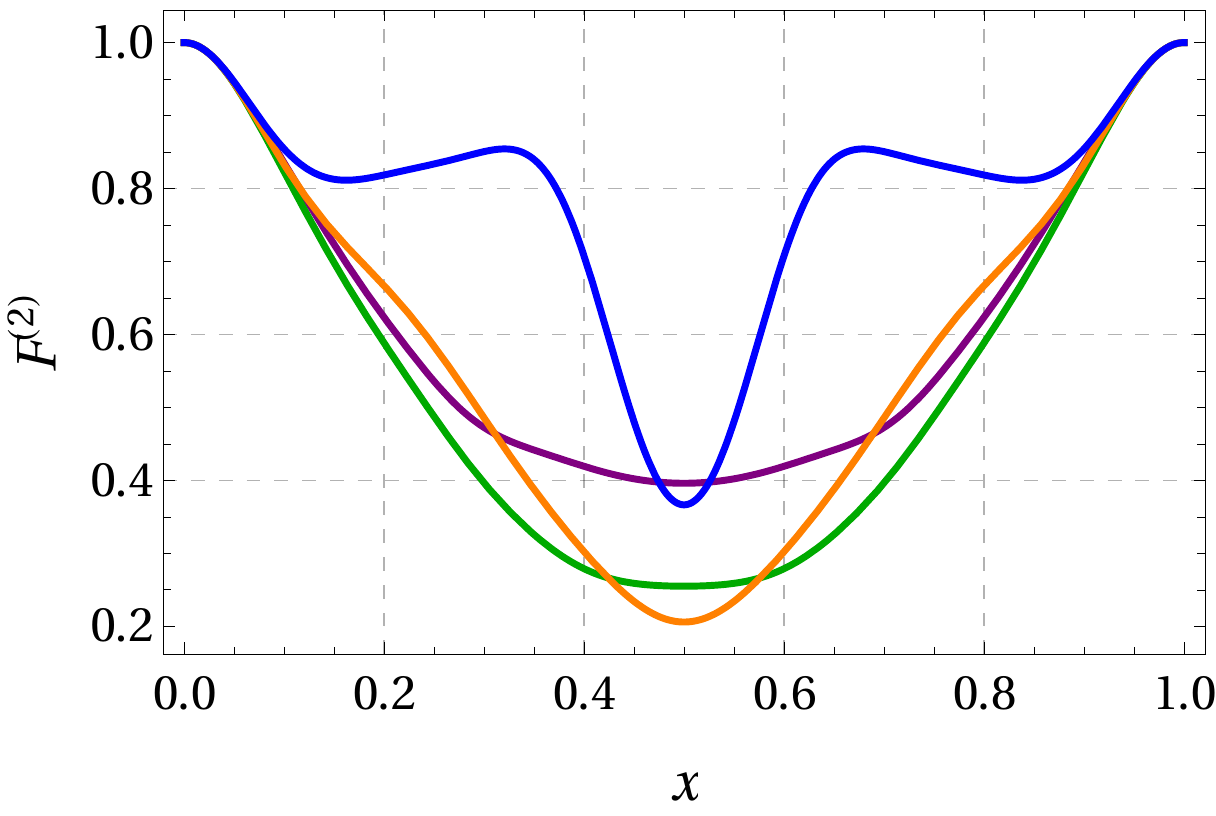}};
    \node at (0,-4.7) {\includegraphics[width=.45\textwidth]{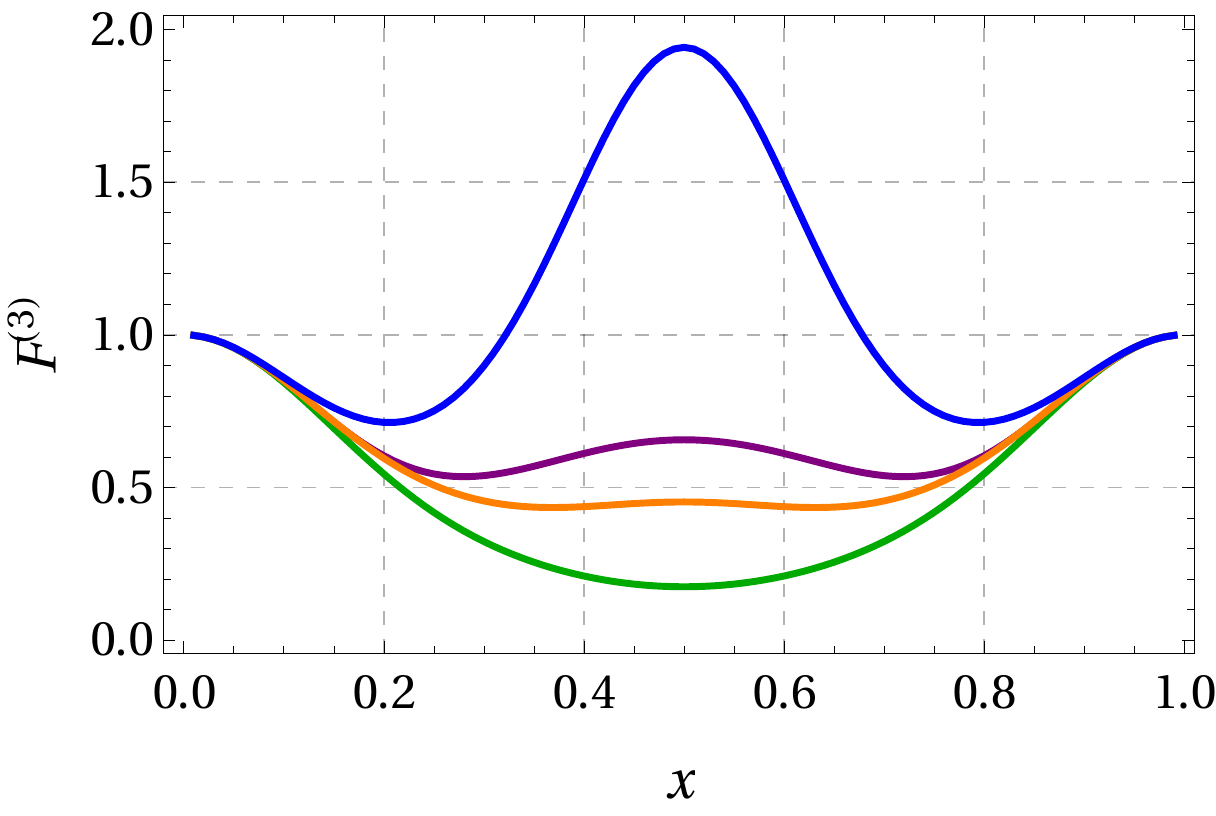}};
    \node at (.47\textwidth,-4.7)  {\includegraphics[width=.45\textwidth]{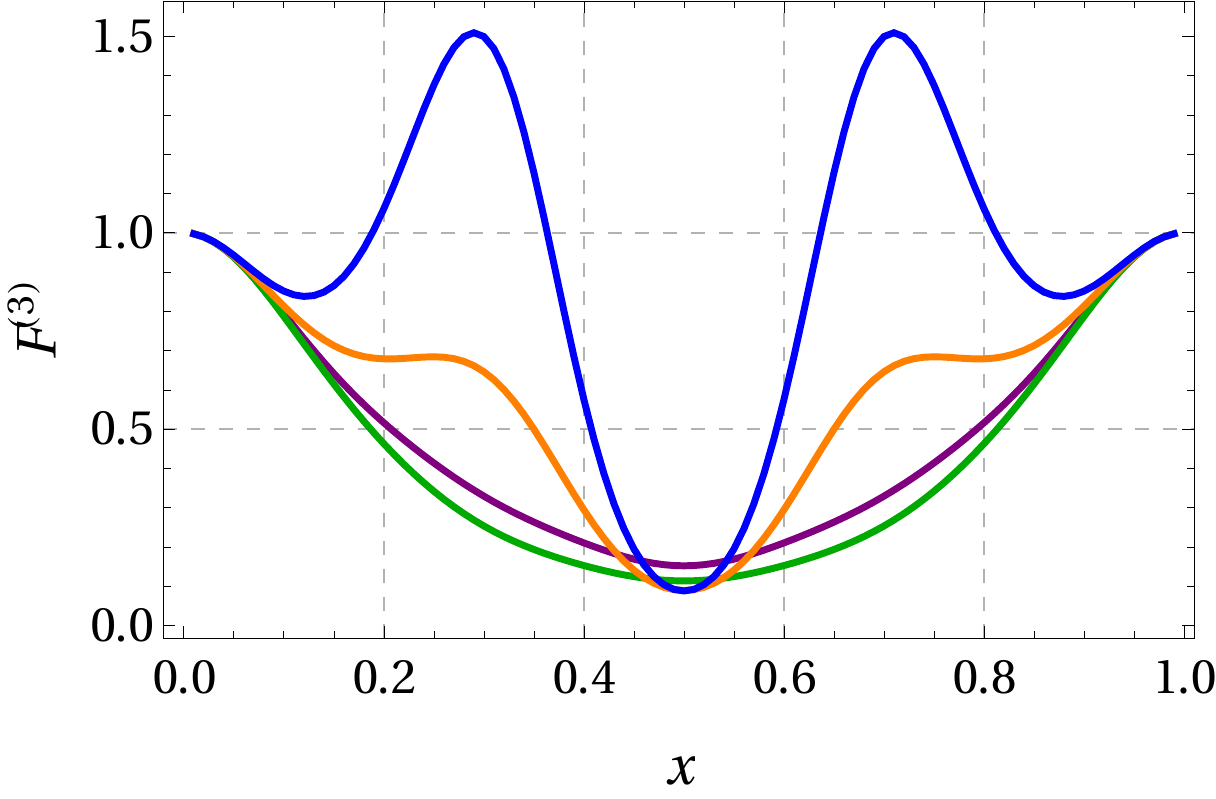}};
    \node at (.25\textwidth,-2.1
    ) {\includegraphics[width=.25\textwidth]{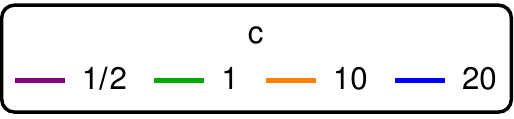}};
    \node at (.4,1.5) {(a)};
    \node at (7.5,1.5) {(b)};
    \node at (.4,-3.2) {(c)};
    \node at (7.5,-3.2) {(d)};
    \end{tikzpicture}  
    \caption{The correlator $F^{(n)}_{\ket{s}}$ for (a) $n=2$, $\ket{s} = L_{-2}^2\ket{0}$ , (b) $n=2$, $\ket{s} = L_{-3}L_{-2}\ket{0}$, (c) $n=3$ $\ket{s} = L_{-2}\ket{0}$, and (d)  $n=3$ $\ket{s} = L_{-3}\ket{0}$ for several values of the central charge.}
    \label{fig:RE2lowB}
\end{figure}

The vacuum module is degenerate at conformal weight $h=4$ and $h=5$. In addition to the states $L_{-4}\ket{0}$ and $L_{-5}\ket{0}$ there are the states $L_{-2}^2\ket{0}$ and $L_{-3}L_{-2}\ket{0}$, respectively. Their correlators $F^{(2)}_{\ket{s}}$ are shown in figure \ref{fig:RE2lowB} (a) and (b) for different values of the central charge. Interestingly, although their small subsystem behaviour is given by \eqref{eq:RElowx} and, hence, it is the same as for $L_{-4}\ket{0}$ and $L_{-5}\ket{0}$, respectively, their general behaviour is rather different at large central charge! Their oscillation period is not proportional to the conformal weight but proportional to the level of the lowest Virasoro generator appearing in it.

Already these two examples show that in particular at large central charge the behaviour of the R\'enyi entropy and, hence, also of the entanglement entropy of descendant states does not only depend on their conformal weight, i.e. the \textit{energy of the state}, but also significantly on their building structure. In particular, theories with a (semi-)classical gravity dual need large central charge. It is widely believed that black hole microstates in $AdS_3$ correspond to typical high conformal dimension states in the CFT. However, a typical state at conformal dimension $\Delta\gg 1$ is a descendant at level $\Delta/c$ of a primary with conformal dimension $\tfrac{c-1}{c}\Delta$ (see e.g. \cite{Datta:2019jeo}). This means that a typical state will be a descendant at large but finite central charge $c$! The results we present here show that descendants with the same conformal dimension can in fact show very different behaviour when it comes to the entanglement structure.
It will be interesting to further study the large $c$ limit, in particular for non-vacuum descendants, to analyse the holographic effect of these different behaviours.

Finally, in figure \ref{fig:RE2lowB} (c) and (d) we show the correlator $F^{(3)}$ for the first two excited states $L_{-2}\ket{0}$ and $L_{-3}\ket{0}$. They show qualitatively the same behaviour as the respective correlators for $n=2$ (see figure \ref{fig:RE2lowA} (a) and (b)). However, their dependence on the central charge is stronger and the oscillating behaviour starts at lower $c$. For example, $F^{(3)}_{L_{-2}\ket{0}}$ is larger than one at $l=1/2$ for $c\gtrsim14.74945$. 

The stronger dependence on the central charge for larger $n$ is expected. Any $F^{(n)}_{\ket{s}}$ can be expanded as 
\begin{equation}
    F^{(n)}_{\ket{s}} = \sum_{k=-n+1}^{n} A_k^{(n)} c^k\,,
\end{equation}
where all the dependence on the state $\ket{s}$ and the  relative subsystem size $x=l/L$ sits in the coefficients $A_k^{(n)}$ . The negative powers of $c$ originate from the normalization of the state. Positive powers of $c$ follow from the Virasoro commutation relations when using the Ward identities. Therefore, at large central charge we get 
\begin{equation}
     \left.F^{(n)}_{\ket{s}}\right|_{c\gg 1} \approx A_n c^n\,.
\end{equation}

\subsection{Sandwiched R\'enyi divergence}

As argued in section \ref{sec:SRD} it is possible to express the sandwiched R\'enyi divergence~\eqref{eq:SRD} for integer parameters $n$ in terms of a $2 n$ point functions $\mathcal{F}^{(n)}$ \eqref{eq:SRDcorr} if $\rho_1$ is the reduced density matrix of the vacuum. In case of the state $L_{-2}\ket{0}$ we e.g. obtain
\begin{align}
    \mathcal{F}^{(2)}_{L_{-2}\ket{0}} = & \frac{(\cos (4 \pi x)+7) (-512 \cos (4 \pi x)+128 \cos (8 \pi x)+384) \sec ^8(\pi x)}{16384 c}\\
    &+\frac{(\cos (4 \pi x)+7) (847 \cos (4 \pi x)-22
   \cos (8 \pi x)+\cos (12 \pi x)+1222) \sec ^8(\pi x)}{16384}\nonumber
\end{align}
where $x = l/L < 1/2$\,. Expressions for the $L_{-n}\ket{0}$, $n=3,4,5$ can be found in appendix \ref{app:SRDresultsvac}. 

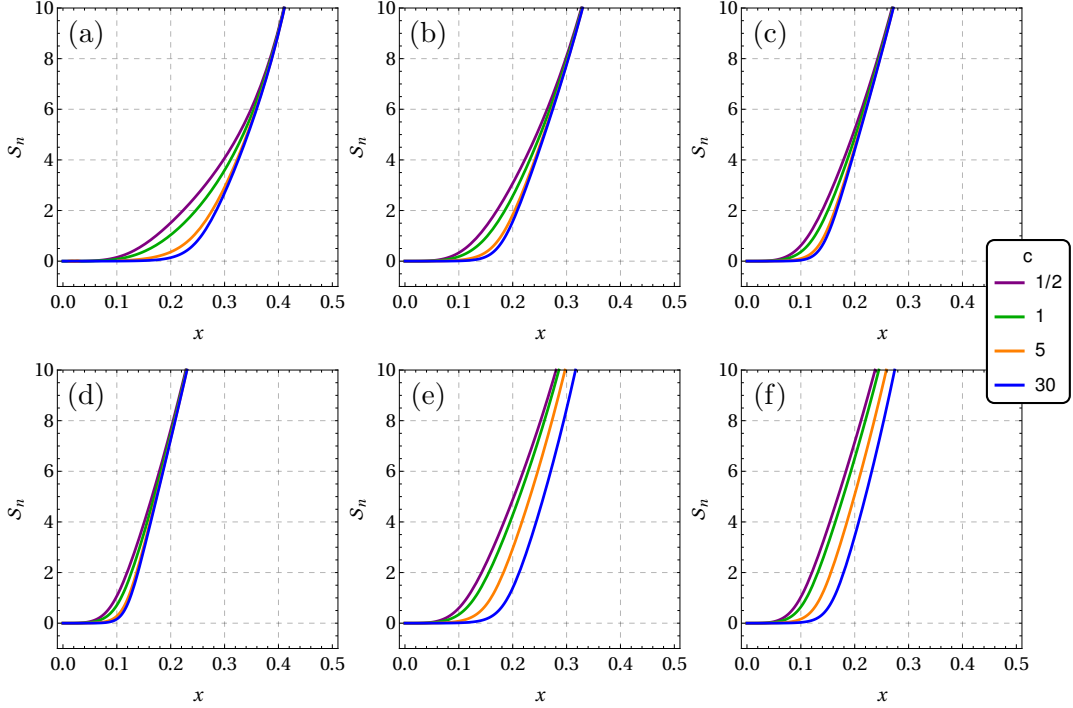
\begin{figure}[t]
    \centering
    \begin{tikzpicture}
    \node at (0,0) {\includegraphics[width=.3\textwidth]{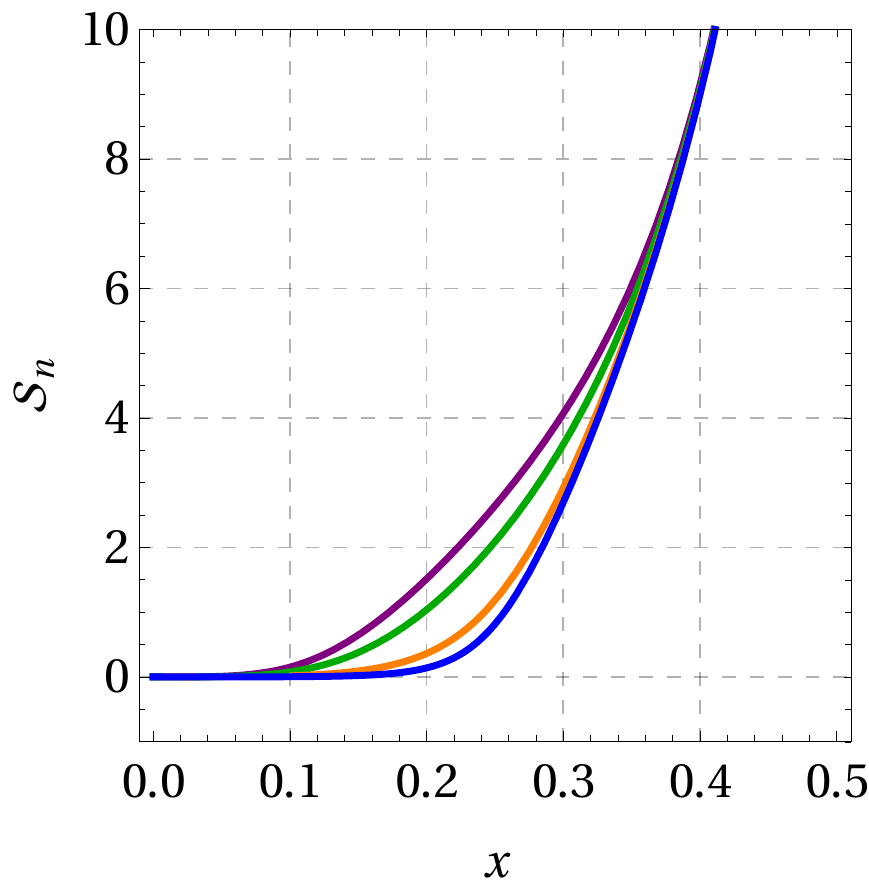}};
    \node at (.3\textwidth,0) {\includegraphics[width=.3\textwidth]{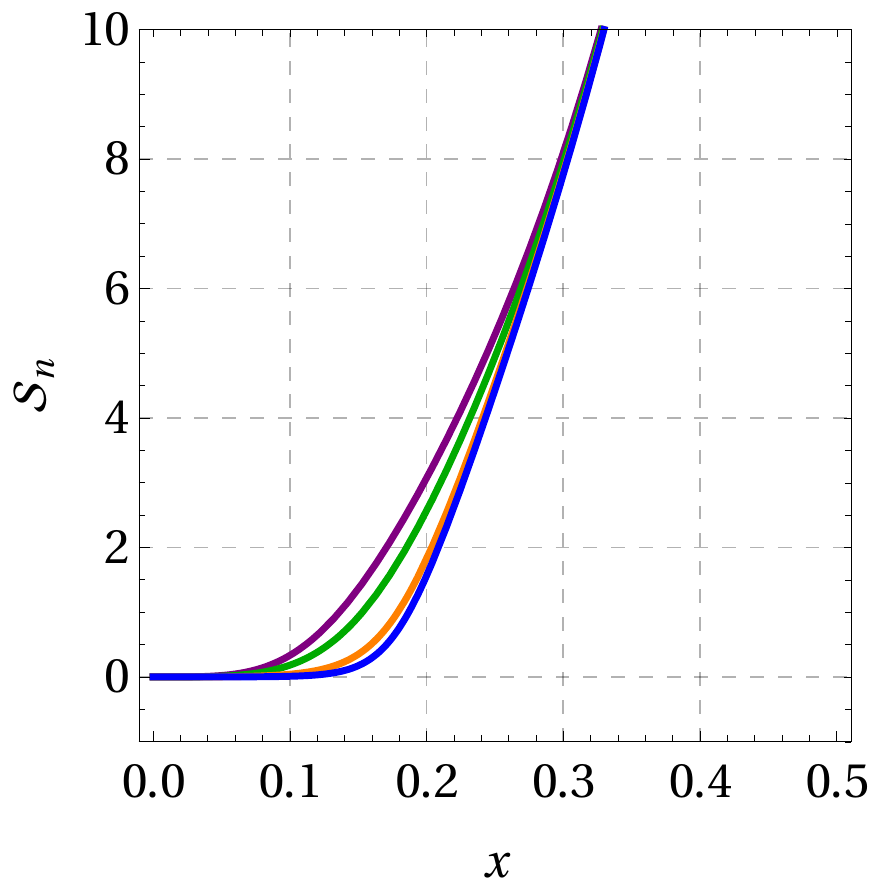}};
    \node at (.6\textwidth,0) {\includegraphics[width=.3\textwidth]{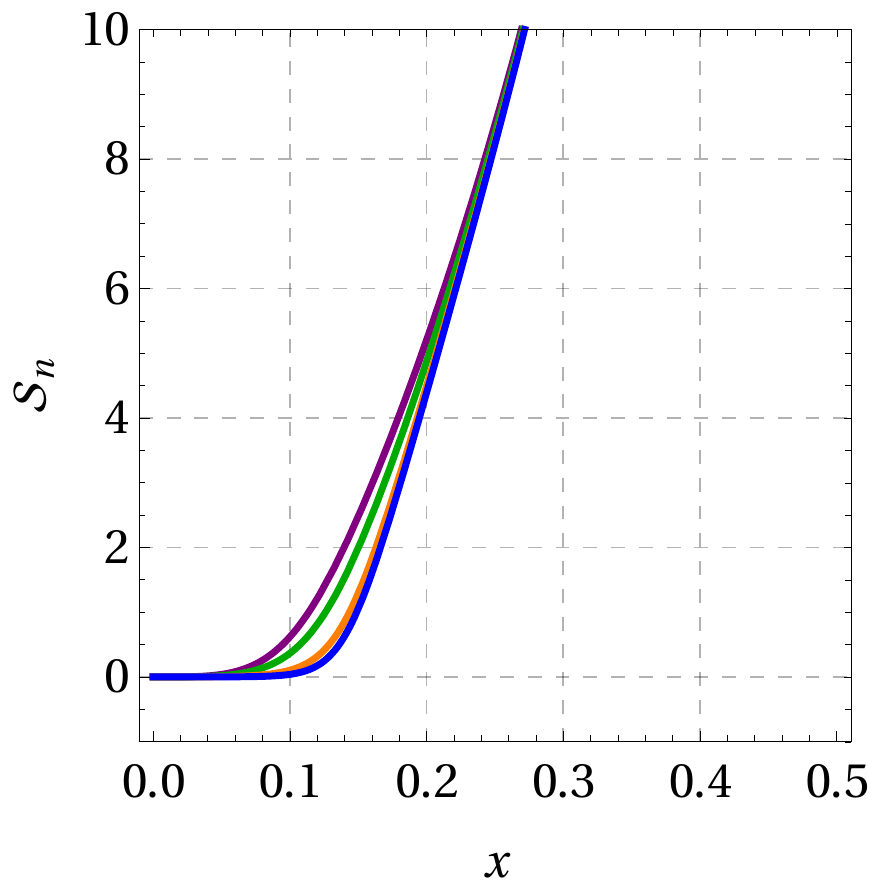}};
    \node at (0,-.32\textwidth) {\includegraphics[width=.3\textwidth]{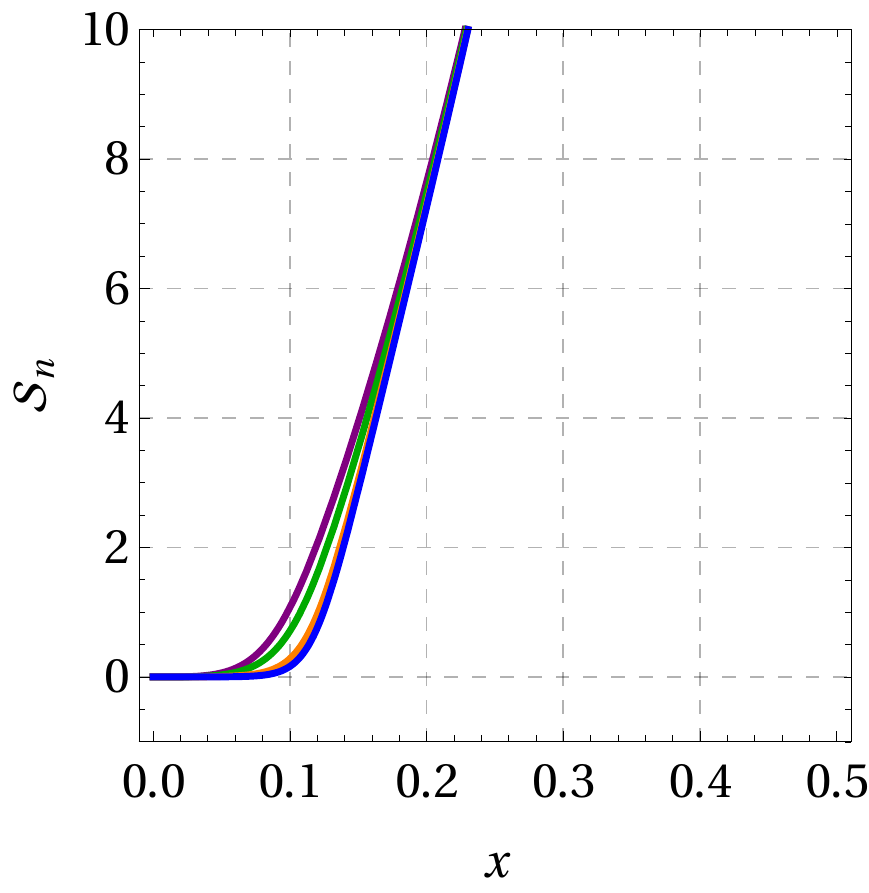}};
    \node at (.3\textwidth,-.32\textwidth) {\includegraphics[width=.3\textwidth]{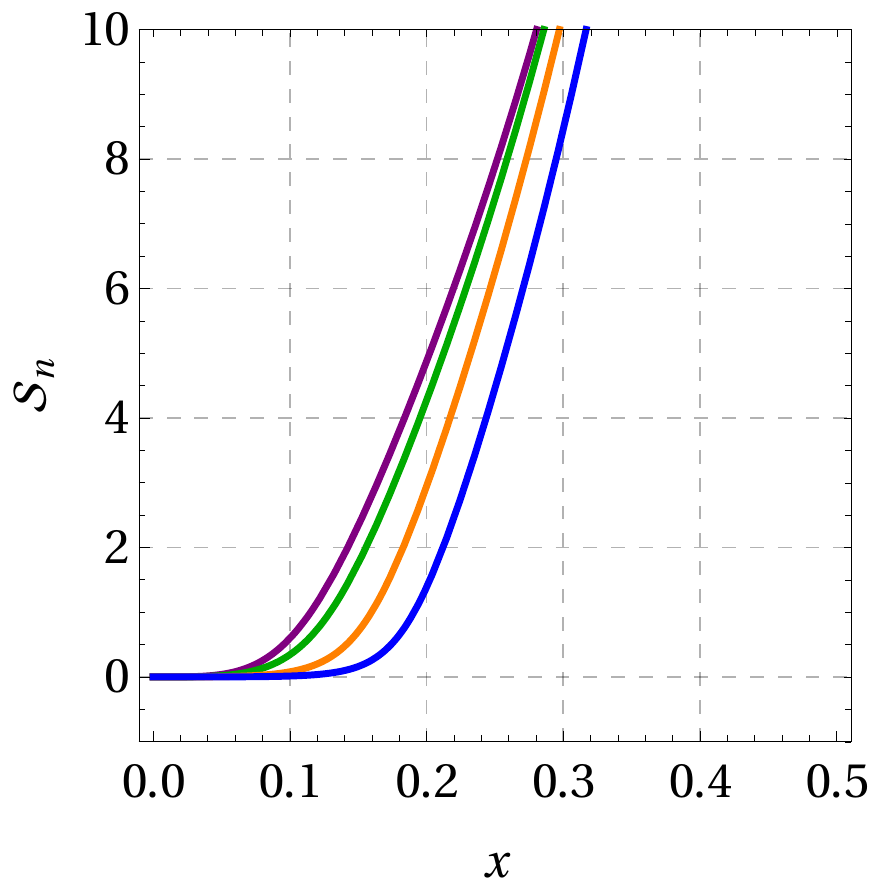}};
    \node at (.6\textwidth,-.32\textwidth) {\includegraphics[width=.3\textwidth]{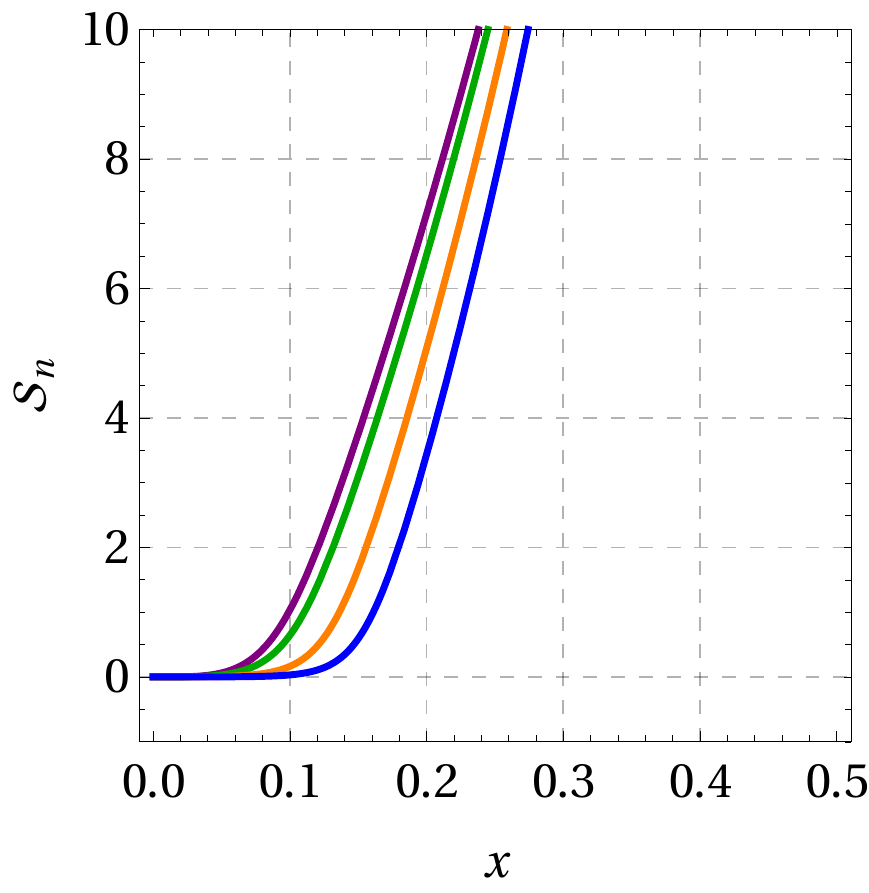}};
    \node at (.75\textwidth,-2) {\includegraphics[width=.075\textwidth]{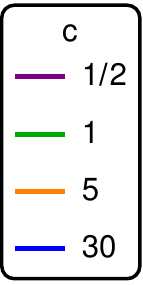}};
    \node at (-1.15,1.8) {(a)};
    \node at (3.35,1.8) {(b)};
    \node at (7.85,1.8) {(c)};
    \node at (-1.15,-3) {(d)};
    \node at (3.35,-3) {(e)};
    \node at (7.85,-3) {(f)};
    \end{tikzpicture}
    \caption{The Sandwiched R\'enyi Divergence for $n=2$ between the reduced groundstate and (a) $L_{-2}\ket{0}$, (b) $L_{-3}\ket{0}$, (c) $L_{-4}\ket{0}$, (d) $L_{-5}\ket{0}$, (e) $L_{-2}^2\ket{0}$, (f) $L_{-3}L_{-2}\ket{0}$ for different values of the central charge $c$.}
    \label{fig:SRE2}
\end{figure}

Again we first want to draw attention to the small subsystem behaviour of the sandwiched R\'enyi divergence. The results for the second SRD between the reduced vacuum state and all states up to conformal weight five show the small subsystem behaviour
\begin{equation}
    \mathcal{S}^{(2)}_{\ket{s}} = \frac{2 h_s^2}{c} \pi^4 x^4 +  \frac{2 h_s^2}{3c} \pi^6 x^6 + O(x^8)\,. \label{eq:SRDsmallx}
\end{equation}

\noindent
Its small subsystem behaviour only depends on the central charge and the conformal weight of the respective state and is independent of the specific structure of the state! 

In case of $n=2$, the SRD diverges at $x=1/2$. We find the behaviour 
\begin{equation}\label{eq:srd_vac_divergence}
 \mathcal{F}^{(2)}_{\ket{s}} = \exp\left(\mathcal{S}^{(2)}_{\ket{s}}\right) = \frac{A_{\ket{s}}}{\pi^{4 h_s}\left(x-\frac12\right)^{4h_s}} \, ,
\end{equation}
where the coefficient $A_{\ket{s}}$ depends on the specifics of the state. For states of the form $L_{-n}\ket{0}$ up to $n=10$ it takes the form 
\begin{equation}
    A_{L_{-n}\ket{0}} = \binom{2n-1}{n-2}^2\,.
\end{equation}

In figure \ref{fig:SRE2} we show the SRD for the first six excited states. All of them show a plateau at small values of $x$ that increases for larger $c$ and shrinks for higher energy. This is expected from the asymptotic result \eqref{eq:SRDsmallx}. Interestingly, although in the asymptotic regimes, i.e. at $x\to 0$ and $x\to1/2$, the second SRD for the states $L_{-2}^2\ket{0}$ and $L_{-3}L_{-2}\ket{0}$ behave similarly to the states $L_{-4}\ket{0}$ and $L_{-5}\ket{0}$ with the same conformal weight they look quite differently for intermediate regimes of $x$. They, in particular, show to be more sensible to the central charge. This shows again that descendant states at the same conformal dimension can behave quite differently, in particular at large central charge.

\begin{figure}[t]
    \centering
    \includegraphics[width=.7\textwidth]{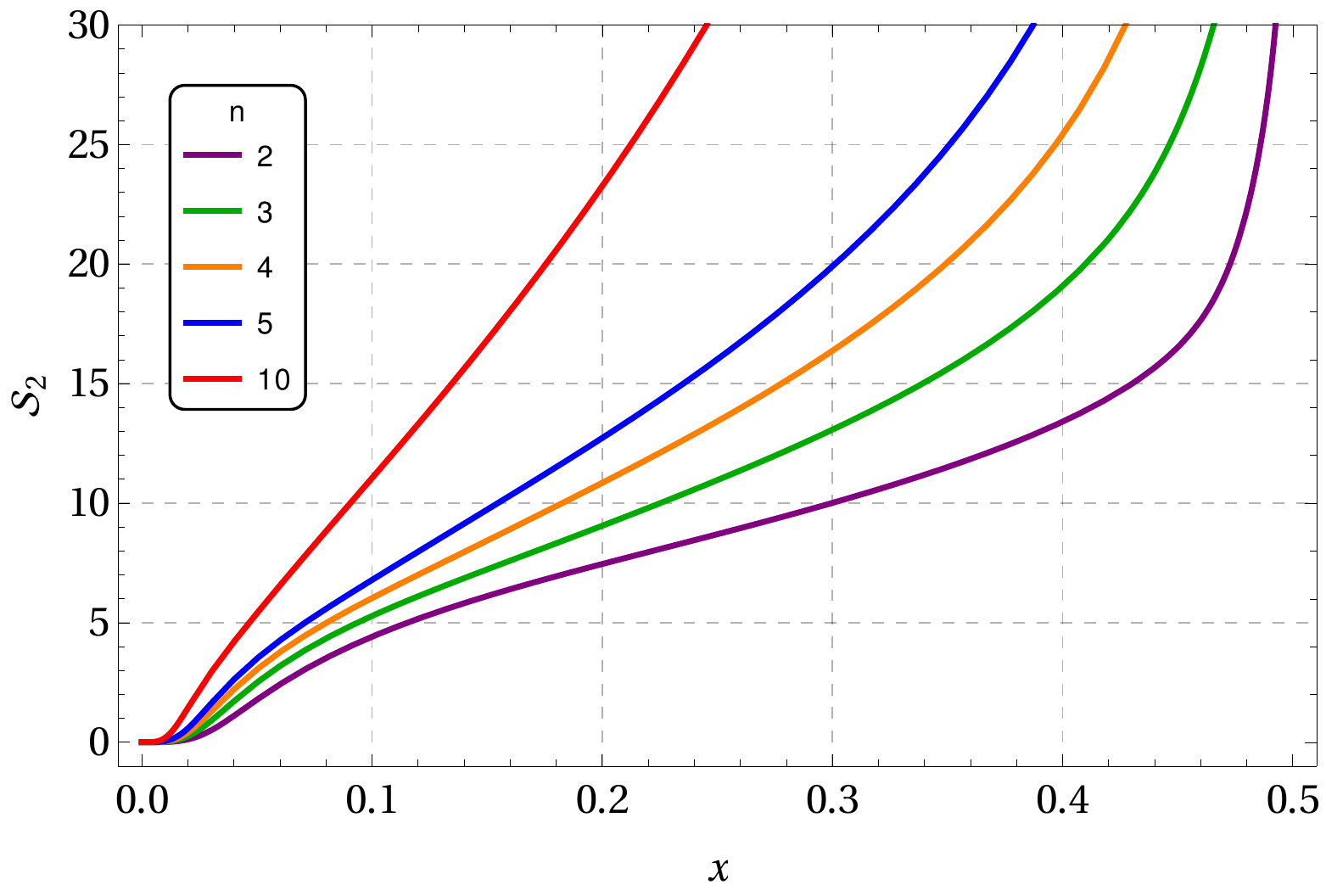}
    \caption{The second Sandwiched R\'enyi Divergence for the states $L_{-n}\ket{0}$, $n=2,3,4,5,10$, at central charge $c=1/1000$\,.}
    \label{fig:nonconvex}
\end{figure}

In all plots so far the second SRD shows to be a convex function of the relative subsystem size $x = l/L$. However, in cases of small central charge it is not! I.e. there are regions of $x$ with $\frac{\partial^2 S^{(2)}}{\partial x^2} <0$. For example, in case of $\ket{s} = L_{-2}\ket{0}$ the second SRD is not convex for $c\lesssim 0.1098$\,. This shows that there are examples where the generalized version of the QNEC is not true! However, conformal field theories with central charges smaller than 1/2 are quite unusual. They cannot be part of the ADE classifiation of rational, unitary, modular invariant CFTs \cite{Cappelli:1986hf} but could e.g. be logarithmic \cite{Nivesvivat:2020gdj}. 
In figure \ref{fig:nonconvex} we show the second SRD for states $L_{-n}\ket{0}$ with $n=2,3,4,5,10$ and $c=1/1000$ to illustrate its non-convexity for all these states. 

\subsection{Trace squared distance}\label{sec:vacSRD}

Again only the expressions for the first few excited states are compact enough to display them explicitly. For example, the TSD between the vacuum and the state $L_{-2}\ket{0}$ is given by
\begin{align}
    T^{(2)}_{L_{-2}\ket{0},\ket{0}} &= \frac{c^2 \sin ^8(\pi x)}{1024}-\frac{1}{512} c \sin ^6(\pi x) (\cos (2 \pi x)+15)+\frac{\sin ^4(\pi x) (\cos (2 \pi x)+7)}{16 c}\nonumber\\
    &\quad+\frac{-32768 \cos
   (\pi x)+8008 \cos (2 \pi x)-228 \cos (4 \pi x)}{32768} \label{eq:TSD21}\\
    &\quad+\frac{120 \cos (6 \pi x)+\cos (8 \pi x)+24867}{32768}\,,\nonumber
\end{align}
where we use the abbreviation $x = \frac{l}{L}$ again. 
Some other explicit expressions can be found in appendix \ref{app:TSDvacResults}. 

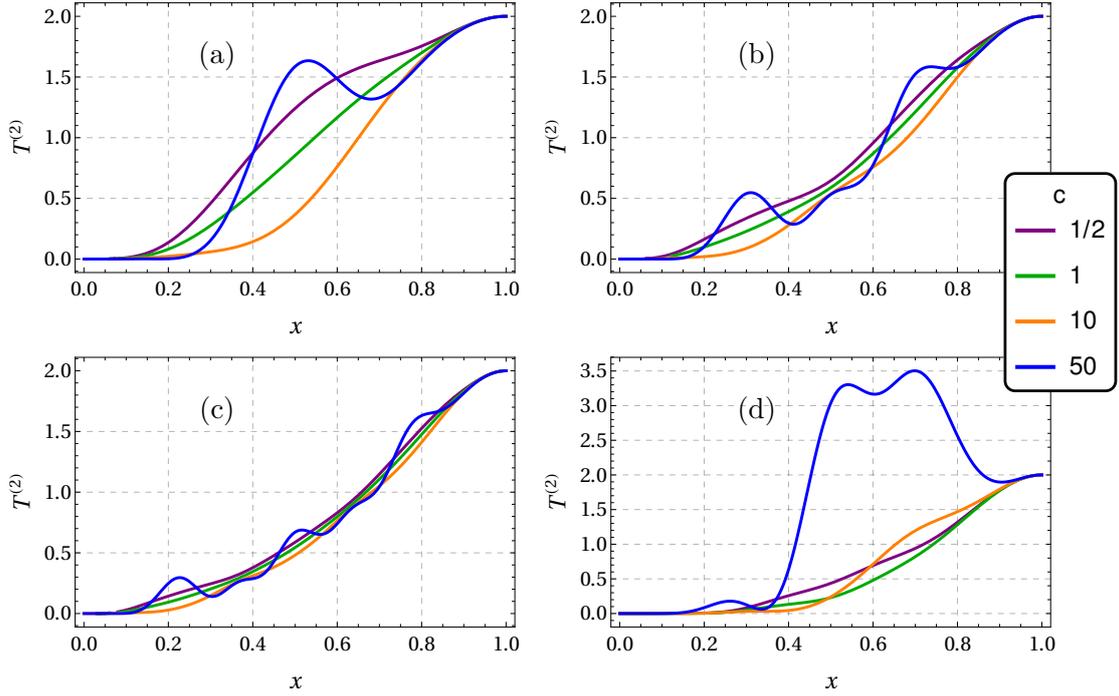
\begin{figure}[t]
    \centering
    \begin{tikzpicture}
    \node at (0,0) {\includegraphics[width=.45\textwidth]{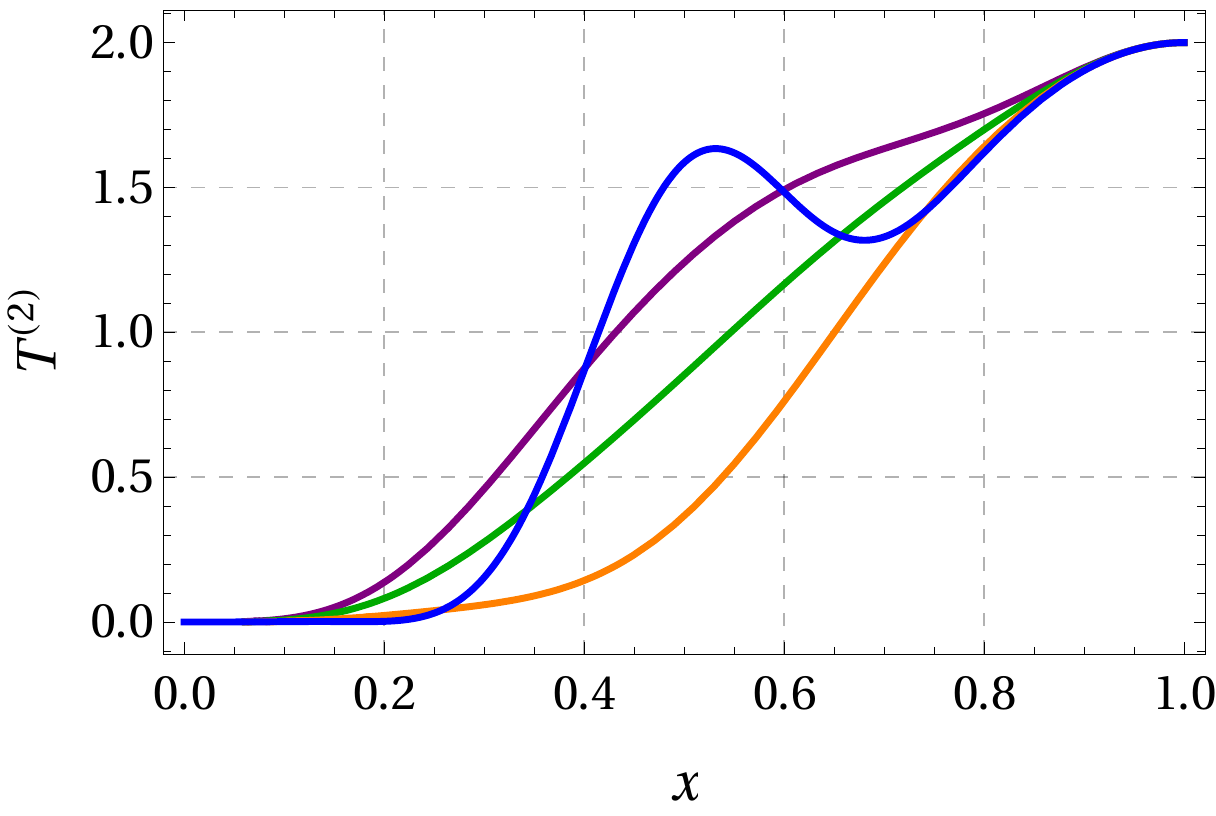}};
    \node at (.47\textwidth,0)  {\includegraphics[width=.45\textwidth]{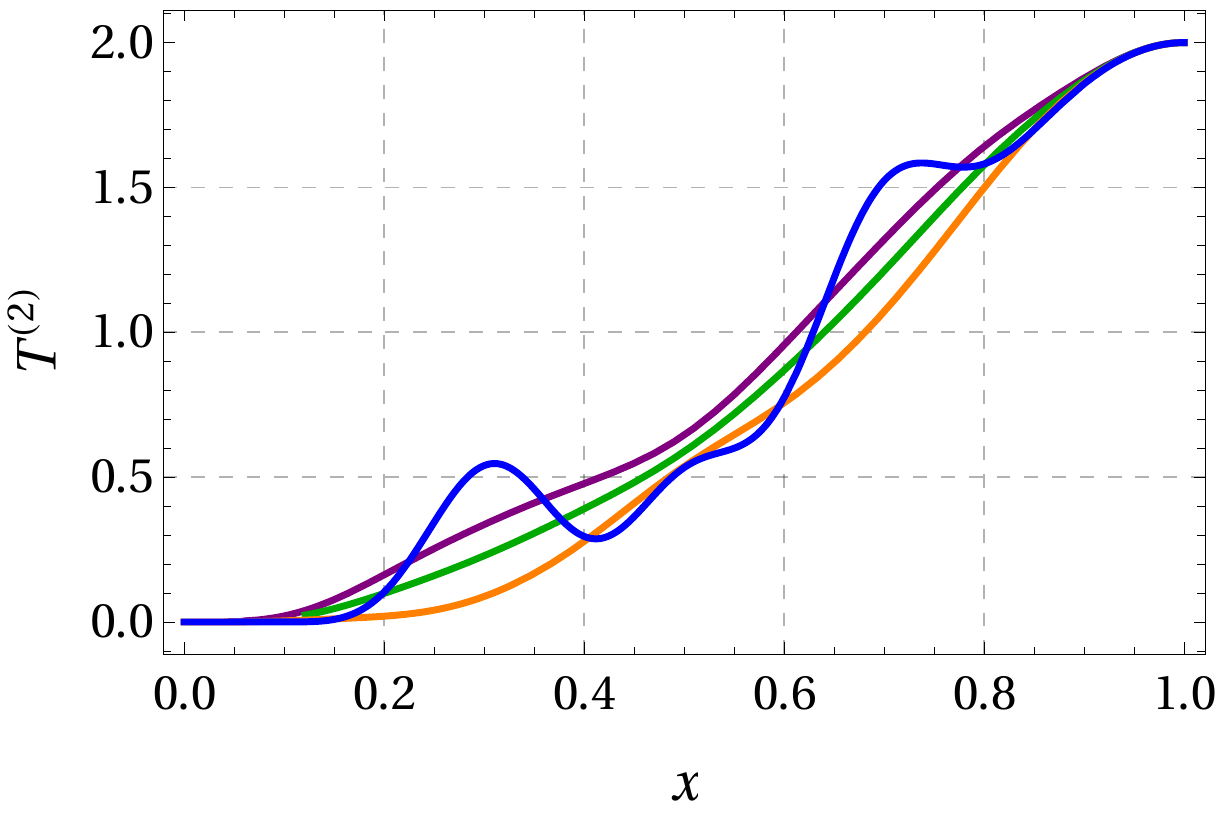}};
    \node at (0,-4.7) {\includegraphics[width=.45\textwidth]{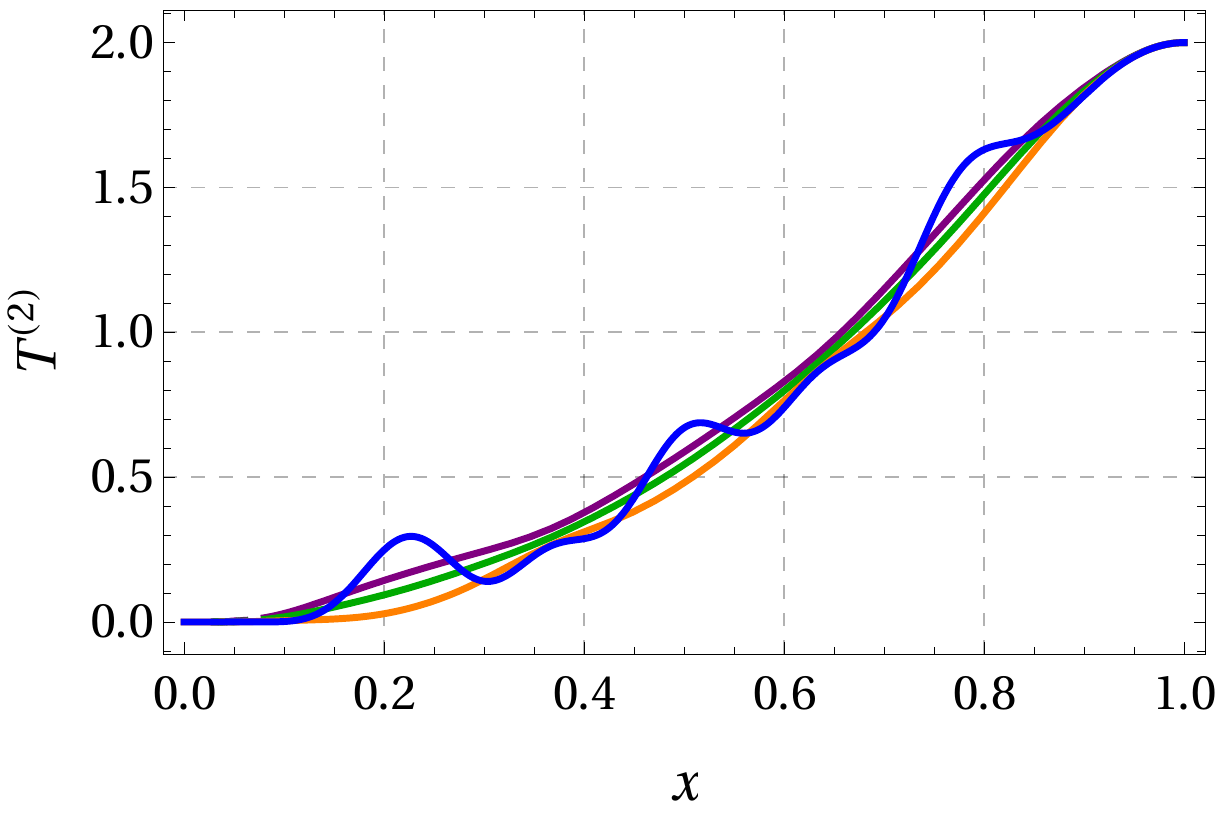}};
    \node at (.47\textwidth,-4.7)  {\includegraphics[width=.45\textwidth]{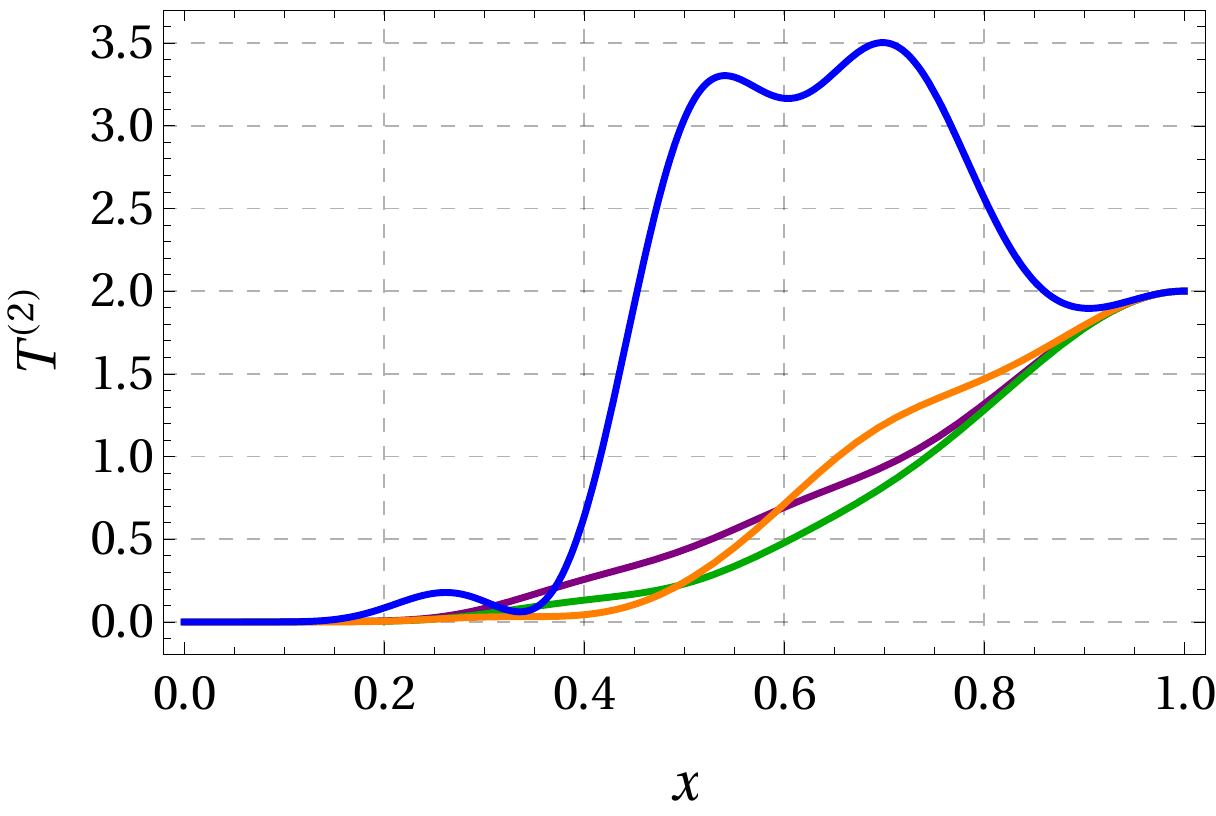}};
    \node at (.7\textwidth,-1.5) {\includegraphics[width=.1\textwidth]{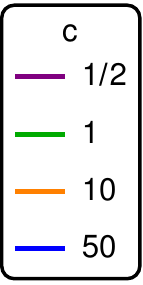}};
    \node at (-.6,1.5) {(a)};
    \node at (6.5,1.5) {(b)};
    \node at (-.6,-3.2) {(c)};
    \node at (6.5,-3.2) {(d)};
    \end{tikzpicture}
    \caption{The Trace Squared Distance between the reduced states of (a) the vacuum and $L_{-2}\ket{0}$, (b) the vacuum and $L_{-3}\ket{0}$, (c) the vacuum and $L_{-4}\ket{0}$, and (d) the states $L_{-3}\ket{0}$ and $L_{-2}\ket{0}$\, for different values of the central charge $c$.}
    \label{fig:TSDA}
\end{figure}

In the limit $x\to 0$ the reduced states have no support and, hence, must be trivial. Consequently, the trace square distance vanishes in this limit independently of the original states we choose. We checked the leading order in $x\ll1$ for all states up to conformal weight five and find the behaviour
\begin{equation}\label{eq:tsd_vacuum_smallx}
    T^{(2)}_{s_1,s_2} =  \frac{2+c}{16 c}  (h_1-h_2)^2 \pi^4 x^4 + O(x^6)\,.
\end{equation}

\noindent
We can see that to leading order, $x^4$, the TSD depends on the central charge and the difference in conformal weight of the two states. We also see that for large central charge the dependence on $c$ is negligible. 

In case of $h_1 -h_2 =0$ the TSD starts at order $x^8$ for small $x$. We e.g. obtain
\begin{align}
    T^{(2)}_{L_{-2}^2\ket{0},L_{-4}\ket{0}} & = \frac{ (2 c+1)^2 \left(25 c^3+420 c^2+2444 c+4752\right) \pi ^8 x^8}{1600 c (c+8)^2} + O(x^{10}) \label{eq:TSDsmallxdeg1}\\
    T^{(2)}_{L_{-3}L_{-2}\ket{0},L_{-5}\ket{0}} & = \frac{9  c \left(25 c^3+420 c^2+2444 c+4752\right) \pi ^8 x^8}{1024 (c+6)^2}+ O(x^{10})\,.\label{eq:TSDsmallxdeg2}
\end{align}

\noindent
Albeit one common factor, the latter expression do not seem to show a straightforward dependence on the states. It also shows that the large $c$ behaviour is more subtle because the $x^8$ coefficient diverges as $c\to\infty$\,.

In the opposite limit $x\to1$ the TSD can be computed easily because the states become pure. One obtains
\begin{align}
    \lim_{x\to1} T^{(2)}_{\ket{s_1},\ket{s_2}} &= \frac{\Tr(\ket{s_1}\bra{s_1}^2)+\Tr(\ket{s_2}\bra{s_2}^2)-2 \Tr(\ket{s_1}\bra{s_1}\ket{s_2}\bra{s_2})}{\Tr(\ket{0}\bra{0}^2)}\\
    &=2 \left(1- |\bra{s_1}s_2\rangle|^2\right)\equiv \mathcal{T}\,.
\end{align}

\noindent 
We can see that $0\le \lim_{x\to1} T^{(2)}(\rho_1,\rho_2)\le 2$ where we get the first equal sign iff $s_1=s_2$ and the second one iff the two states are orthogonal to each other. 

The explicit results up to conformal weight five show that the expansion around $x=1$ is given by
\begin{equation}
    T^{(2)}_{\ket{s_1},\ket{s_2}} = \mathcal{T} \left(1 - \frac{h_1+h_2}{4} \pi^2 (x-1)^2  + O\!\left((x-1)^4\right) \right)\,.
\end{equation}

\noindent
We can see that the behaviour of the TSD close to $x=1$ depends on the sum of conformal weights $h_1 + h_2$\,. 
This is in contrast to the small $x$ behaviour that depends on the difference. 
Let us, for example, consider the second TSD between the vacuum and $L_{-2}\ket{0}$ (see the explicit expression in \eqref{eq:TSD21}) and the second TSD between the vacuum and $L_{-3}\ket{0}$\, (see the explicit expression in \eqref{eq:TSD31}). 
From the difference of conformal weight we get $$T^{(2)}_{L_{-2}\ket{0},\ket{0}}(x) < T^{(2)}_{L_{-3}\ket{0},\ket{0}}(x)$$ for small $x$. 
However, from the sum of conformal weights we obtain $$T^{(2)}_{L_{-2}\ket{0},\ket{0}}(x) > T^{(2)}_{L_{-3}\ket{0},\ket{0}}(x)$$ for $x$ close to one. 
We immediately can conclude that there must be an odd number of values $x\in (0,1)$, which in particular means at least one, with $$T^{(2)}_{L_{-2}\ket{0},\ket{0}}(x) = T^{(2)}_{L_{-3}\ket{0},\ket{0}}(x).$$ 

\begin{figure}[t]
    \centering
    \begin{tikzpicture}
    \node at (0,0) {\includegraphics[width=.45\textwidth]{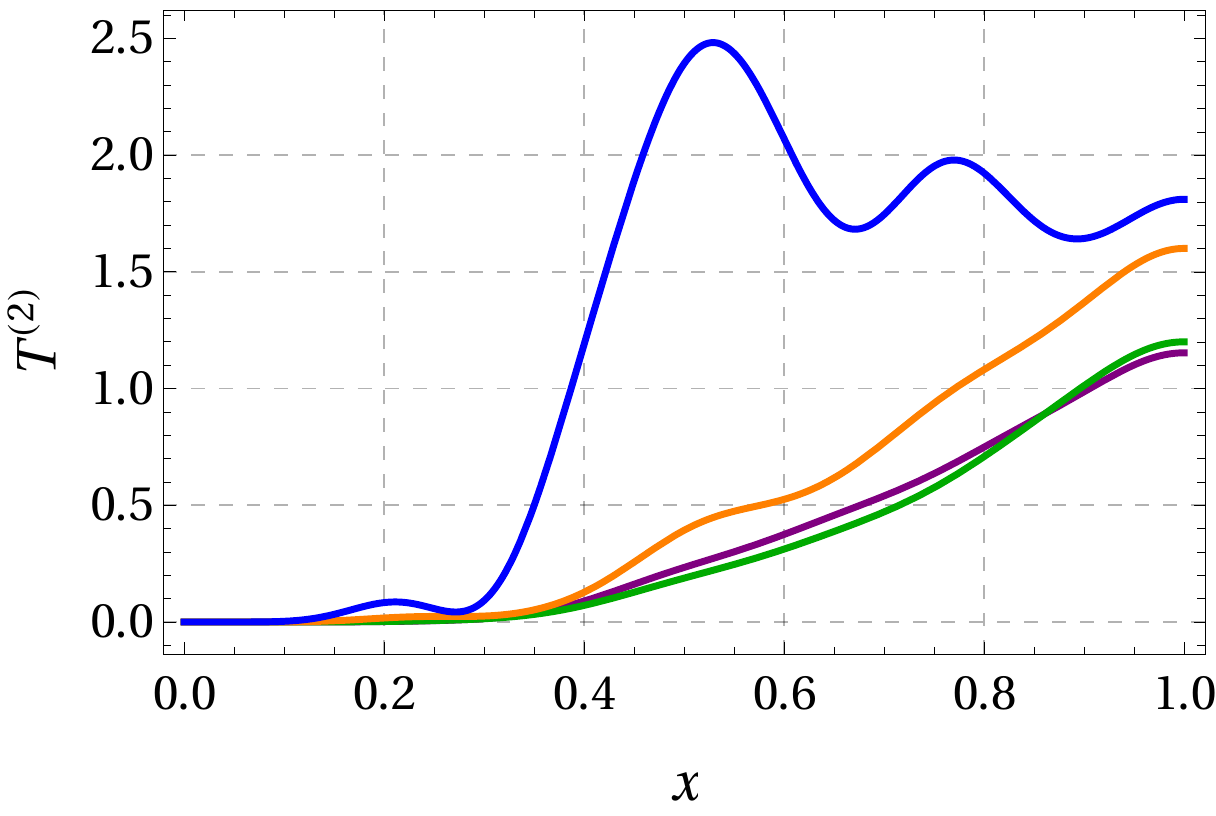}};
    \node at (.47\textwidth,0)  {\includegraphics[width=.45\textwidth]{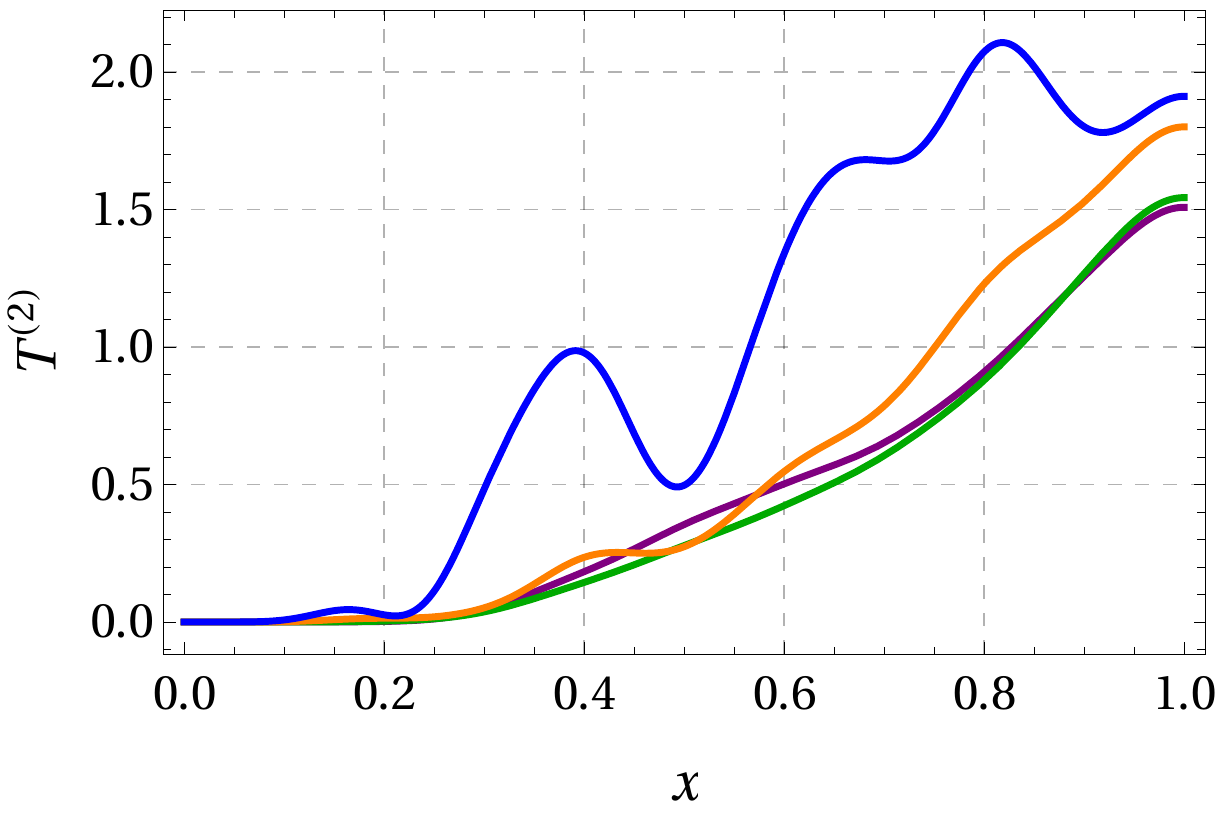}};
    \node at (.25\textwidth,-2.3
    ) {\includegraphics[width=.25\textwidth]{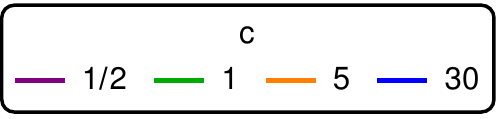}};
    \node at (-.6,1.5) {(a)};
    \node at (6.5,1.5) {(b)};
    \end{tikzpicture}  
    \caption{The Trace Square Distance between the degenerate states at (a) $h_s= 4$, i.e. $L_{-4}\ket{0}$ and $L_{-2}^2\ket{0}$, and (b) $h_s = 5$, i.e. $L_{-5}\ket{0}$ and $L_{-3}L_{-2}\ket{0}$, for different values of the central charge $c$. }
    \label{fig:TSDB}
\end{figure}

We also visualise some of the results. In figure \ref{fig:TSDA} we show the second TSD between the vacuum $\ket{0}$ and $L_{-n}\ket{0}$ for $n=2,3,4$, and between the first two excited states in the vacuum module, $L_{-2}\ket{0}$ and $L_{-3}\ket{0}$\,. In all these examples only for small enough $c$ the TSD is a monotonic function for $x\in[0,1]$\,. At larger $c$ the function starts to meander and can get even bigger than 2, the maximum value of the TSD between pure states. However, the reduced density matrices are not pure and it is not a contradiction per se that the TSD behaves like this. Still, it is hard to interpret the quantity as a meaningful measure of distinguishability for large values of $c$ at intermediate values of the relative subsystem size $x=l/L$.  

In figure \ref{fig:TSDB} we show the TSD between the two degenerate states at conformal dimension $h_s=4$ and $h_s=5$ for different values of $c$. As expected from the results \eqref{eq:TSDsmallxdeg1} and \eqref{eq:TSDsmallxdeg2} we see a quite large flat region at small $x$. At $x\to1$ they converge to the TSD of the respective pure states. In the regions in between they show qualitatively the same behaviour as the other TSDs. For larger central charge they start to meander and at very large $c$ the term proportional to $c^2$ dominates, s.t. the TSD becomes very large, too.

\section{Theory dependent results} \label{sec:nonuniversal}

For non-vacuum descendant states, using relation \eqref{eq:rec1} recursively allows to express the correlation function of chiral descendants $f_{\ket{s_i}}$ as a differential operator acting on the correlation function of the respective  primary fields 
\begin{equation}\label{eq:diff}
    \langle \prod_{i=1}^N f_{\ket{s_i}}(z_i) \rangle = \mathcal{D} \, \langle \prod_{i=1}^N f_{\ket{\Delta_i}}(z_i) \rangle \, .
\end{equation}

\noindent
In general, $\mathcal{D}$ depends on the central charge of the CFT, on the conformal weights of the primary fields, and on the insertion points. As a differential operator it acts on the holomorphic coordinates. 
In appendix~\ref{app:PrimDesCorr} we provide a code to compute it analytically in Mathematica.
If the correlation function of the primaries is known, then it is possible to compute the descendant correlator through~\eqref{eq:diff}.

The correlators in \eqref{eq:RFE}, \eqref{eq:SRDcorr}, and \eqref{eq:TSDcorr} can be written as linear combinations of correlation functions of descendants with coefficients that follow from the respective conformal transformations, i.e.~the uniformization map \eqref{eq:uniformization} in case of the R\'enyi entropy and the trace square distance, and the  usual M\"obius transformations \eqref{eq:Moebius} followed by a rotation in case of the sandwiched R\'enyi divergence. Combining this with \eqref{eq:diff} we can write each of the correlators as
\begin{equation}
    D \bar{D} \langle \prod_{i=1}^N f_{\ket{\Delta_i}}(z_i) \rangle\,,
\end{equation}
with differential operators $D,\bar{D}$. Since we only consider chiral descendants $\bar{D}$ is simply given by the anti-chiral part of the transformation of primaries,
\begin{equation}
    \bar{D} = \prod_{k=1}^n \bar{v}_{0;(k,l)}^{\bar{h}_k}\bar{v}_{0;(k,-l)}^{\bar{h}_k}\,.
\end{equation}

\noindent
E.g. for the correlator of the $n$th R\'enyi entropy \eqref{eq:RFE} we simply get $\bar{D} = \sin^{4\bar{h}}(\pi x)$ from the uniformization map.

In the following sections we explicitly show the expressions of the differential operators $D\bar{D}$ for the simplest descendant state $L_{-1}\ket{\Delta}$. We will then consider results for higher descendants by acting with the operators on particular primary four-point functions in two specific CFTs, the Ising model and the three-state Potts model.

The Ising model is one of the simplest CFTs \cite{DiFrancesco:1997nk}.
It is a unitary minimal model with central charge $c=1/2$ and contains three primary operators: the identity, the energy density $\varepsilon$ and the spin field $\sigma$, whose chiral conformal weights are $0$, $1/2$, $1/16$ respectively.
The $2n$-point correlation functions on the plane of the $\varepsilon$ and $\sigma$ operators are known~\cite{DiFrancesco:1997nk} and, in particular, the four-point correlator of the energy density reads
\begin{equation}\label{eq:isingen}
    \left\langle \varepsilon(z_1,\bar{z}_1) \ldots \varepsilon(z_4 ,\bar{z}_4) \right\rangle = \left| \frac{1}{(z_{12}z_{34})^2} + \frac{1}{(z_{13}z_{24})^2} + \frac{1}{(z_{23}z_{14})^2}  \right| 
\end{equation}
while the four-point correlator of the spin is given by
\begin{equation}\label{eq:isingsig}
    \left\langle \sigma(z_1,\bar{z}_1) \ldots \sigma(z_4 ,\bar{z}_4) \right\rangle = \frac{1}{\sqrt{2}} \frac{1}{|z_{14} z_{23}|^{1/4} } \frac{\sqrt{1 + |\eta| + |1-\eta|}}{|\eta|^{1/4}} \,,
\end{equation}
where $z_{ij} = z_i - z_j$ and $\eta = z_{12}z_{34}/z_{13}z_{24}$ is the cross ratio.
Given these expressions, it is possible to study the R\'enyi entanglement entropy  and the quantum measures for various descendants of $\varepsilon$ and $\sigma$.

The three-state Potts model is the unitary minimal model with $c= 4/5$~\cite{DiFrancesco:1997nk}.
It can e.g. be realized as a particular class of the more general $N$-state clock model which enjoys $\mathbb{Z}_N$ symmetry.
For $N=2$ one recovers the Ising model, while the case $N=3$ is equivalent to the three-state Potts model~\cite{Fradkin:1980th,Fateev:1985mm,Fateev:1987vh,Dotsenko:1984if}.
Its operator content is richer than that of the Ising model.
In particular, it contains six primary operators with conformal weight $0$, $2/5$, $7/5$, $3$, $1/15$, and $2/3$.
The dimensions of the thermal operator $\varepsilon$ and the spin field $\sigma$ are $2/5$ and $1/15$ respectively. 
Again, a number of correlation functions between operators of the three-states Potts model are known (e.g.~\cite{Fateev:1985mm,Dotsenko:1984if}) and, since we will focus on descendants of the energy operator in the following, we provide here the four-point correlation function of the energy density \cite{Dotsenko:1984if}:
\begin{align}\label{eq:pottsen}
    \left\langle \varepsilon(z_1,\bar{z}_1) \ldots \varepsilon(z_4 ,\bar{z}_4) \right\rangle &=  \frac{1}{|z_{13} z_{24}|^{8/5}} \left[ \frac{1}{| \eta (1-\eta) |^{8/5}} \left| _2F_1\left( -\tfrac85, -\tfrac15; -\tfrac25; \eta \right) \right|^2 \right. \nonumber\\
    & \phantom{=} \left. -   \frac{ \Gamma\left(-\tfrac25\right)^2 \Gamma\left(\tfrac65\right)\Gamma\left(\tfrac{13}5\right)}{\Gamma\left(\tfrac{12}{5}\right)^2 \Gamma\left(-\tfrac15\right)\Gamma\left(-\tfrac85\right)} |\eta(1-\eta)|^{6/5} \left| _2F_1\left( \tfrac65, \tfrac{13}{5}; \tfrac{12}{5}; \eta \right) \right|^2  \right]
\end{align}
where $ _2F_1 $ is the hypergeometric function.

\subsection{R\'enyi entanglement entropy}

Let us first consider $F_{\ket{s}}^{(2)}$ with $\ket{s} = L_{-1}\ket{\Delta}$.
As discussed above we can write
\begin{equation}\label{eq:ree1prim}
F_{\ket{s}}^{(2)} = \bar{D}^{F^{(2)}} D_{L_{-1}}^{F{(2)}} \, \left\langle f_{\ket{\Delta}}(e^{-\frac12 i\pi x}) f_{\ket{\Delta}}( e^{\frac12 i\pi x} ) f_{\ket{\Delta}}(- e^{-\frac12 i\pi x}) f_{\ket{\Delta}}( - e^{\frac12 i\pi x} ) \right\rangle_\mathbb{C}
\end{equation}
with $\bar{D}_{L_{-1}}^{F^{(2)}}= \sin ^{4 \bar{h}}(\pi  x)$ and $D_{L_{-1}}^{F^{(2)}}$ can be computed to be
\begin{align}
    D_{L_{-1}}^{F(2)} =& \, \frac{1}{64} \sin ^{4 h}(\pi x)  \Big[ 4 h^2 (3 \cos (2 \pi  x)+5)^2 + \frac{16 \sin ^4(\pi  x)}{h^2} \partial_1 \partial_2 \partial_3 \partial_4 \nonumber\\
    &+h e^{-\frac{7}{2} i \pi  x} \left(3+e^{2 i \pi  x}\right)^2 \left(-2 e^{2 i \pi  x}+3 e^{4 i \pi  x}-1\right) \left( \partial_2 - \partial_4 \right) \nonumber\\
    &+ h e^{-\frac{9}{2} i \pi  x} \left(1+3 e^{2 i \pi  x}\right)^2 \left(2 e^{2 i \pi  x}+e^{4 i \pi  x}-3\right) \left( \partial_3 - \partial_1 \right) \nonumber\\
    &+ 8 \sin ^2(\pi  x) (3 \cos (2 \pi  x)+5) \left( \partial_1\partial_2 + \partial_3\partial_4 - \partial_2\partial_3 - \partial_1\partial_4 \right)\nonumber\\
    & -e^{-3 i \pi  x} \left(2 e^{2 i \pi  x}+e^{4 i \pi  x}-3\right)^2 \partial_2 \partial_4 -e^{-5 i \pi  x} \left(2 e^{2 i \pi  x}-3 e^{4 i \pi  x}+1\right)^2 \partial_1 \partial_3 \nonumber\\
    & + \frac{1}{h}e^{-\frac{7}{2} i \pi  x} \left(-1+e^{2 i \pi  x}\right)^3 \left(3+e^{2 i \pi  x}\right) \left( \partial_1 \partial_2 \partial_4 - \partial_2 \partial_3\partial_4 \right) \nonumber\\
    & + \frac{1}{h} e^{-\frac{9}{2} i \pi  x} \left(-1+e^{2 i \pi  x}\right)^3 \left(1+3 e^{2 i \pi  x}\right) \left( \partial_1\partial_3\partial_4 - \partial_1\partial_2\partial_3 \right) \Big]\,,
\end{align}
where $\partial_n$ is the partial differentiation w.r.t.~the $n$-th insertion point.
Unfortunately already at level 2, the general expressions are too cumbersome to express them here explicitly.

Given the four-point correlation functions~\eqref{eq:isingen}, \eqref{eq:isingsig}, \eqref{eq:pottsen}, we can compute $F_{L_{-1}\ket{\Delta}}^{(2)}$ from eq.~\eqref{eq:ree1prim} for $h= 1/2, \, 1/16$ in the Ising model and $h=2/5$ in the three-states Potts model.
We performed the same computations for descendants up to level 3 and show the results in figure~\ref{fig:REEprim}; some analytic expressions are given in appendix~\ref{app:REEresultsIsing} and~\ref{app:REEresultsPotts}.

\begin{figure}[tb]
    \centering
    \begin{tikzpicture}
        \node at (0,0) {\includegraphics[width=.32\textwidth]{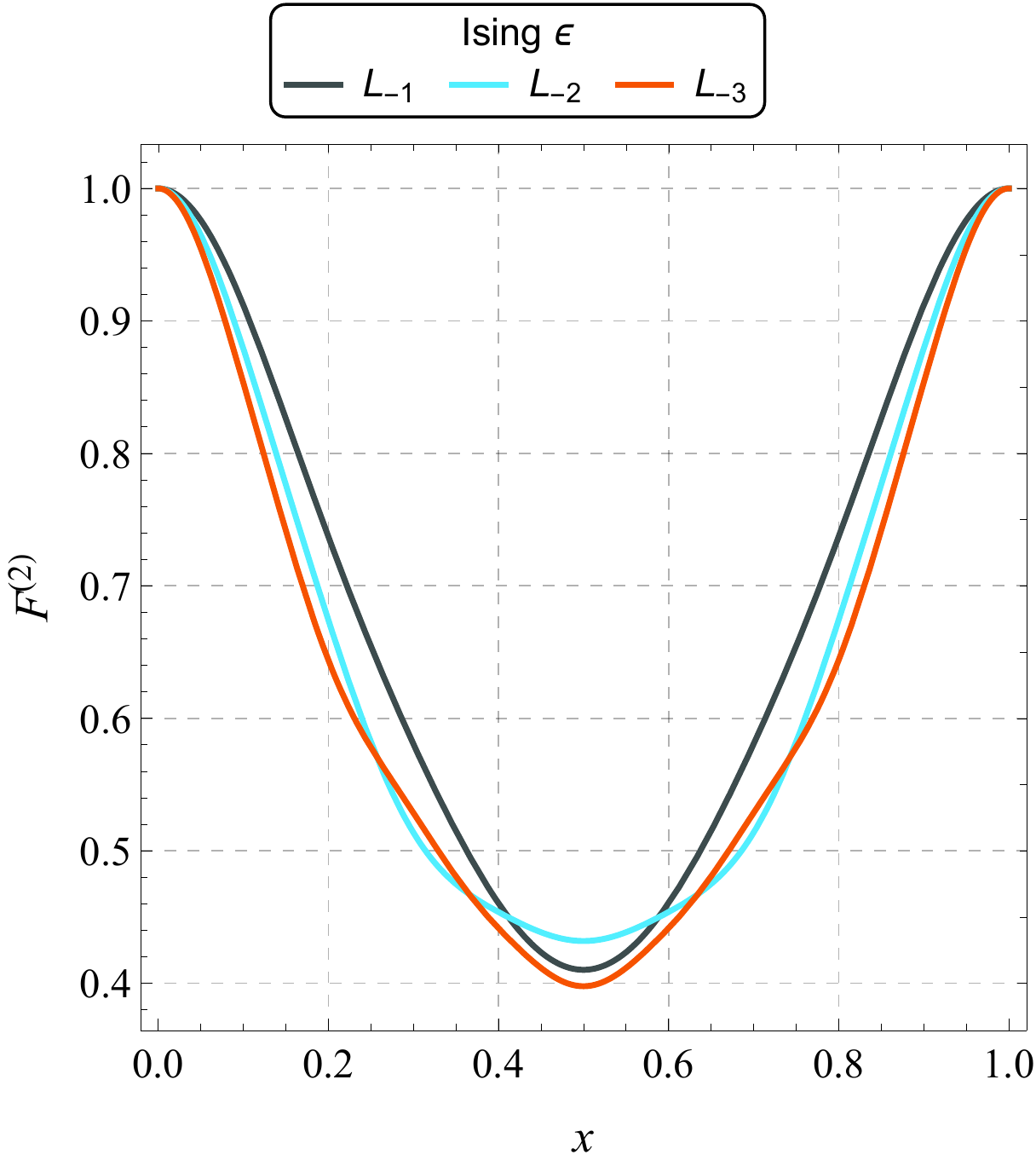}};
    \node at (.33\textwidth,0)  {\includegraphics[width=.32\textwidth]{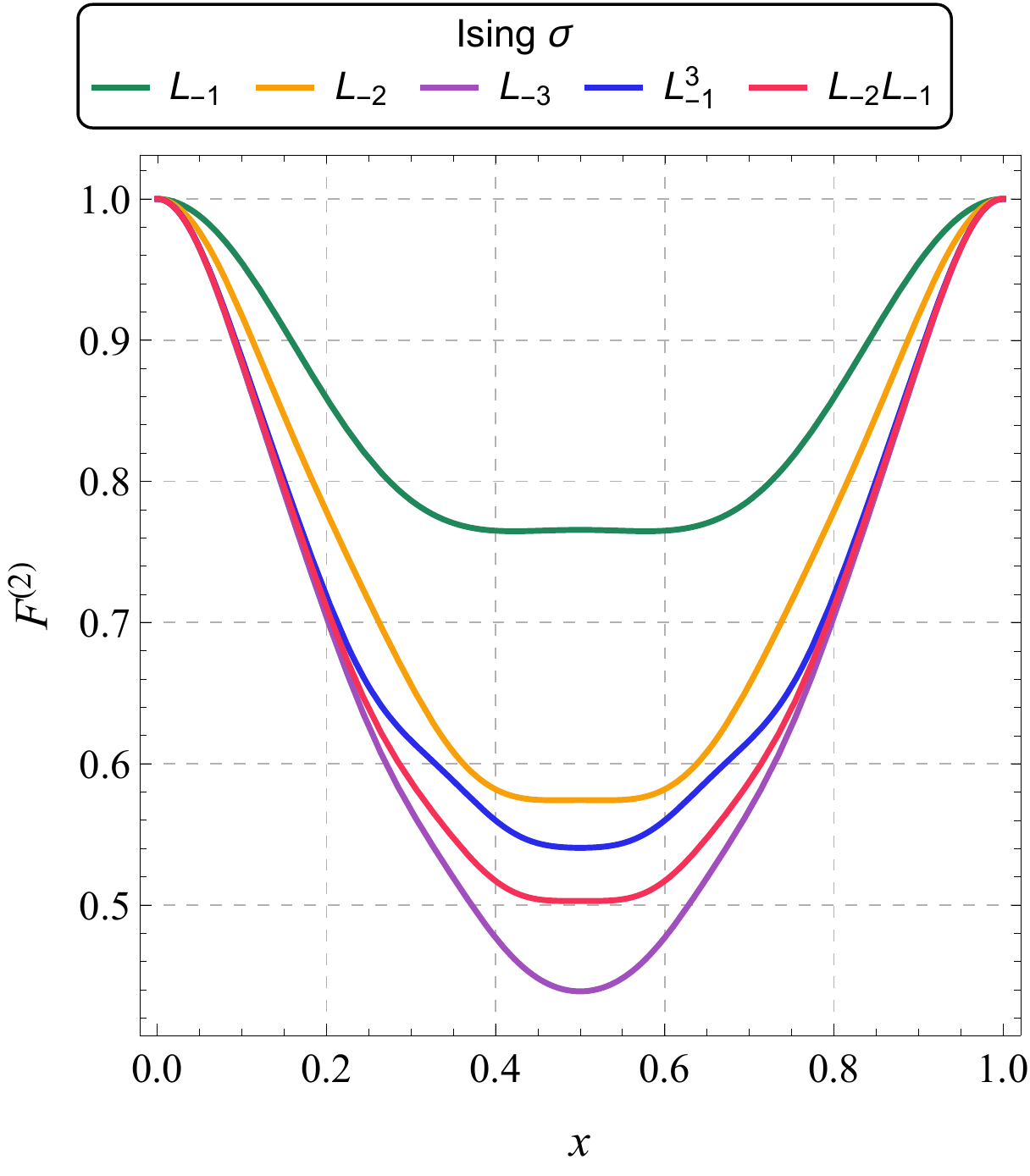}};
    \node at (.66\textwidth,0)  {\includegraphics[width=.32\textwidth]{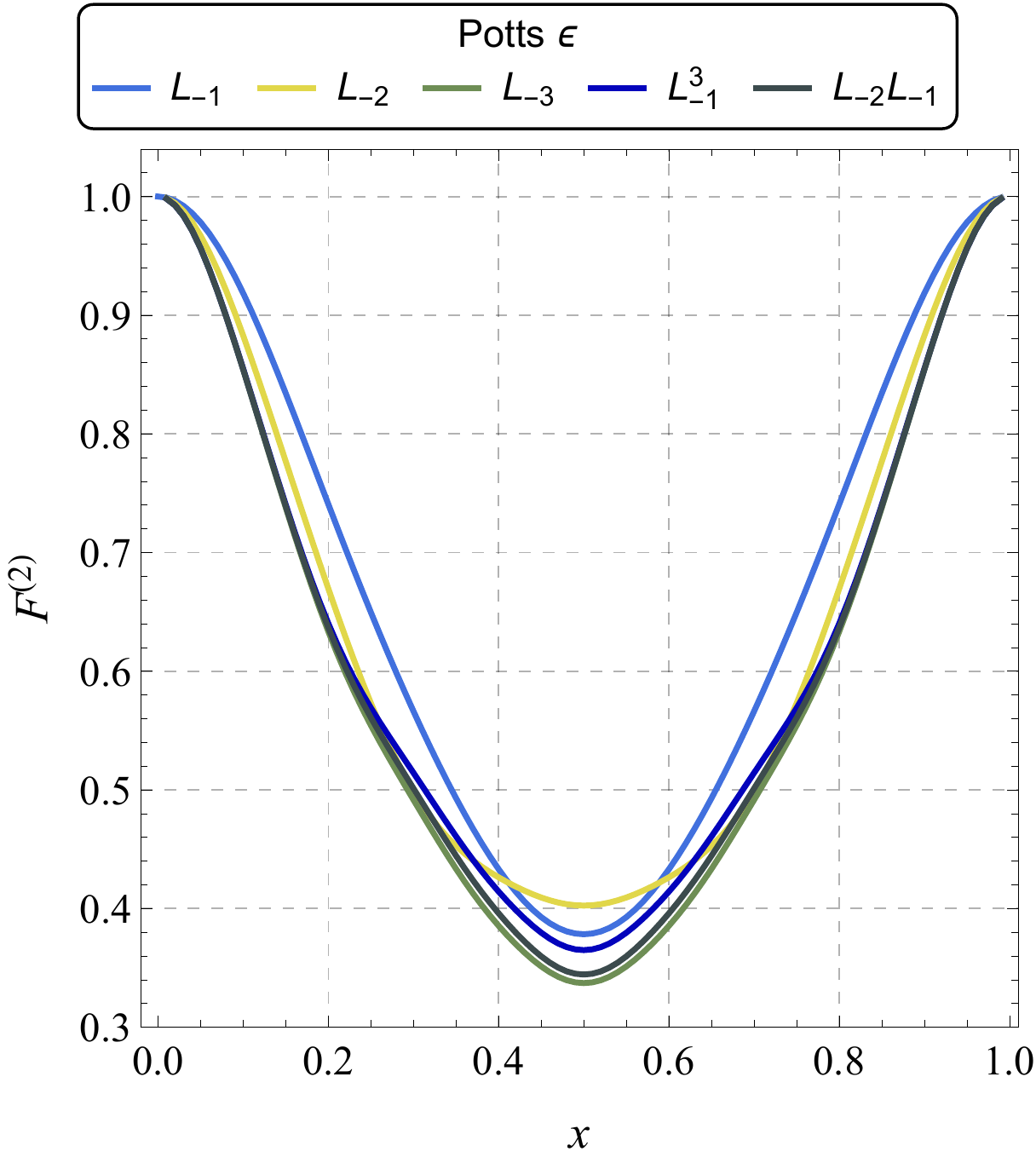}};
    \end{tikzpicture}
    \caption{The correlator $F^{(2)}_{\ket{s}}$ for different descendants of $\ket{s} = \ket{\varepsilon}$ and $\ket{s} = \ket{\sigma}$ in the Ising model and $\ket{s} = \ket{\varepsilon}$ in the Potts model.}
    \label{fig:REEprim}
\end{figure}

In the Ising model, there is only one physical state in the module of the energy operator at each level up to level 3.
A consequence is that $F_{L_{-2}\ket{\varepsilon}}^{(2)} = F_{L_{-1}^2\ket{\varepsilon}}^{(2)}$, even though $D^{F^{(2)}}_{L_{-2}} \neq D^{F^{(2)}}_{L_{-1}^2}$.
The same happens at level 3 for the different descendant states $L_{-3}\ket{\varepsilon}$, $L_{-1}^3\ket{\varepsilon}$ and $L_{-2}L_{-1}\ket{\varepsilon}$. As expected we see this in our result. 
For $\sigma$ descendants, again there is only one physical state at level 2 and $F_{L_{-2}\ket{\sigma}}^{(2)} = F_{L_{-1}^2\ket{\sigma}}^{(2)}$, but at level 3 there are two physical states and $L_{-3}\ket{\sigma}$, $L_{-1}^3\ket{\sigma}$ and $L_{-2}L_{-1}\ket{\sigma}$ produce different REEs as shown in figure~\ref{fig:REEprim}.
Notice that the REEs for the different descendants of $\sigma$ at level 3 have a similar behaviour for small values of $x$, but are clearly distinguishable for $x \sim 1/2$.

For descendants states of the energy density of the three-states Potts model there is again only one physical state at level 2 and two physical states at level 3.
Similarly to the case of descendants of $\sigma$ in Ising, we found that $F_{L_{-2}\ket{\varepsilon}}^{(2)} = F_{L_{-1}^2\ket{\varepsilon}}^{(2)}$ but the different descendants that we considered at level 3 produced different REEs, as plotted in figure~\ref{fig:REEprim}.
Notice that also in Potts the behaviour for small $x$ is given by the level and not by the state configuration, while all the curves are distinguishable for $x\sim 1/2$.
In particular, $F_{L_{-1}^3\ket{\varepsilon}}^{(2)}$ behaves more like $F_{L_{-1}\ket{\varepsilon}}^{(2)}$ than $F_{L_{-3}\ket{\varepsilon}}^{(2)}$ for $x\sim 1/2$, while the plot of $F_{L_{-2}L_{-1}\ket{\varepsilon}}^{(2)}$ is very similar to $F_{L_{-3}\ket{\varepsilon}}^{(2)}$.

If we expand the analytic results for energy descendants in both the Ising and Potts model for small $x$, we find the behaviour
\begin{equation}\label{eq:reesmallx}
    F_{L_{-n}\ket{\varepsilon}}^{(2)} = 1 - \frac{n + 2h_\varepsilon}{2} (\pi x)^2 + O(x^4) \quad h_\varepsilon= \left\{\begin{matrix}1/2& \text{Ising}\\ 2/5 & \text{Potts}\end{matrix}\right.\,, \quad n=1,2,3\,.
\end{equation}

\noindent
This is in general expected, since for small subsystem size $z_1 \sim z_2$ and $z_3 \sim z_4$ and to first order the four-point function is $(h=\bar{h}=\Delta/2)$
\begin{equation}
    \left\langle f_{\ket{\Delta}}(z_1, \bar{z}_1) f_{\ket{\Delta}}(z_2, \bar{z}_2) f_{\ket{\Delta}}(z_3, \bar{z}_3)  f_{\ket{\Delta}}(z_4, \bar{z}_4) \right\rangle_\mathbb{C} \simeq \frac{1}{| z_{12} z_{34}|^{4 h}}\,.
\end{equation}

\noindent
Then, using this correlation function in~\eqref{eq:ree1prim} as well as in the corresponding equations for higher descendants and taking the small $x$ limit we reproduce precisely eq.~\eqref{eq:reesmallx}, which is the clear generalization of eq.~\eqref{eq:RElowx} in agreement with~\cite{Alcaraz:2011tn}.
However, the leading behaviour of $F^{(2)}_{L_{-n} \ket{\sigma}}$ is different from the one outlined in~\eqref{eq:reesmallx}.
This happens because in the OPE of two Ising spin operator there is an additional contribution, that is absent in the OPE of two energy operators or subleading in the case of Potts.
Indeed, consider in general the OPE between two primary fields 
\begin{equation}\label{eq:ope_light}
    f_{\ket{\Delta_i}} (z_1,\bar{z}_1) f_{\ket{\Delta_i}}(z_2,\bar{z}_2) = \frac{1}{|z_{12}|^{4 h_i}} + \frac{C^k_{ii} \, f_{\ket{\Delta_k}}(z_2,\bar{z}_2)}{|z_{12}|^{4 h_i - 2 h_k}} \ldots  \, ,
\end{equation}
where we included the contribution from the lightest primary field $f_{\ket{\Delta_k}}$ in the module of $f_{\ket{\Delta_i}}$.
Then, to this order the four-point function for $z_1 \sim z_2$ and $z_3 \sim z_4$ becomes
\begin{align}\label{eq:corr_light}
    \left\langle f_{\ket{\Delta_i}}(z_1, \bar{z}_1) \ldots f_{\ket{\Delta_i}}(z_4, \bar{z}_4) \right\rangle_\mathbb{C} &\simeq \frac{1}{| z_{12} z_{34}|^{4 h_i}} + \frac{(C^{k}_{ii})^2}{|z_{12}z_{34}|^{4h_i - 2 h_k}} \frac{1}{|z_{24}|^{4 h_k}}
\end{align}
so that
\begin{equation}\label{eq:reesmallxsubleading}
    F_{L_{-n}\ket{\Delta_i}}^{(2)} = 1 - \frac{n + 2 h_i}{2} (\pi x)^2 + \left(C^{k}_{ii}\right)^2 \! \left( \frac{ c (n-1)^2  + 4 n h_i  + 2 n^2 (h_k -1) h_k }{ c (n-1)^2 + 4 n h_i }\right)^2 \! \left( \frac{\pi x}{2} \right)^{4 h_k} + \ldots \, .
\end{equation}
The second term is in general a subleading contribution, e.g.~in the Potts model $\varepsilon \times \varepsilon = \mathbb{I} + X$ with X having dimension $7/5$.
However, due to the fusion rule $\sigma \times \sigma = \mathbb{I} + \varepsilon$ in Ising, in this case $h_k = 1/2$, and we see that the second term in~\eqref{eq:reesmallxsubleading} contributes to leading order.
Indeed, eq.~\eqref{eq:reesmallxsubleading} with $C_{\sigma\sigma}^\varepsilon = \frac12$ correctly predicts the small $x$ behaviour of $F^{(2)}_{L_{-n}\ket{\sigma}}$ for $n=1,2,3$ that we computed (see appendix~\ref{app:REEresultsIsing}).

Some results of the REE in the Ising and three-states Potts models were already considered in~\cite{Palmai:2014jqa,Taddia:2016dbm,Taddia_2013}; we checked that our code produces the same analytic results studied in these references.

\subsection{Sandwiched R\'enyi divergence}

Consider now the correlator $\mathcal{F}^{(2)}_{\ket{s}}$ related to the SRD as in eq.~\eqref{eq:SRDcorr} with $\ket{s} = L_{-1} \ket{\Delta}$.
Then, we find
\begin{equation}\label{eq:srdprimlvl1}
    \mathcal{F}^{(2)}_{\ket{s}} = \bar{D}^{\mathcal{F}(2)} D_{L_{-1}}^{\mathcal{F}(2)}  \, \left\langle f_{\ket{\Delta}}(e^{- i\pi x}) f_{\ket{\Delta}}( e^{ i\pi x} ) f_{\ket{\Delta}}(- e^{- i\pi x}) f_{\ket{\Delta}}( - e^{ i\pi x} ) \right\rangle_\mathbb{C}\,. 
\end{equation}
From the anti-chiral part of the conformal transformation we now obtain
\begin{equation}
    \bar{D}^{\mathcal{F}(2)} = 2^{4 \bar{h}}  \sin ^{4 \bar{h}}(\pi  x)
\end{equation}
and the differential operator acting on the holomorphic coordinates reads
\begin{align}
    D_{L_{-1}}^{\mathcal{F}(2)} &= \frac{2^{4h}}{h^2} e^{-2 i \pi x} \sin ^{4 h}(\pi 
   x) \left[ 4 h^4 e^{2 i \pi  (h+1) x} \left(e^{-2 i \pi  x}\right)^h \right. \nonumber\\
   &\phantom{=} \left. + 2 h^3 e^{i \pi  x} \left(1 - e^{2 i \pi  x}\right) ( \partial_1 + \partial_4 - \partial_2 - \partial_3 )  \right. \nonumber\\
   &\phantom{=} \left. + h^2 \left(e^{2 i \pi  x} - 1\right)^2 ( \partial_1\partial_4 + \partial_2\partial_3 -\partial_1\partial_2  -\partial_1\partial_3 -\partial_2\partial_4 - \partial_3\partial_4 ) \right. \nonumber\\
   &\phantom{=} \left. +4 i h e^{2 i \pi  x} \sin ^3(\pi  x) ( \partial_1\partial_2\partial_3 + \partial_2\partial_3\partial_4 - \partial_1\partial_3\partial_4 - \partial_1\partial_2\partial_4 ) \right. \nonumber\\
   &\phantom{=} \left. + \frac{1}{4} \left(e^{2 i \pi  x} -1 \right)^4 e^{2 i \pi  (h-1) x} \left(e^{-2 i \pi  x}\right)^h \partial_1\partial_2\partial_3\partial_4 \right] \, .
\end{align}

\noindent 
We explicitly study the results for descendants up to level 3. The general expressions for $D$ are, however, again too cumbersome to show them here. 
With the four-point functions \eqref{eq:isingen}, \eqref{eq:isingsig}, \eqref{eq:pottsen} we compute $\mathcal{S}^{(2)}_{\ket{s}}$ for the descendants of the energy and spin primary states in Ising and of the energy state in Potts.
The results are plotted in figure \ref{fig:SRDprim} and some closed expressions are given for descendants of the energy state of Ising in appendix~\ref{app:SRDising}.

\begin{figure}[tb]
    \centering
    \begin{tikzpicture}
        \node at (0,0) {\includegraphics[width=.32\textwidth]{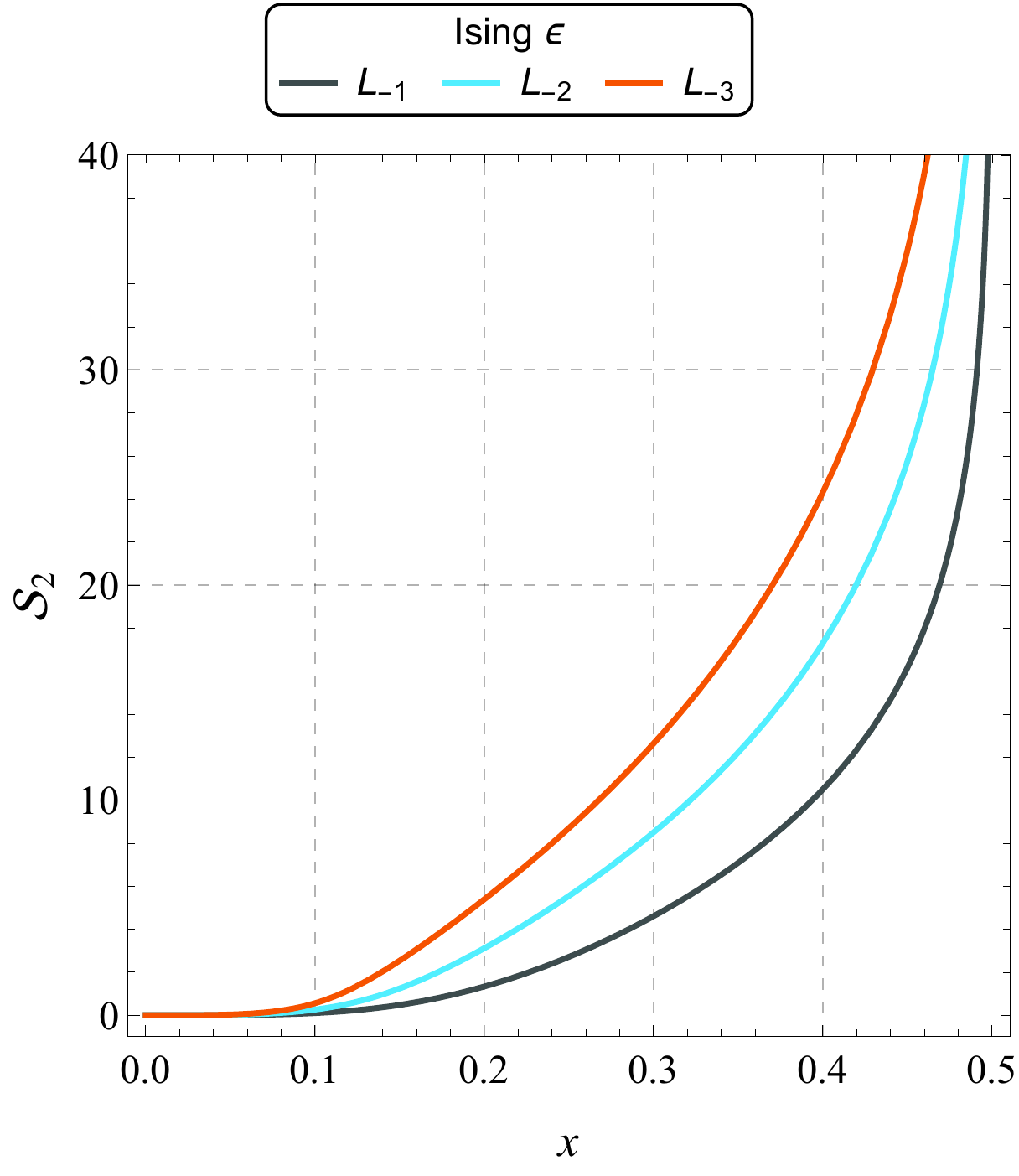}};
    \node at (.33\textwidth,0)  {\includegraphics[width=.32\textwidth]{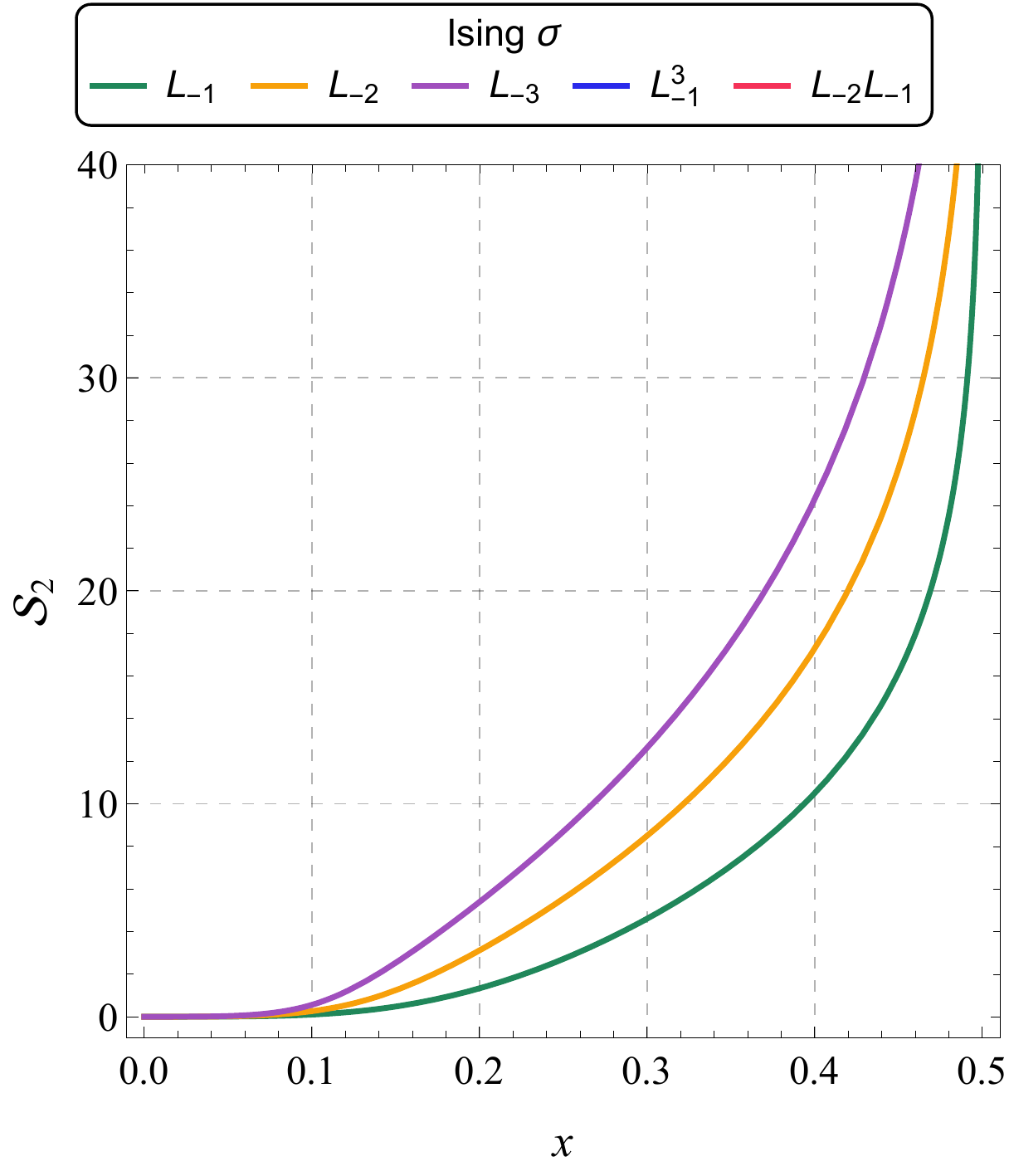}};
    \node at (.66\textwidth,0)  {\includegraphics[width=.32\textwidth]{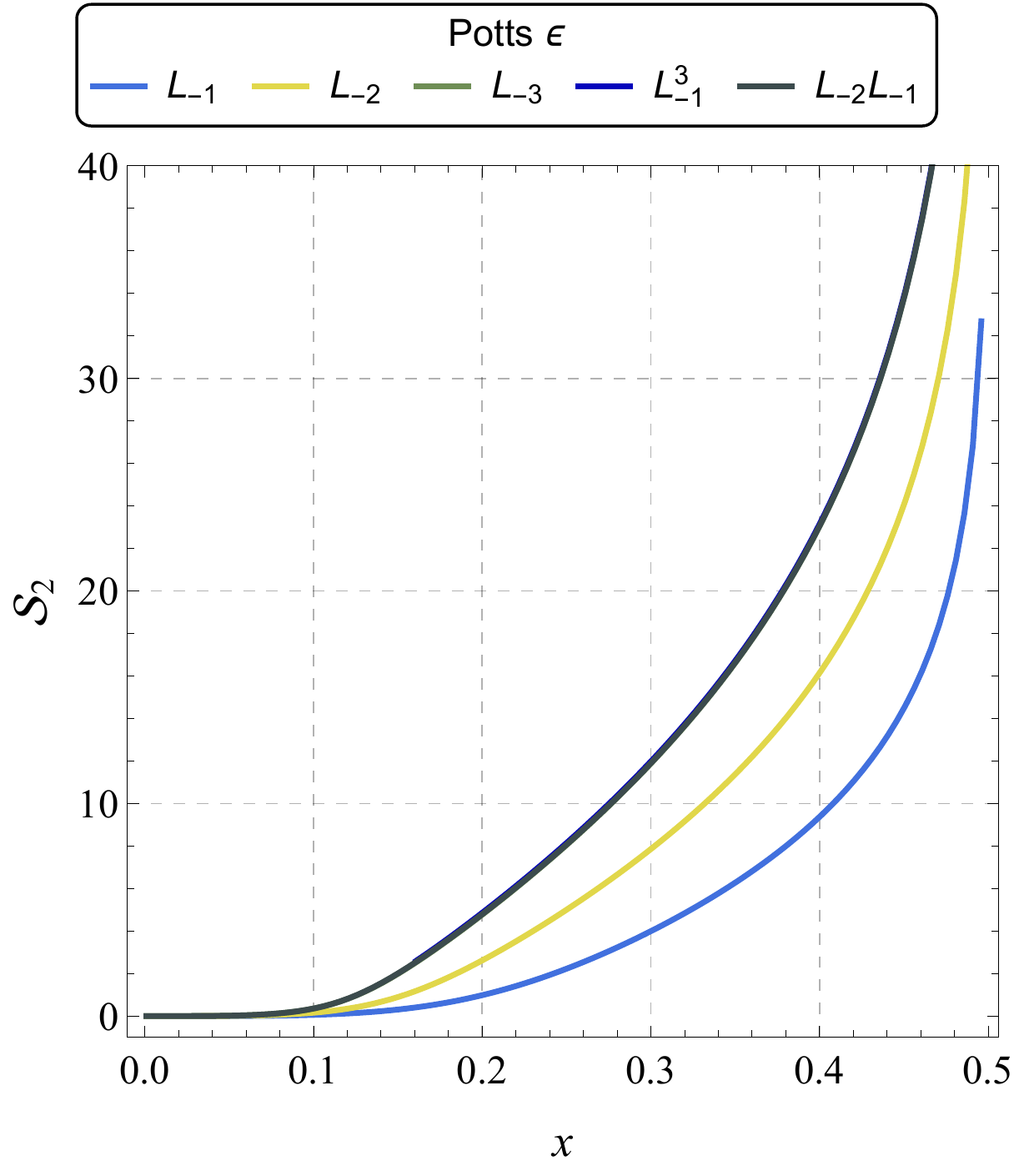}};
    \end{tikzpicture}
    \caption{The Sandwiched R\'enyi Divergence between the reduced groundstate and different descendants of $\ket{\varepsilon}$ and $\ket{\sigma}$ in the Ising model and $\ket{\varepsilon}$ in the Potts model.}
    \label{fig:SRDprim}
\end{figure}

As expected, the SRDs start from 0 and diverge at $x=1/2$. 
We also see from the plots that for higher level descendants the SRD grows more rapidly.
In the Ising model degenerate descendants of $\varepsilon$ at level~2 and~3 produce the same SRDs, while for degenerate descendants of $\sigma$ at level~3 we found three different expressions. 
However, the differences between the plotted results are so small that the three curves at level~3 overlap in figure~\ref{fig:SRDprim}.
The same happens for descendants of $\varepsilon$ in the Potts model.

Now, let us check the limit of small subsystem size.
Consider the OPE between two primary fields ($h_i = \bar{h}_i = \Delta_i/2$)
\begin{equation}\label{eq:ope_srd}
    f_{\Delta_i} (z_1,\bar{z}_1) f_{\Delta_i}(z_2,\bar{z}_2) = \frac{1}{|z_{12}|^{4 h_i}} + \frac{2 h_i c^{-1} \, T(z_2)}{z_{12}^{2h_i -2} \bar{z}_{12}^{2 h_i}} + \frac{2 h_i c^{-1} \, \bar{T}(\bar{z}_2)}{z_{12}^{2h_i} \bar{z}_{12}^{2 h_i - 2}} + \ldots \,,
\end{equation}
where for now we only included the leading contributions from the vacuum module.
Then, if we insert this OPE in the four-point function for $z_1 \sim z_2$ and $z_3\sim z_4$ we obtain
\begin{align}\label{eq:opecorr_srd}
    \left\langle f_{\ket{\Delta}}(z_1, \bar{z}_1) \ldots f_{\ket{\Delta}}(z_4, \bar{z}_4) \right\rangle_\mathbb{C} &\simeq \frac{1}{| z_{12} z_{34}|^{4 h}} + \frac{2 h^2 c^{-1}}{|z_{12}z_{34}|^{4 h}} \frac{z_{12}^2 z_{34}^2}{z_{24}^4}  + \frac{2 h^2 c^{-1}}{|z_{12}z_{34}|^{4 h}} \frac{\bar{z}_{12}^2 \bar{z}_{34}^2}{\bar{z}_{24}^4} \, .
\end{align}

\noindent 
With this expression we can study the limit $x\to 0$ in~\eqref{eq:srdprimlvl1} and similar expressions for higher level descendants.
We find 
\begin{equation}\label{eq:srdsmallxlaw}
    \mathcal{S}^{(2)}_{L_{-n}\ket{\Delta}} = \frac2c \left( n^2 + 2n h + 2 h^2 \right) ( \pi x)^4 + \ldots \, .
\end{equation}

\noindent 
Expanding our analytic results for descendants of the energy in Ising and Potts for $x\to 0$ we found perfect agreement with eq.~\eqref{eq:srdsmallxlaw}.
For $\sigma$ descendants, however, the leading order contribution to the SRD in the limit $x\to 0$ is different.
Indeed, if we think of the OPE as in~\eqref{eq:ope_light} with the correlator~\eqref{eq:corr_light}, then we find the following leading contribution in the SRD for $n=1,2,3$
\begin{equation}\label{eq:srdsmallxsigma}
    \mathcal{S}^{(2)}_{L_{-n}\ket{\Delta_i}} = \left(C^{k}_{ii}\right)^2 \left( \frac{ c (n-1)^2  + 4 n h_i  + 2 n^2 (h_k -1) h_k }{ c (n-1)^2 + 4 n h_i }\right)^2 \left( \pi x \right)^{4 h_k} + \ldots \, .
\end{equation}

\noindent 
Since $h_k = 1/2$ for $\ket{\Delta_i} = \ket{\sigma}$ in the Ising model, we see that the contribution from the $\varepsilon$ channel dominates over the one from the energy momentum tensor in~\eqref{eq:srdsmallxlaw}.
We checked that~\eqref{eq:srdsmallxsigma} with $C^\varepsilon_{\sigma\sigma} = 1/2$ correctly reproduce the $x\to 0$ limit of our results.

It is interesting to consider also the opposite limit $x\to 1/2$ and see how the SRDs scale with the singularity.
In this case, it is enough to consider the first contribution in the OPE~\eqref{eq:ope_srd}, but making the appropriate changes as with our insertion points $x\to 1/2$ means $z_1 \sim z_4$ and $z_2 \sim z_3$.
Then, for $n=1,2,3$ we find the following expression
\begin{equation}
    \mathcal{S}^{(2)}_{L_{-n}\ket{\Delta}} = \log \left( \frac{A_n}{ \pi^{-4 (2h + n)} \left( x -\frac12 \right) ^{-4 (2h + n)}} \right) + \ldots
\end{equation}
with
\begin{equation}
    A_n = (-1)^{8 h} \left( \frac{(n-1)(3n-5)(3n-4) \frac{c}{2} + 4 \left( \frac{6^n}{3} -1 \right) h + 2(n+1)^2 h^2 }{ c(n-1)^2 + 4n h } \right)^2 \, .
\end{equation}
Notice that for $h\to 0$ we recover the same scaling as in~\eqref{eq:srd_vac_divergence}.

In all the examples that we considered, the SRD proved to be a convex function of $x$, providing further evidence to the validity of the R\'enyi QNEC in two dimensions \cite{Moosa:2020jwt} for large enough central charge.

\subsection{Trace square distance}

Consider now the trace square distance between a primary state $\ket{\Delta}$ and its first descendants $L_{-1}\ket{\Delta}$. Then
\begin{equation}\label{eq:TSDprim}
T_{L_{-1}\ket{\Delta},\ket{\Delta}}^{(2)} =  \bar{D}^{T(2)} D_{L_{-1}}^{T(2)} \, \left\langle f_{\ket{\Delta}}(e^{-\frac12 i\pi x}) f_{\ket{\Delta}}( e^{\frac12 i\pi x} ) f_{\ket{\Delta}}(- e^{-\frac12 i\pi x}) f_{\ket{\Delta}}( - e^{\frac12 i\pi x} ) \right\rangle_\mathbb{C} \, ,
\end{equation}
where again the differential operator on the anti-holomorphic coordinates is simply given by the transformation factor
\begin{equation}
    \bar{D}^{T(2)} = \sin ^{4 \bar{h}}(\pi  x)
\end{equation}
while the differential operator on the holomorphic coordinates is given by:
\begin{align}
    D_{L_{-1}}^{T(2)} &= \frac{1}{64} \sin ^{4 h}(\pi  x) \left[ 4 (3 h \cos (2 \pi  x)+5 h-4)^2 \right. \\
    &\phantom{=} \left. + 2 e^{-\frac{5}{2} i \pi  x} \left(2 e^{2 i \pi  x}-3 e^{4 i \pi  x}+1\right) (3 h \cos (2 \pi  x)+5 h-8) \partial_1 \right. \nonumber\\
    &\phantom{=} \left. + 2 e^{-\frac{3}{2} i \pi  x} \left(2 e^{2 i \pi  x}+e^{4 i \pi  x}-3\right) (3 h \cos (2 \pi  x)+5 h-8) \partial_2 \right. \nonumber\\
    &\phantom{=} \left. + h e^{-\frac{9}{2} i \pi  x} \left(1+3 e^{2 i \pi  x}\right)^2 \left(2 e^{2 i \pi  x}+e^{4 i \pi  x}-3\right) \partial_3 \right.\nonumber\\
    &\phantom{=} \left. - h e^{-\frac{7}{2} i \pi  x} \left(3+e^{2 i \pi  x}\right)^2 \left(-2 e^{2 i \pi  x}+3 e^{4 i \pi  x}-1\right)\partial_4 \right. \nonumber\\
    &\phantom{=} \left. + 8 \sin ^2(\pi  x) (3 \cos (2 \pi  x)+5) \left( \partial_3\partial_4 - \partial_2\partial_3 - \partial_1\partial_4 \right) \right. \nonumber\\
    & \phantom{=} \left. -e^{-3 i \pi  x} \left(2 e^{2 i \pi  x}+e^{4 i \pi  x}-3\right)^2 \partial_2\partial_4 -4 e^{-i \pi  x} (2 i \sin (2 \pi  x)+\cos (2 \pi  x)-1)^2 \partial_1\partial_3 \right. \nonumber\\
    &\phantom{=} \left. + \frac{8}{h} \sin ^2(\pi  x) (3 h \cos (2 \pi  x)+5 h-8) \partial_1\partial_2 \right. \nonumber\\
    &\phantom{=} \left. + \frac{16}{h} e^{\frac{i \pi  x}{2}} \sin ^3(\pi  x) (\sin (\pi  x)+2 i \cos (\pi  x)) \partial_2\partial_3\partial_4 \right. \nonumber\\
    &\phantom{=} \left. + \frac{16}{h} e^{-\frac{1}{2} i \pi  x} \sin ^4(\pi  x) (1-2 i \cot (\pi  x)) \left( \partial_1\partial_3\partial_4 - \partial_1\partial_2\partial_3 \right) \right.\nonumber\\
    &\phantom{=} \left. \frac{1}{h}e^{-\frac{7}{2} i \pi  x} \left(e^{2 i \pi  x} -1 \right)^3 \left(3+e^{2 i \pi  x}\right) \partial_1\partial_2\partial_4 + \frac{16}{h^2} \sin ^4(\pi  x) \partial_1\partial_2\partial_3\partial_4 \right]\nonumber
\end{align}

\noindent 
Again, we limit ourselves to display this result, which is the simplest, since for higher descendants the expressions become much more involved.
As in the previous cases, we computed $T^{(2)}_{L_{-n}\ket{\Delta},\ket{\Delta}}$ as in~\eqref{eq:TSDprim} for $n=1,2,3$ and for the degenerate states at level~2 and~3.
Then, by using the four-point functions~\eqref{eq:isingen}, \eqref{eq:isingsig}, and \eqref{eq:pottsen} we obtained analytic expressions for the TSD between the primary state and its descendants for the energy and spin operators in the Ising model and for the energy in the three states Potts model.
Figure~\ref{fig:TSDprim} shows the plots of the results, while in appendix~\ref{app:TSDising} and~\ref{app:TSDpotts} we provide some explicit expressions.

\begin{figure}[tb]
    \centering
    \begin{tikzpicture}
        \node at (0,0) {\includegraphics[width=.32\textwidth]{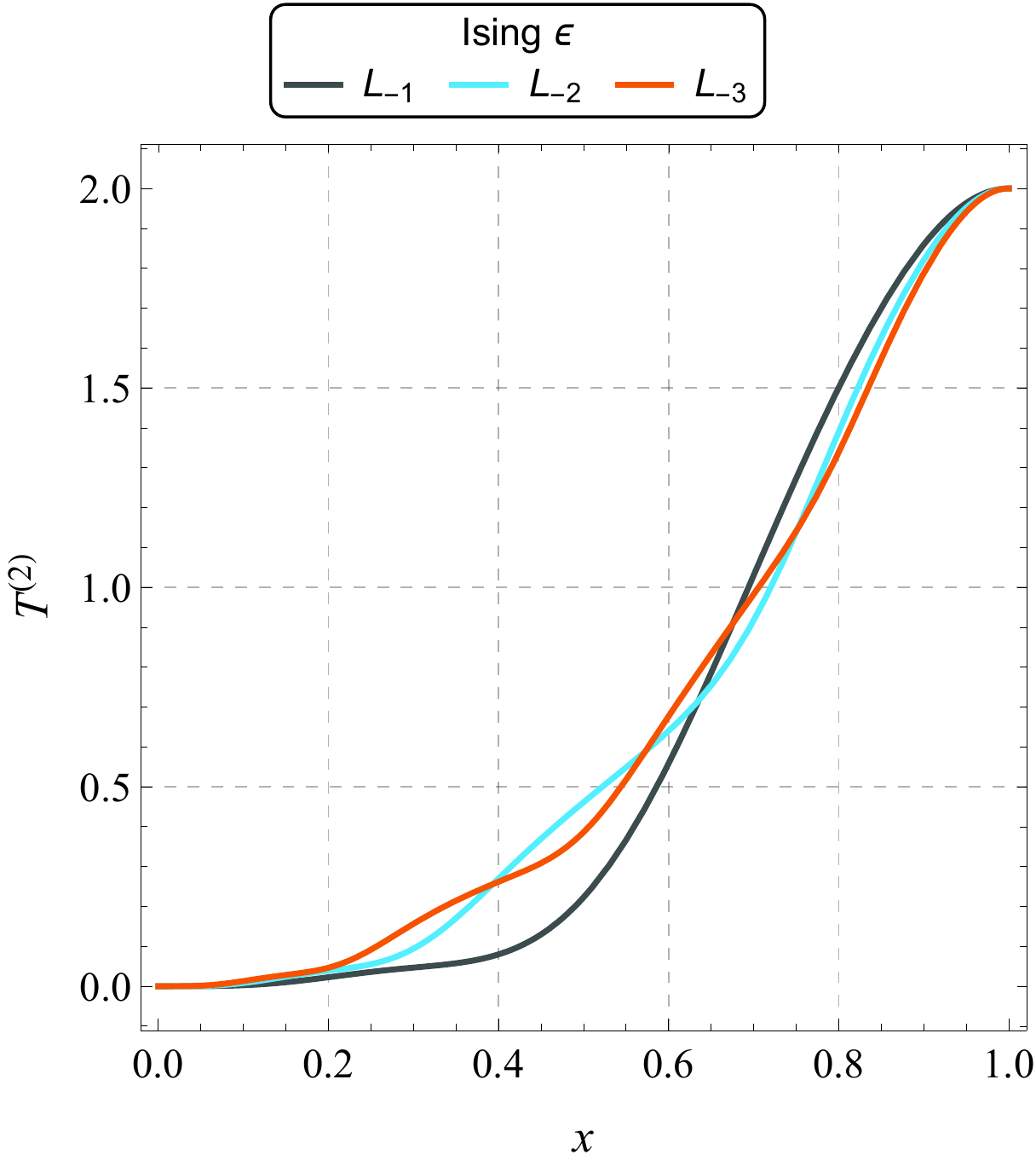}};
    \node at (.33\textwidth,0)  {\includegraphics[width=.32\textwidth]{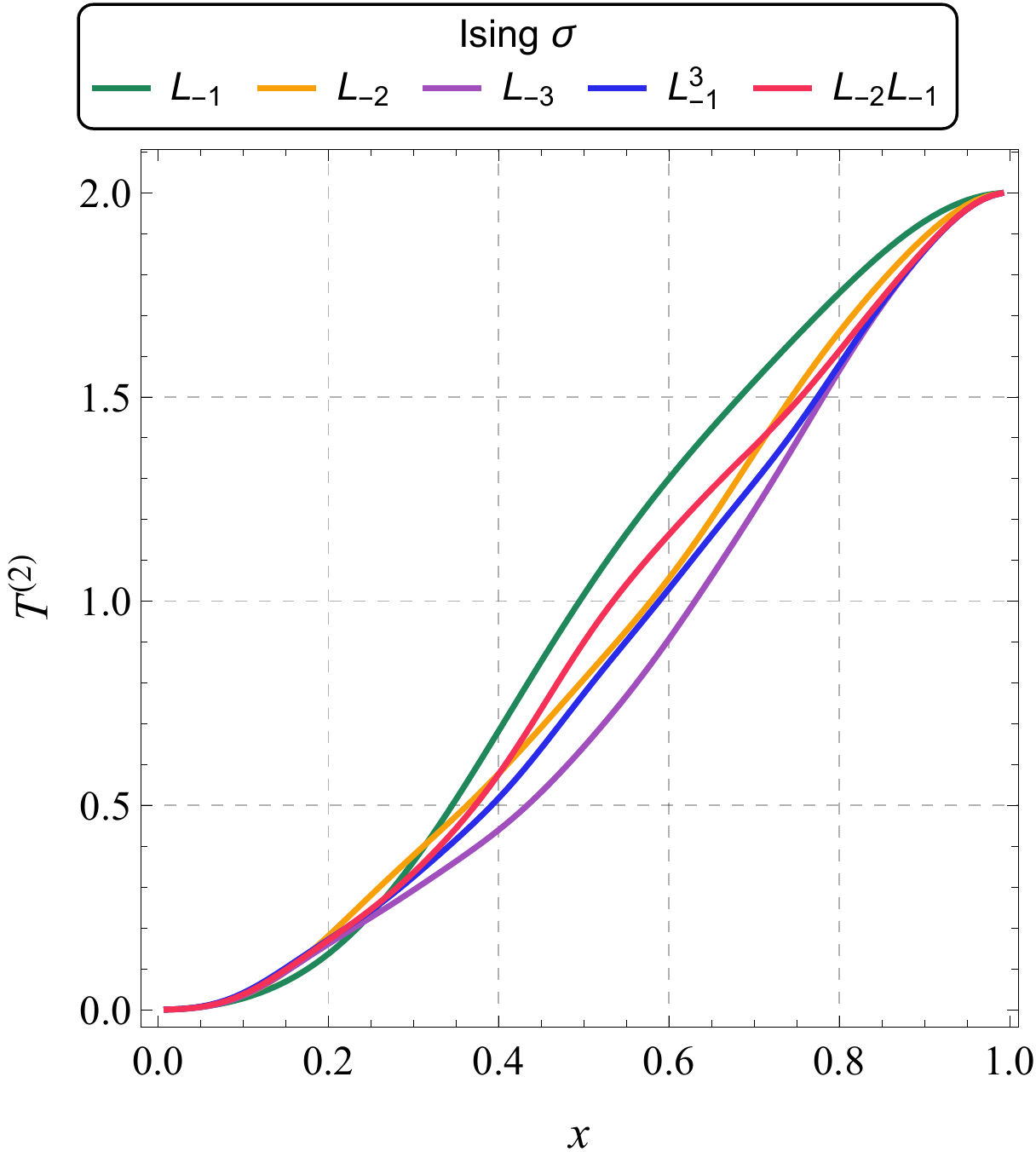}};
    \node at (.66\textwidth,0)  {\includegraphics[width=.32\textwidth]{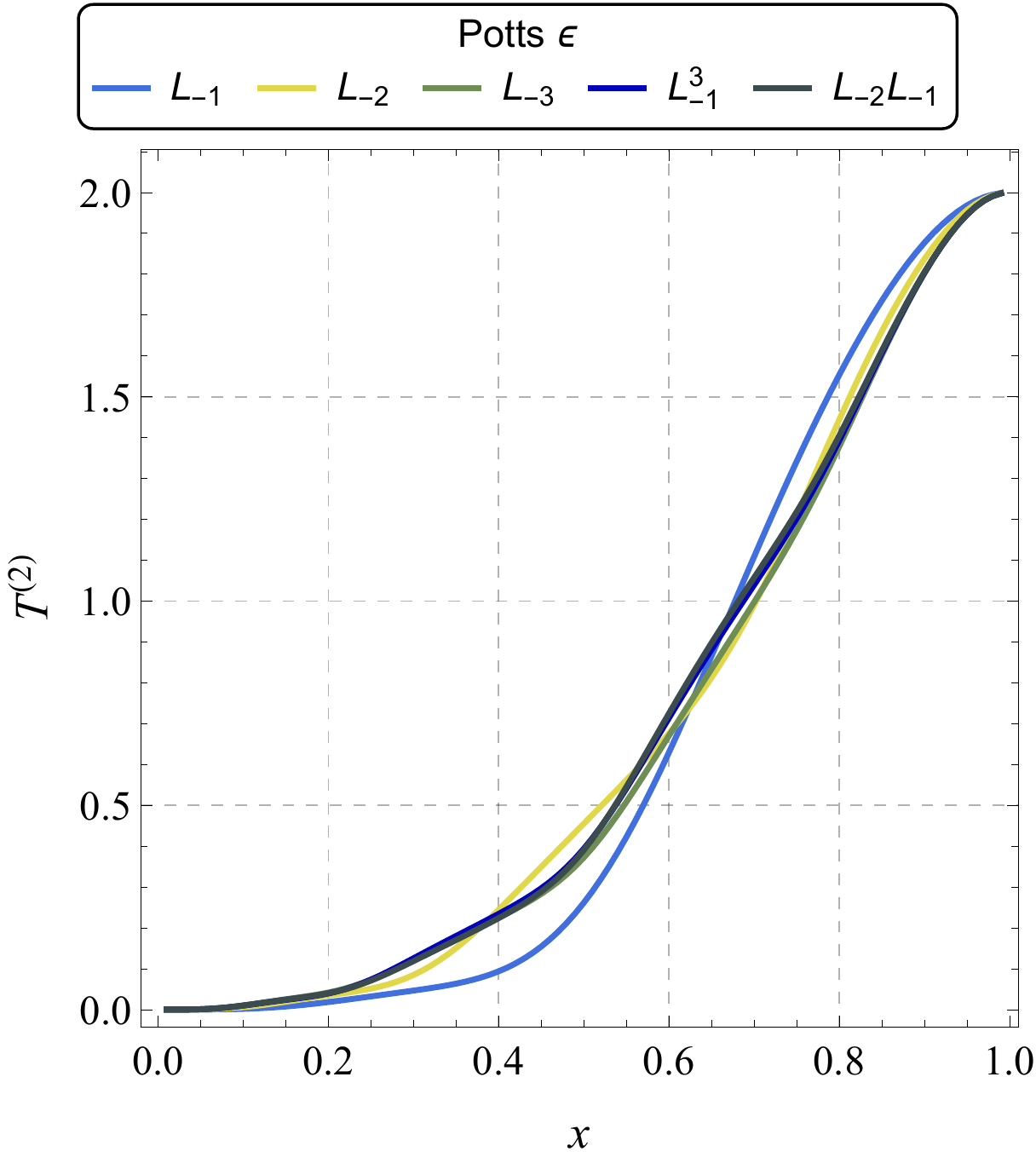}};
    \end{tikzpicture}
    \caption{TSD between different descendants and their primary state $\ket{\varepsilon}$ and  $\ket{\sigma}$ in the Ising model and $\ket{\varepsilon}$ in the Potts model.}
    \label{fig:TSDprim}
\end{figure}

In the Ising model we find that degenerate states of the energy density produce the same TSD w.r.t.~the primary state up to level~3. This again is as expected.
For spin descendants instead this is not true at level~3, with $T^{(2)}_{L_{-3}\ket{\Delta},\ket{\Delta}} \neq T^{(2)}_{L_{-1}^3\ket{\Delta},\ket{\Delta}} \neq T^{(2)}_{L_{-2}L_{-1}\ket{\Delta},\ket{\Delta}}$.
However, in the small and large subsystem size limits we see that these different expressions have the same behaviour, while they differ the most around $x\sim 1/2$.
In the Potts model, TSDs between degenerate states at level~3 and the energy density are again different, but from the plots we see that the difference is barely visible, and in particular for $x\to0$ and $x\to 1$ it is negligible.

If we study the small subsystem size limit, we can generically predict the behaviour of the TSD.
Consider for instance the OPE between two primary states as given by~\eqref{eq:ope_srd} and the correlator as in~\eqref{eq:opecorr_srd}. 
Then, we find the following behaviour in the limit $x\to 0$ for $n=1,2,3$
\begin{equation}\label{eq:tsd_smallx}
    T^{(2)}_{L_{-n}\ket{\Delta},\ket{\Delta}} = \frac{2+c}{16 c} n^2 (\pi x)^4 + O(x^6)
\end{equation}
in agreement with the vacuum result in~\eqref{eq:tsd_vacuum_smallx} and in perfect agreement with the analytic results that we found in Ising and Potts models for energy descendants.
However, for $\sigma$ descendants in the Ising model the next to leading order contribution as $x\to 0$ does not come from the energy momentum tensor but from the energy field $\varepsilon$ in the OPE.
Indeed, consider again the OPE as in~\eqref{eq:ope_light} with the correlator~\eqref{eq:corr_light}, then the contribution to the TSD as $x\to 0$ for $n=1,2,3$ reads
\begin{align}\label{eq:tsd_smallx_nonvacchannel}
    T^{(2)}_{L_{-n}\ket{\Delta},\ket{\Delta}} = \left(C_{hh}^{h_k}\right)^2 \left( \frac{2 n^2 (h_k-1)h_k }{c (n-1)^2 + 4 n h} \right)^2 \left(\frac{\pi x}{2} \right)^{4 h_k} + \ldots \, .
\end{align}

\noindent
We see that this term dominates over the one one outlined in~\eqref{eq:tsd_smallx} for $h_k < 1$, which is the case for the Ising spin.
We checked that~\eqref{eq:tsd_smallx_nonvacchannel} with $C^\varepsilon_{\sigma\sigma}$ perfectly matches the small $x$ behaviour of the results for $\sigma$ in appendix~\ref{app:TSDising}.

Consider now the large subsystem size limit $x \to 1$. 
Then, with our coordinates we have $z_1 \sim z_4$ and $z_2 \sim z_3$ and by taking the OPE similarly as in~\eqref{eq:ope_srd} but with appropriate insertion points we find the behaviour
\begin{equation}
    T^{(2)}_{L_{-n}\ket{\Delta},\ket{\Delta}} = 2 - (2h + \frac{n}{2}) \pi^2 (x-1)^2 + \ldots
\end{equation}
that agrees with the $x\to 1$ limit of the explicit results we found for descendants of the energy in Ising and Potts.
Again, for $\sigma$ descendants we need to take into account the contribution from the lightest field in the OPE.
We then find 
\begin{equation}\label{eq:tsd_largex}
    T^{(2)}_{L_{-n}\ket{\Delta},\ket{\Delta}} = 2 \left(C^{h_k}_{hh}\right)^2 C_n \left( \frac{\pi}{2} \right)^{4 h_k} (x-1)^{4 h_k} \, ,
\end{equation}
where
\begin{align}
    C_n &= \frac{c^2 (n-1)^4 + 4 c n(n-1) \left( 2(n-1) h + (1 - 2^{n-1}) h_k \right)   }{\left( c(n-1)^2 + 4 n h \right)^2} \nonumber\\
    &\phantom{=} + \frac{(4n)^2 h^2 - (2n)^3 h h_k + 2n^4(h_k -1)^2 h_k^2}{\left( c(n-1)^2 + 4 n h \right)^2} \, .
\end{align}
For $\sigma$ in the Ising model $h_k = 1/2$ and we see that the contribution from the $\varepsilon$ channel sums up with the leading correction in~\eqref{eq:tsd_largex}.
Once this is taken into account, we correctly match the large $x$ limit of the $\sigma$ expressions in appendix~\ref{app:TSDising}.

\section{Conclusion and outlook}

In this work we showed how to systematically compute the R\'enyi entanglement entropy, the sandwiched R\'enyi divergence and the trace square distance of generic descendant states reduced to a single interval subsystem in a conformal field theory. In practice the computations can be performed with the help of computer algebra programs and with the implementation of a recursive function that computes any correlator of descendants as a (differential) operator acting on the correlator of the respective primaries.
We explicitly computed the aforementioned quantum measures for rather low excitation in the vacuum module and for excitations of primaries in the Ising model and the three-state Potts model. 

In particular, from the results in the vacuum module we saw that degenerate descendant states only show equal behaviour for small subsystem sizes. At large central charge any of the above quantities behaved very different for degenerate states, as outlined already in sec.~\ref{sec:renyivac}.
This may be a hint that even more generally the holographic R\'enyi entanglement entropy can be very different between degenerate descendant states.
This analysis goes beyond the scope of the present paper, but can be tackled with the code we presented.

We also checked explicitly if predictions from the generalized version of QNEC \cite{Lashkari:2018nsl,Moosa:2020jwt} are true for descendant states, namely that the sandwiched R\'enyi divergence is a convex function of subsystem size. In the Ising model and Potts model in all the cases we checked, the SRD is a convex function. Nonetheless, we could show that for small but positive central charge, the SRD of descendant states in fact becomes non-convex. However, as already stated in section \ref{sec:vacSRD} theories with central charge smaller than 1/2 are quite unusual. 

Many of the analytic expressions that we obtained are too large to show them explicitly. However, showing the results in the small subsystem size limit is possible and they are always in agreement with the expectations from taking the respective limits in the operator product expansion. We again want to state that one very particular result in this limit is that the differences of degenerate states is not visible. Only with larger and larger subsystem size the difference between degenerate states becomes visible (e.g. in the numerous plots we show). 

The existing code that led to our results is openly accessible and can be used to compute the former quantities for more descendant states or in different models. One could for example consider quasiprimary states, i.e. $sl_2$ invariant descendant states in the module and check if they behave special compared to generic descendant states. Other interesting states to study might be those that correspond to currents of the KdV charges (see e.g. \cite{Sasaki:1987mm,Brehm:2019fyy}). 
The code can also be modified easily to compute other (quantum information theoretical) quantities as long as it is possible to express them in terms of correlation functions. There is e.g. a so-called R\'enyi relative entropy (e.g. considered in \cite{Sarosi:2016oks}) that could be computed with the methods presented here. 

There are also various directions to exploit to improve the code, e.g. the possibility to use symmetries in the construction that might speed up the computations significantly. A faster and more efficient code allows to compute higher R\'enyi indices or higher descendants within reasonable time and without too much memory consumption. 

\subsubsection*{Acknowledgments}
We thank Stefan Theisen for comments on the draft of the paper.
MB is supported by the International Max Planck Research School for Mathematical and Physical Aspects of Gravitation, Cosmology and Quantum Field Theory.

\newpage
\appendix

\section{About the action of conformal transformations}

\subsection{Mathematica code to obtain the v's}

\label{app:matv}

\begin{verbatim}
Poly[m_, f_] := Sum[v[j] t^(j + 1) , {j, 1, m - 1}] D[f, t]
PolyToPowerNN[NN_, m_, f_] := If[NN == 1, Poly[m, f],
                                 PolyToPowerNN[NN - 1, m, Poly[m, f]]]
lhs[MM_] := v[0] t + v[0] Sum[1/i! PolyToPowerNN[i, MM, t], {i, 1, MM}]
Equ[NN_] := Block[{tmp},
                tmp=CoefficientList[lhs[NN]
                            -Sum[a[i] t^i,{i,1,NN}],t][[2;;NN+1]];
                {Table[tmp[[i]] == 0, {i, NN}], Table[v[i], {i,0,NN-1}]}
                 ]
    
ListOfVs[NN_] := Block[{tmp},
                    tmp = Equ[NN];
                    Table[v[n],{n,0,NN-1}]/.Solve[tmp[[1]],tmp[[2]]][[1]]]
\end{verbatim}

\subsubsection{Example: Coefficients up to $j=5$}

\begin{align}
    v_0 & = a_1\,,\\
    v_1 & = \frac{a_2}{a_1}\,,\\
    v_2 & = \frac{a_1 a_3 -a_2^2}{a_1^2}\,,\\
    v_3 & = \frac{3a_2^3 - 5 a_1 a_2 a_3 +2a_1^2a_4}{2a_1^3}\,,\\
    v_4 & = -\frac{16a_2^4 -37a_1a_2^2a_3 + 9 a_1^2a_3^2 +18a_1^2a_2 a_4 -6 a_1^3a_5}{6a_1^4}\,,\\
    v_5 & = \frac{31 a_2^5 -92 a_1 a_2^3a_3 +48 a_1^2a_2^2a_4 -7 a_1^2a_2 \left(-7a_3^2+3a_1a_5\right) +3a_1^3\left(-7a_3a_4 +2a_1a_6\right)}{6a_1^5}\,.
\end{align}

\subsection{Local action for the uniformization map}
\label{app:uniformization}

For the local action of the uniformization map
\begin{equation}
    w(z) = \left(\frac{z e^{ -i\pi \frac{l}{L}} - 1}{z -  e^{-i\pi\frac{l}{L}} }\right)^{\frac1n}
\end{equation}
we need the transformation of local coordinates. We choose the standard local coordinates on the $k$th sheet around a point $z_k$
\begin{equation}
   \varphi_{z_k}(\rho) = \rho + z_k\,.
\end{equation}
which are mapped to
\begin{equation}
    \tilde{\beta}_{z_k}(\rho) = \left(\frac{\left(\rho+z_k\right)e^{ -\frac{i\pi l}{L}} - 1}{\rho+z_k -  e^{ -\frac{i\pi l}{L}} }\right)^{\frac1n}\,.
\end{equation}
    
\noindent 
The standard local coordinates on the plane around $w(z_k)$ are simply
\begin{equation}
    \beta_{w(z_k)}(\rho) = \rho + w(z_k)\,.
\end{equation}
    
\noindent
Now, the local coordinate change $\eta_{z_k}$ should satisfy
\begin{equation}
    \tilde{\beta}_{z_k}(\rho) = \beta_{w(z_k)}(\eta_{z_k}(\rho))\,
\end{equation}
and, hence, 
\begin{equation}
    \eta_{z_k}(\rho) = \beta_{w(z_k)}^{-1} \!\left(\tilde{\beta}_{z_k}(\rho)\right)\,.
\end{equation}

\noindent
Since we deal with the standard local coordinates this is straight forward to compute
\begin{equation}
    \eta_{z_k}(\rho) = \tilde{\beta}_{z_k}(\rho) - w(z_k) =  \left(\frac{\left(\rho+z_k\right)e^{ -\frac{i \pi l}{L}} - 1}{\rho+z_k -  e^{ -\frac{i \pi l}{L}} }\right)^{\frac1n} -  \left(\frac{z_k e^{ -\frac{i \pi l}{L}} - 1}{z_k -  e^{ -\frac{i \pi l}{L}} }\right)^{\frac1n}\,,
\end{equation}
and hence for the actual insertion points $z_k = 0_k$ we get
\begin{equation}
    \eta_{0_k}(\rho) = e^{\frac{2\pi i(k-1)}{n} } \left(\left(\frac{\rho \,e^{\frac{i \pi l}{L}} - 1}{\rho -  e^{ \frac{i \pi l}{L}} }\right)^{\frac1n} - e^{\frac{i \pi l}{nL}}\right)\,.
\end{equation}

\noindent 
Expanding this around $\rho$ allows us to solve for the coefficients $v_j$ appearing in the local action $\Gamma_{w(0,k)} \equiv \Gamma_{k,l}$. Up to $j=5$ they are given by
\begin{align}
    v_0 &= \frac{ e^{\frac{2 \pi i (k-1)}{n} + \frac{\pi i (1-n) l }{n L}}}{n} \left(e^{2\pi i \frac{l}{L}} -1\right)\,,\\
    v_1 &= \cos\left(\frac{\pi l}{L}\right) + i \frac{\sin\left(\frac{\pi l}{L}\right)}{n}\,,\\
    v_2 &= \frac{1-n^2}{3n^2} \sin^2\!\left(\frac{\pi l}{L}\right)\,,\\
    v_3 &= \frac{v_2}{2n} \left(n \cos\left(\frac{\pi l}{L}\right) -i \sin\left(\frac{\pi l}{L}\right)\right)\,,\\
    v_4 &= \frac{v_2}{30n^2} \left(n^2-4 + 4(n^2+1) \cos\left(\frac{2\pi l}{L}\right) - 10 i\, n \sin\left(\frac{2 \pi l}{L}\right)\right)\,,\\
    v_5 &= -\frac{v_2 \sin\left(\frac{\pi l}{L}\right) }{18 n^3} \left((11n^3+61n) \sin\left(\frac{2\pi l}{L}\right) + i (61 n^2 + 11) \cos\left(\frac{2\pi l}{L}\right) + i (29n^2-11)\right)\,.
\end{align}

\noindent
Note that for the \textit{dual fields} we basically have to take the composition of the uniformization map with the inversion, i.e. we have to repeat the latter arguments for $w(1/z)$. Let us denote the local coordinate change by $\theta_{0_k}(\rho)$. It is given by 
\begin{equation}
    \theta_{0_k}(\rho) = \eta_{0_k}(\rho)|_{l\to-l}\,,
\end{equation}
s.t. the respective local action is given by $\Gamma_{w(1/z)} = \Gamma_{w(z)} \Gamma_{1/z} \equiv \Gamma_{k,-l}$\,.

\section{About computing correlation functions of descendants}

We use the Mathematica implementation of Virasoro by M. Headrick that can be downloaded from \url{http://people.brandeis.edu/~headrick/Mathematica/}.  

\subsection{Any \texorpdfstring{$N$}{N}-point function of vacuum descendants}\label{app:VacDesCorr}

The Mathematica code to compute any correlator of descendants of the vacuum is

\begin{verbatim}
VacNptFct[stat_] := Module[{states, ntrivial, TMP, tmp0, tmp1, tmp2},
  (*reorders the states s.t. the descendants with more Virasoro 
  generators are mostleft. This makes the recursion faster:*)
  
  states = Sort[stat,
            Length[GetSequence[#1[[1]]]]<Length[GetSequence[#2[[1]]]] &];
            
  (*Checks at which position there are non-trivial descendants:*)
  
  ntrivial = Position[Sign[level[states[[All, 1]]]], 1] // Flatten; 
  (*When there are no non-trivial descendants the function returns 1. 
  For only one descendant it returns 0 due to translation invariance. 
  In any other case it uses the recursion:*)
  
  Which[Length[ntrivial] == 0, 1, Length[ntrivial] == 1, 0, True,
   TMP = Table[
     RecStep[states[[ntrivial[[1]], 1]], states[[ntrivial[[i]], 1]], 
      states[[ntrivial[[1]], 2]], states[[ntrivial[[i]], 2]]], {i, 2, 
      Length[ntrivial]}];
   Sum[Sum[
     TMP[[j - 1, nn, 1]] VacNptFct[
       Table[Which[
         i == ntrivial[[1]], {CutFirst[states[[i, 1]]], 
          states[[i, 2]]}, 
         i == ntrivial[[j]], {TMP[[j - 1, nn, 2]], states[[i, 2]]}, 
         True, states[[i]]], {i, Length[states]}]], {nn, 
      Length[TMP[[j - 1]]]}], {j, 2, Length[ntrivial]}]
   ]
  ]
\end{verbatim}
where we define the functions
\begin{verbatim}
RecStep[ L[n1__], L[n2__], z1_, z2_] := 
 Module[{tmp1, test}, 
  tmp1 = List @@ 
     Expand[Sum[- Coeff[-{n1}[[1]], n, z1, z2] L[n - 1] ** L[n2] ** 
         vac, {n, 0, -Plus[n2] + 1}]] ** vac;
  tmp1 = ReArr[tmp1];
  test = 1;
  While[test == 1, test = 0;
   For[nnn = 1, nnn < Length[tmp1], nnn++, 
    If[GetSequence[tmp1[[nnn, 2]]] == GetSequence[tmp1[[nnn + 1, 2]]],
      test = 1; tmp1[[nnn, 1]] = tmp1[[nnn, 1]] + tmp1[[nnn + 1, 1]]; 
     tmp1 = Delete[tmp1, nnn + 1]; Break;]]];
  tmp1 = Simplify[tmp1]
  ]

ReArr[a_ L[m__]] := {a, L[m]}
ReArr[ L[m__]] := {1, L[m]}
ReArr[a_ ] := {a, 1}
SetAttributes[ReArr, Listable];

Coeff[m_, n_, zi_, zj_] := (-1)^n Binomial[n + m - 2, n] (zj - zi)^(
  1 - m - n)
  
CutFirst[ L[n__]] :=  L[{n}[[2 ;; Length[{n}]]] /. List -> Sequence]
\end{verbatim}

\noindent 
The function \verb+VacNptFct+ takes as arguments a list of descendants together with their coordinates. The descendants are given in the form $\verb+L+[-n_1,...,-n_k]$, where $n_i \ge n_{i+1}$, $n_i\in \mathbb{N}$. The coordinates can either be variables or specific values. For example 
\begin{verbatim}
    VacNptFct[{{L[-2],z},{L[-2],w}}]
\end{verbatim}
gives the result for the two-point function of the energy momentum tensor, $\frac{c/2}{(z-w)^4}$.

\subsection{Any \texorpdfstring{$N$}{N}-point function of descendants of primaries}\label{app:PrimDesCorr}

Given a correlator of descendants of primaries, we compute the differential operator acting on the correlator of primaries with the function \verb|NPtFct|:
\begin{verbatim}
NPtFct[stat_] :=
 Which[
  (* checks the input is given in the correct form :*)
  And @@ Table[Length[stat[[i]]] != 2, {i, Length[stat]}], 
  "The number of fields and coordinates do not match!",
  (* If there is only one descendant then it returns 0 due to 
     translational invariance *)
  Length[stat] == 1, 0,
  True, Module[{states, virpos, derivative, tmp, rec, noone, pr},
   (*reorders the states s.t. 
     the descendants with more Virasoro generators are most left. 
     This makes the recursion faster :*)
   states = 
    Sort[stat, 
     Length[GetSequence[#1[[1]]]] <= Length[GetSequence[#2[[1]]]] &];
   pr = FindPermutation[stat, states];
   virpos = Position[Sign[level[states[[All, 1]]]], 1] // Flatten;
   derivative = 
    Table[Length[GetSequence[states[[virpos[[i]], 1]]]] == 
      level[states[[virpos[[i]], 1]]], {i, 1, Length[virpos]}];
   (*When there are no non-
     trivial descendants the function returns corrp[...]. 
     If the descendants are only level 1 descendants, 
     it returns the appropriate derivatives acting on corrp[...]. 
     In any other case it uses the recursion :*)
   Which[
    Length[virpos] == 0, corrp[stat[[All, 2]] /. List -> Sequence],
    And @@ derivative, 
    Derivative[level[stat[[All, 1]]] /. List -> Sequence][corrp][
     stat[[All, 2]] /. List -> Sequence],
    True, noone = Position[derivative, False] // Flatten;
    rec = 
     Drop[Table[i, {i, 1, Length[states]}], {virpos[[noone[[1]]]]}];
      tmp = 
     Table[RecStep[states[[virpos[[noone[[1]]]], 1]], states[[i, 1]], 
       states[[virpos[[noone[[1]]]], 2]], states[[i, 2]]], {i, rec}];
    Sum[tmp[[i, j, 1]] NPtFct[
       Permute[
        ReplacePart[
         states, {{virpos[[noone[[1]]]], 1} -> 
           CutFirst[states[[virpos[[noone[[1]]]], 1]]], {rec[[i]], 
            1} -> tmp[[i, j, 2]]}], Ordering[PermutationList[pr]]]
       ], {i, 1, Length[tmp]}, {j, 1, Length[tmp[[i]]]}]]]]
\end{verbatim}
where we define the functions:
\begin{verbatim}
RecStep[ L[n1__] ** prim[p1_], L[n2__] ** prim[p2_], z1_, z2_] := 
 Module[{tmp1, test}, 
  tmp1 = List @@ 
    Expand[Sum[-Coeff[-{n1}[[1]],n,z1,z2] L[n-1]**L[n2]**prim[p2],
            {n, 0, -Plus[n2] + 1}]];
  tmp1 = ReArr[tmp1];
  test = 1;
  While[test == 1, test = 0;
   For[nnn = 1, nnn < Length[tmp1], nnn++, 
    If[GetSequence[tmp1[[nnn, 2]]] == GetSequence[tmp1[[nnn + 1, 2]]],
      test = 1; tmp1[[nnn, 1]] = tmp1[[nnn, 1]] + tmp1[[nnn + 1, 1]]; 
     tmp1 = Delete[tmp1, nnn + 1]; Break;]]];
  tmp1 = Simplify[tmp1]
  ]

RecStep[ L[n1__] ** prim[p1_], prim[p2_], z1_, z2_] := 
 Module[{tmp1, test}, 
  tmp1 = List @@ 
    Expand[-Sum[
       Coeff[-{n1}[[1]], n, z1, z2] L[n - 1] ** prim[p2], {n, 0, 1}]];
  tmp1 = ReArr[tmp1];
  test = 1;
  While[test == 1, test = 0;
   For[nnn = 1, nnn < Length[tmp1], nnn++, 
    If[GetSequence[tmp1[[nnn, 2]]] == GetSequence[tmp1[[nnn + 1, 2]]],
      test = 1; tmp1[[nnn, 1]] = tmp1[[nnn, 1]] + tmp1[[nnn + 1, 1]]; 
     tmp1 = Delete[tmp1, nnn + 1]; Break;]]];
  tmp1 = Simplify[tmp1]
  ]
  
ReArr[a_ L[m__] ** prim[p_]] := {a, L[m] ** prim[p]}
ReArr[L[m__] ** prim[p_]] := {1, L[m] ** prim[p]}
ReArr[a_ prim[p_]] := {a, prim[p]}
ReArr[prim[p_]] := {1, prim[p]}
SetAttributes[ReArr, Listable];

CutFirst[ L[n__] ** prim[p_]] :=  
 L[{n}[[2 ;; Length[{n}]]] /. List -> Sequence] ** prim[p]
 
GetSequence[L[m__] ** prim[p_]] := {m}
GetSequence[prim[p_]] := {}
SetAttributes[GetSequence, Listable];
\end{verbatim}

\noindent
The function \verb|NPtFct| takes as arguments a list of $N$ lists, where in the innermost lists the first entry is the descendant and the second entry is the coordinate.
The descendants are given as $\verb|L[|-n_1,...,-n_k\verb|]**prim[p]|$, where again $n_i \ge n_{i+1}$, $n_i\in \mathbb{N}$ and \verb|prim[p]| denotes the primary state.
For instance, 
\begin{verbatim}
    tp = NPtFct[{ {L[-2] ** prim[p], z}, {L[-1, -1] ** prim[p], w} }]
\end{verbatim}
produces the output
\begin{equation*}
\frac{6 \text{corrp}[z,w] h[p]}{(w-z)^4}-\frac{2 (1+2 h[p]) \text{corrp}^{(0,1)}[z,w]}{(w-z)^3}+\frac{(2+h[p]) \text{corrp}^{(0,2)}[z,w]}{(w-z)^2}+\frac{\text{corrp}^{(0,3)}[z,w]}{-w+z}
\end{equation*}
where \verb|h[p]| is the conformal dimension of \verb|prim[p]| and the function \verb|corrp|, which is a function of the insertion points, denotes the correlator of primaries. The derivatives acting on it are displayed in the Mathematica language.
If we know the explicit expression of \verb|corrp|, we can further simplify the output; in our example we can for instance write:
\begin{verbatim}
    corrp[z1_, z2_] := 1/(z1 - z2)^(2 h[p])
    tp // Simplify
    Clear[corrp]
\end{verbatim}
to get the explicit result
\begin{equation*}
    6 (-w+z)^{-2 (2+h[p])} h[p] \left(3+5 h[p]+2 h[p]^2\right) \, .
\end{equation*}

\section{Explicit results}

\subsection{R\'enyi entanglement entropy}

\subsubsection{Vacuum module}\label{app:REresultsVac}

The second R\'enyi entanglement entropy for $L_{-3}\ket{0}$, $L_{-4}\ket{0}$, and $L_{-5}\ket{0}$ are 

\begin{align}\label{eq:RFE[-3]}
    F^{(2)}_{L_{-3}\ket{0}} &= \frac{\sin ^8(\pi x) \cos ^4(\pi x)}{64} c^2 \\
    &\quad+\frac{c \sin ^4(\pi x) \cos ^2(\pi x) (255 \cos (2 \pi x)+90 \cos (4 \pi x)+17 \cos (6 \pi 
   l)+1686)}{8192} \nonumber\\
   &\quad+\frac{\sin ^4(\pi x) (8391 \cos (2 \pi x)+1890 \cos (4 \pi x)+361 \cos (6 \pi x)+7790)}{16384 c}\nonumber\\
   &\quad+\frac{3032808 \cos (2 \pi x)+819919 \cos (4 \pi x)-27612 \cos (6 \pi x)}{8388608}\nonumber\\
   &\quad+\frac{386 \cos (8 \pi x)+8436 \cos (10 \pi x)+289
   \cos (12 \pi x)+4554382}{8388608}\nonumber
\end{align}

\begin{align}
    F^{(2)}_{L_{-4}\ket{0}} &= \quad\frac{c^2 \sin ^8(\pi x) (3 \cos (2 \pi x)+2)^4}{6400} \label{eq:RFE[-4]}\\
    &\quad+\frac{c \sin ^4(\pi x) (3 \cos (2 \pi x)+2)^2 (12760 \cos (2 \pi x)+7396 \cos (4 \pi
    l)+2152 \cos (6 \pi x))}{6553600}\nonumber\\
    &\quad+\frac{c \sin ^4(\pi x) (3 \cos (2 \pi x)+2)^2 (1263 \cos (8 \pi x)+140269)}{6553600}\nonumber \\
    &\quad+\frac{\sin ^4(\pi x) (5444642 \cos (2 \pi x)+2684168 \cos (4 \pi x)+913973 \cos (6 \pi x))}{6553600 c}\nonumber \\
    &\quad+\frac{\sin ^4(\pi x) (286934 \cos (8 \pi x)+59049 \cos (10 \pi x)+3718434)}{6553600 c}\nonumber \\
    &\quad+\frac{16641312784 \cos (2 \pi x)+4954285000 \cos (4 \pi  x)+1976400688 \cos (6 \pi x)}{53687091200}\nonumber \\
   &\quad+\frac{-121298020 \cos (8 \pi x)-5870960 \cos (10 \pi x)+13794296 \cos (12 \pi x)}{53687091200}\nonumber \\
   &\quad+\frac{18977328 \cos (14 \pi x)+1595169
   \cos (16 \pi  x)+30207894915}{53687091200}\nonumber
\end{align}

\begin{align}
   F^{(2)}_{L_{-5}\ket{0}} &= \quad\frac{c^2 (-16777216 \cos (2 \pi  x)-51380224 \cos (4 \pi  x)-79691776 \cos (6 \pi  x))}{3435973836800} \label{eq:RFE[-5]}\\
   &\quad+\frac{c^2 (-98566144 \cos (8 \pi  x)+121634816 \cos (10 \pi 
   x)+17825792 \cos (12 \pi  x))}{3435973836800}\nonumber \\
   &\quad+\frac{c^2 (8388608 \cos (14 \pi  x)-8388608 \cos (16 \pi  x)-33554432 \cos (18 \pi  x))}{3435973836800}\nonumber \\
   &\quad+\frac{c^2 (16777216 \cos (20 \pi x)+123731968)}{3435973836800}\nonumber \\
   &\quad+\frac{c (-4076806144 \cos (2 \pi  x)-9140649984 \cos (4 \pi  x))}{3435973836800}\nonumber \\
   &\quad+\frac{c (-14113284096 \cos (6 \pi  x)-18862759936 \cos (8 \pi  x))}{3435973836800}\nonumber \\
   &\quad+\frac{c (18117304320 \cos (10 \pi  x)-27205632 \cos (12 \pi  x)+101367808 \cos (14 \pi  x))}{3435973836800}\nonumber \\
   &\quad+\frac{c (114972672 \cos (16 \pi  x)-28581888 \cos (18 \pi x))}{3435973836800}\nonumber \\
   &\quad+\frac{c (74997760 \cos (20 \pi  x)+27840645120)}{3435973836800}\nonumber \\
   &\quad+\frac{-19727178304 \cos (2 \pi  x)-27011932672 \cos (4 \pi  x)-22303010688 \cos (6
   \pi  x)}{3435973836800 c}\nonumber \\
   &\quad+\frac{-10523886336 \cos (8 \pi  x)+8735760000 \cos (10 \pi  x)+2720936448 \cos (12 \pi  x)}{3435973836800 c}\nonumber \\
   &\quad+\frac{1016710944 \cos (14 \pi  x)+879348416 \cos (16
   \pi  x)}{3435973836800 c}\nonumber \\
   &\quad+\frac{919004192 \cos (18 \pi  x)+65294248000}{3435973836800 c}\nonumber \\
   &\quad+\frac{967144492584 \cos (2 \pi  x)+295129895330 \cos (4 \pi  x)+135116995760
   \cos (6 \pi  x)}{3435973836800}\nonumber \\
   &\quad+\frac{61785824936 \cos (8 \pi  x)-5774059280 \cos (10 \pi  x)-260030763 \cos (12 \pi  x)}{3435973836800}\nonumber \\
   &\quad+\frac{179443820 \cos (14 \pi  x)+597122990 \cos
   (16 \pi  x)+506168268 \cos (18 \pi  x)}{3435973836800}\nonumber \\
   &\quad+\frac{83814025 \cos (20 \pi  x)+1981464169130}{3435973836800}\nonumber
\end{align}

\subsubsection{Ising model}\label{app:REEresultsIsing}
Up to level 3 descendants of the energy density operator we find the following results for the $n=2$ R\'enyi entanglement entropy:
\begin{align}
    F^{(2)}_{L_{-1}\ket{\varepsilon}} &= \frac{(\cos (2 \pi  x)+7) (1558 + 439 \cos (2 \pi  x)+26 \cos (4 \pi  x)+25 \cos (6 \pi  x))}{16384} \\
    F^{(2)}_{L_{-2}\ket{\varepsilon}} &= \frac{\cos (2 \pi  x)+7}{67108864} (6085442 + 1693410 \cos (2 \pi  x)+514952 \cos (4 \pi  x) \nonumber\\
    &\phantom{ = \frac{\cos (2 \pi  x)+7}{67108864} } +49813 \cos (6 \pi  x)+9270 \cos (8 \pi  x)+35721 \cos (10 \pi  x)) \\
    &= F^{(2)}_{L_{-1}^2\ket{\varepsilon}} \nonumber\\
    F^{(2)}_{L_{-3}\ket{\varepsilon}} &= \frac{\cos (2 \pi  x)+7}{17179869184} (1523423468 + 432147835 \cos (2 \pi  x)+111740030 \cos (4 \pi  x) \nonumber\\ 
    & \phantom{=\frac{\cos (2 \pi  x)+7}{17179869184} } +65921129 \cos (6 \pi  x)+7438836 \cos (8 \pi  x)+1584475 \cos (10 \pi  x) \nonumber\\ 
    & \phantom{=\frac{\cos (2 \pi  x)+7}{17179869184} } +626850 \cos
   (12 \pi  x)+4601025 \cos (14 \pi  x)) \\
   &= F^{(2)}_{L_{-1}^3\ket{\varepsilon}} = F^{(2)}_{L_{-2}L_{-1}\ket{\varepsilon}} \nonumber
\end{align}
where the common prefactor is due to the factorization of the holomorfic and antiholomorfic parts of the correlator.
Even though $\mathcal{D}^{F(2)}_{L_{-1}^2} \neq \mathcal{D}^{F(2)}_{L_{-2}}$, at level~2 we find the same entanglement entropy for the different descendants and the same happens at level~3.
This reflects the existence of only one physical state at level~2 and~3.

For $\sigma$ descendants:
\begin{align}
    F^{(2)}_{L_{-1}\ket{\sigma}} &= \frac{435 + 60 \cos (2 \pi  x)+17 \cos (4 \pi  x)}{512} \displaybreak[0]\\
    F^{(2)}_{L_{-2}\ket{\sigma}} &= \frac{1560707 + 438088 \cos (2 \pi  x)+75420 \cos (4 \pi  x)}{2097152} \nonumber\\
    &\phantom{=} +\frac{8312 \cos (6 \pi  x)+14625 \cos (8 \pi  x)}{2097152} \\
    &= F^{(2)}_{L_{-1}^2\ket{\sigma}} \displaybreak[0]\nonumber\\
    F^{(2)}_{L_{-3}\ket{\sigma}} &= \frac{42511910 + 16535144 \cos (2 \pi  x)+2825131 \cos (4 \pi  x)+1123684 \cos (6 \pi  x)}{63438848} \nonumber\\
   &\phantom{=} + \frac{179114 \cos (8 \pi  x)+141364 \cos (10 \pi  x)+122501 \cos (12 \pi 
   x)}{63438848} \displaybreak[0]\\
   F^{(2)}_{L_{-1}^3\ket{\sigma}} &= \frac{ 3968670881070 + 1175831066472 \cos (2 \pi  x)+306581016863 \cos (4 \pi  x)}{5585604968448} \nonumber\\
   &\phantom{=} + \frac{102222772068 \cos (6 \pi  x)+11235770850 \cos (8 \pi  x)}{5585604968448} \nonumber\\
   &\phantom{=} + \frac{5235592500 \cos (10 \pi  x)+15827868625
   \cos (12 \pi  x)}{5585604968448} \displaybreak[0]\\
   F^{(2)}_{L_{-2}L_{-1}\ket{\sigma}} &= \frac{120071187054 + 40660967528 \cos (2 \pi  x)+9353937343 \cos (4 \pi  x)}{173946175488} \nonumber\\
   &\phantom{=} + \frac{2427785700 \cos (6 \pi  x)+860494498 \cos (8 \pi  x)}{173946175488} \nonumber\\
   &\phantom{=} + \frac{152046004 \cos (10 \pi  x)+419757361 \cos (12 \pi 
   x)}{173946175488}
\end{align}
In this case we have one physical state at level~2, while two physical states at level~3 and we thus find different expressions for the REEs for degenerate states at level~3.

\subsubsection{Three-state Potts model}\label{app:REEresultsPotts}
For the first descendant of the energy density in the three-states Potts model we find: 
\begin{align}
    F^{(2)}_{L_{-1}\ket{\varepsilon}} &= \tfrac{1}{64} \, F\!\left(-\tfrac{8}{5},-\tfrac{1}{5};-\tfrac{2}{5};\eta \right){}^2 (\cos (2 \pi  x)+7)^2 \nonumber\\
    &\phantom{=} +\tfrac{1}{166400}\big\{ F\left(-\tfrac{8}{5},-\tfrac{1}{5};-\tfrac{2}{5};\eta \right) \sin ^4(\pi  x) \big[9 \sin (\pi  x) \left(49 F\!\left(\tfrac{12}{5},\tfrac{19}{5};\tfrac{18}{5};\eta \right) \sin ^3(\pi  x) \right.  \nonumber\\
    &\phantom{=} \left. +260 F\! \left(\tfrac{7}{5},\tfrac{14}{5};\tfrac{13}{5};\eta \right) \sin (2 \pi 
   x)\right)+5200 F\!\left(-\tfrac{3}{5},\tfrac{4}{5};\tfrac{3}{5};\eta \right) \cos (\pi  x) \nonumber\\
   & \phantom{=}  +520 F\!\left(\tfrac{2}{5},\tfrac{9}{5};\tfrac{8}{5};\eta \right)
   (31 \cos (2 \pi  x)+41)\big] \big\} \nonumber\\
   &\phantom{=} -\left(111411200\ 2^{2/5} \Gamma \left(-\tfrac{8}{5}\right) \Gamma \left(\tfrac{17}{10}\right) \Gamma
   \left(\tfrac{12}{5}\right)\right)^{-1} \nonumber\\
   & \phantom{=} \times \bigg\{\Gamma \left(-\tfrac{2}{5}\right) \Gamma \left(\tfrac{3}{10}\right) \Gamma \left(\tfrac{13}{5}\right) F\! \left(\tfrac{6}{5},\tfrac{13}{5};\tfrac{12}{5};\eta \right) \sin ^{\tfrac{28}{5}}(\pi  x) \nonumber\\
   &\phantom{=}  \times \big[ 34 F\! \left(\tfrac{6}{5},\tfrac{13}{5};\tfrac{12}{5};\eta
   \right) (14196 \cos (2 \pi  x)+15129 \cos (4 \pi  x)+7667) \nonumber\\
   &\phantom{=} +13 \sin ^2(\pi  x) \big(9016 F\!\left(\tfrac{26}{5},\tfrac{33}{5};\tfrac{32}{5};\eta \right) \sin
   ^6(\pi  x)  \nonumber\\
   &\phantom{=} +17 F\! \left(\tfrac{11}{5},\tfrac{18}{5};\tfrac{17}{5};\eta \right) (8977 \cos (\pi  x)+6479 \cos (3 \pi  x))  \nonumber\\
   &\phantom{=}  +79488 F\! \left(\tfrac{21}{5},\tfrac{28}{5};\tfrac{27}{5};\eta \right) \sin ^4(\pi  x) \cos (\pi  x) \nonumber\\
   &\phantom{=}  +99 F\! \left(\tfrac{16}{5},\tfrac{23}{5};\tfrac{22}{5};\eta \right)
   \sin ^2(\pi  x) (1373 \cos (2 \pi  x)+1003)\big)\big] \bigg\}
\end{align}
where $F \equiv \; _2F_1 $ is the hypergeometric function and $\eta = \sin^2 \left( \frac{\pi x}{2} \right)$.
For higher level descendants the expressions are more involved, and we limit ourselves to show this simplest example.

\subsection{Sandwiched R\'enyi divergence}

\subsubsection{Vacuum module}\label{app:SRDresultsvac}

Some explicit expressions for the SRD between the vaccum and light states:

\begin{align}
    F^{(2)}_{L_{-3}\ket{0}} =&\quad \frac{e^{12 i \pi x} (-202752 \cos (2 \pi x)-3775424 \cos (4 \pi x)+356352 \cos (6 \pi x)}{2048 c \left(1+e^{2 i \pi x}\right)^{12}}\\
   &+\frac{888064 \cos (8 \pi x)-184320 \cos (10 \pi x)-137\cos (12 \pi x)+27648 \cos (14 \pi x)}{2048 c \left(1+e^{2 i \pi 
   x}\right)^{12}}\nonumber\\
   &+\frac{14272 \cos (16 \pi x)+3072 \cos (18 \pi x)+288 \cos (20 \pi x)+2874176)}{2048 c \left(1+e^{2 i \pi x}\right)^{12}}\nonumber\\
   &+\frac{e^{12 i \pi x} (-86237760 \cos (2 \pi x)+49130280 \cos (4 \pi x)-14301120 \cos (6 \pi x))}{2048 \left(1+e^{2 i \pi x}\right)^{12}}\nonumber\\
   &+\frac{e^{12 i \pi x} (2900567 \cos (8 \pi x)-122592\cos (10 \pi x)+12228 \cos (12 \pi x))}{2048 \left(1+e^{2 i \pi x}\right)^{12}}\nonumber\\
   &+\frac{e^{12 i \pi x} (-1568 \cos (14 \pi x)-2062 \cos (16 \pi x)-416 \cos (18 \pi x))}{2048 \left(1+e^{2 i \pi x}\right)^{12}}\nonumber\\
   &+\frac{e^{12 i \pi x} (276 \cos (20 \pi x)+160 \cos (22 \pi 
   l)+25 \cos (24 \pi x)+57010590)}{2048 \left(1+e^{2 i \pi x}\right)^{12}}\nonumber
\end{align}

\begin{align}
    F^{(2)}_{L_{-4}\ket{0}} =&\quad\frac{e^{16 i \pi x} (-2183086080 \cos (2 \pi x)-14065619072 \cos (4 \pi x))}{163840 c \left(1+e^{2 i \pi x}\right)^{16}}\\
   &+\frac{e^{16 i \pi x} (3329505280 \cos (6 \pi x)+3373021952 \cos (8 \pi x)-1144576000 \cos
   (10 \pi x))}{163840 c \left(1+e^{2 i \pi 
   x}\right)^{16}}\nonumber\\
   &+\frac{e^{16 i \pi x} (-7513472 \cos (12 \pi x)-26378240 \cos (14 \pi x)+22635008 \cos (16 \pi x))}{163840 c \left(1+e^{2 i \pi x}\right)^{16}}\nonumber\\
   &+\frac{e^{16 i \pi x} (21217280 \cos (18 \pi x)+10149760 \cos (20 \pi 
   l)+3225600 \cos (22 \pi x))}{163840 c \left(1+e^{2 i \pi x}\right)^{16}}\nonumber\\
   &+\frac{e^{16 i \pi x} (693504 \cos (24 \pi x)+92160 \cos (26 \pi x)+5760 \cos (28 \pi x)+10666626560)}{163840 c \left(1+e^{2 i \pi 
   x}\right)^{16}}\nonumber\\
   &+\frac{e^{16 i \pi x} (-1927558100400 \cos (2 \pi x)+1107347224880 \cos (4 \pi x))}{163840 \left(1+e^{2 i \pi x}\right)^{16}}\nonumber\\
   &+\frac{e^{16 i \pi x} (-420523178000 \cos (6 \pi x)+102577128040 \cos
   (8 \pi x))}{163840 \left(1+e^{2 i \pi x}\right)^{16}}\nonumber\\
   &+\frac{e^{16 i \pi x} (-14129186800 \cos (10 \pi x)+1083586960 \cos (12 \pi x))}{163840 \left(1+e^{2 i \pi x}\right)^{16}}\nonumber\\
   &+\frac{e^{16 i \pi x} (-18763600 \cos (14 \pi x)+2105260 \cos (16 \pi x)-2374000 \cos (18 \pi 
   l))}{163840 \left(1+e^{2 i \pi x}\right)^{16}}\nonumber\\
   &+\frac{e^{16 i \pi x} (-2296400 \cos (20 \pi x)-722000 \cos (22 \pi x)+205400 \cos (24 \pi x))}{163840 \left(1+e^{2 i \pi x}\right)^{16}}\nonumber\\
   &+\frac{e^{16 i \pi x} (286800 \cos (26 \pi x)+120720 \cos (28 \pi x)+25200 \cos (30 \pi 
   l))}{163840 \left(1+e^{2 i \pi x}\right)^{16}}\nonumber\\
   &+\frac{e^{16 i \pi x} (2205 \cos (32 \pi x)+1161961353975)}{163840 \left(1+e^{2 i \pi x}\right)^{16}}\nonumber
\end{align}

\subsubsection{Ising model}\label{app:SRDising}
We present here some of the correlation functions related to the SRD computation.
For simplicity we show only the results for $\varepsilon$ descendants in the Ising model:
\begin{align}
    \mathcal{F}^{(2)}_{L_{-1}\ket{\varepsilon}} &= \frac{(\cos (4 \pi  x)+7)  }{4096 \cos ^8(\pi 
   x)} (954 -776 \cos (2 \pi  x)+319 \cos (4 \pi  x) +4 \cos (6 \pi  x) \nonumber\\
   & \phantom{= \frac{(\cos (4 \pi  x)+7)  }{4096 \cos ^8(\pi 
   x)}} +6 \cos (8 \pi  x)+4 \cos (10 \pi  x)+\cos (12 \pi  x)) \displaybreak[0]\\
   \mathcal{F}^{(2)}_{L_{-2}\ket{\varepsilon}} &= \frac{(\cos (4 \pi  x)+7)}{1048576  \cos ^{12}(\pi  x)} (1546754 -2155216 \cos (2 \pi  x)+864034 \cos (4 \pi  x) \nonumber\\
   & \phantom{= \frac{(\cos (4 \pi  x)+7)}{1048576  \cos ^{12}(\pi  x)}} -139296 \cos (6 \pi  x)+13320 \cos (8 \pi  x)+224 \cos (10 \pi  x) \nonumber\\
   & \phantom{= \frac{(\cos (4 \pi  x)+7)}{1048576  \cos ^{12}(\pi  x)}} +469 \cos (12 \pi  x)+456
   \cos (14 \pi  x) +246 \cos (16 \pi  x) \nonumber\\
   & \phantom{= \frac{(\cos (4 \pi  x)+7)}{1048576  \cos ^{12}(\pi  x)}} +72 \cos (18 \pi  x)+9 \cos (20 \pi  x)) \\
   & = \mathcal{F}^{(2)}_{L_{-1}^2\ket{\varepsilon}} \nonumber\displaybreak[0]\\
   \mathcal{F}^{(2)}_{L_{-3}\ket{\varepsilon}} &=  \frac{(\cos (4 \pi  x)+7)}{67108864\cos ^{16}(\pi  x)} (938450676-1469899184 \cos (2 \pi  x)+710758371 \cos (4 \pi  x) \nonumber\\
   & \phantom{=  \frac{(\cos (4 \pi  x)+7)}{67108864\cos ^{16}(\pi  x)}} -201143980 \cos (6 \pi  x)+32581122 \cos (8 \pi  x) \nonumber\\
   & \phantom{=  \frac{(\cos (4 \pi  x)+7)}{67108864\cos ^{16}(\pi  x)}} -2510220 \cos (10 \pi  x)+99537
   \cos (12 \pi  x)+5080 \cos (14 \pi  x) \nonumber\\
   &  \phantom{=  \frac{(\cos (4 \pi  x)+7)}{67108864\cos ^{16}(\pi  x)}} +13484 \cos (16 \pi  x)+15448 \cos (18 \pi  x)+11059 \cos (20 \pi  x) \nonumber\\
   & \phantom{=  \frac{(\cos (4 \pi  x)+7)}{67108864\cos ^{16}(\pi  x)}} +5260 \cos (22 \pi  x)+1630 \cos (24 \pi  x)+300 \cos
   (26 \pi  x) \nonumber\\
   & \phantom{=  \frac{(\cos (4 \pi  x)+7)}{67108864\cos ^{16}(\pi  x)}} +25 \cos (28 \pi  x)) \\
   & = \mathcal{F}^{(2)}_{L_{-1}^3\ket{\varepsilon}} = \mathcal{F}^{(2)}_{L_{-2}L_{-1}\ket{\varepsilon}}\nonumber
\end{align}

\subsection{Trace square distance}

\subsubsection{Vacuum module}\label{app:TSDvacResults}

Some explicit expressions for the TSD between light states:
\begin{align}
     T^{(2)}_{L_{-3}\ket{0},\ket{0}} &=\frac{1}{64} c^2 \sin ^8(\pi x) \cos ^4(\pi x) \label{eq:TSD31}\\
     &\quad-\frac{c \sin ^6(\pi x) \cos ^2(\pi x) (124 \cos (2 \pi x)+17 \cos (4 \pi x)+243)}{2048}\nonumber\\
     &\quad+\frac{\sin
   ^4(\pi x) (8391 \cos (2 \pi x)+1890 \cos (4 \pi x)+361 \cos (6 \pi x)+7790)}{16384 c}\nonumber\\
   &\quad+\frac{-7864320 \cos (\pi x)+1984232 \cos (2 \pi x)-524288
   \cos (3 \pi x)}{8388608}\nonumber\\
   &\quad+\frac{295631 \cos (4 \pi x)-27612 \cos (6 \pi x)+386 \cos (8 \pi x)}{8388608}\nonumber\\
   &\quad+\frac{8436 \cos (10 \pi x)+289 \cos (12 \pi x)+6127246}{8388608} \nonumber
\end{align}
\begin{align}
    T^{(2)}_{L_{-3}\ket{0},L_{-2}\ket{0}} &= \frac{c^2 \sin ^8(\pi x) (2 \cos (2 \pi x)+1)^2}{1024}\\
    &\quad-\frac{1}{512} c \sin ^6\left(\frac{\pi x}{2}\right) \cos ^4\left(\frac{\pi x}{2}\right) (1571 \cos (\pi x)-194 \cos (2 \pi x)\nonumber\\
   &\hspace{5.5cm}+317 \cos (3 \pi x)+252 \cos (4 \pi x)\nonumber\\
   &\hspace{5.5cm}+143 \cos (5 \pi x)+34 \cos (6 \pi x)\nonumber\\
   &\hspace{5.5cm}+17 \cos (7 \pi x)-604)\nonumber\\
   &\quad+\frac{\sin^4(\pi x) (-3968 \cos (\pi x)+967 \cos (2 \pi x)-128 \cos (3 \pi x))}{16384 c}\nonumber\\
   &\quad+\frac{\sin^4(\pi x) (738 \cos (4 \pi x)+361 \cos (6 \pi x)+4078)}{16384 c}\nonumber\\
   &\quad+\frac{\sin^4\left(\frac{\pi x}{2}\right) (-332488 \cos (\pi x)+332370 \cos (2 \pi x)+1936 \cos (3 \pi x))}{524288}\nonumber\\
   &\quad+\frac{\sin^4\left(\frac{\pi x}{2}\right) (-8856 \cos (4 \pi x)+29776 \cos (5 \pi x)+56205
   \cos (6 \pi x))}{524288}\nonumber\\
   &\quad+\frac{\sin^4\left(\frac{\pi x}{2}\right) (37476 \cos (7 \pi x)+10814 \cos (8 \pi x)+1156 \cos (9 \pi x))}{524288}\nonumber\\
   &\quad+\frac{\sin^4\left(\frac{\pi x}{2}\right) (289 \cos (10 \pi x)+395610)}{524288}\nonumber
\end{align}

\begin{align}
    T^{(2)}_{L_{-4}\ket{0},\ket{0}} &= \frac{c^2 \sin ^8(\pi x) (3 \cos (2 \pi x)+2)^4}{6400}\\
    &\quad-\frac{c \sin ^6(\pi x) (3 \cos (2 \pi x)+2)^2 (15489 \cos (2 \pi x)+4678 \cos (4 \pi x))}{1638400}\nonumber\\
    &\quad-\frac{c \sin ^6(\pi x) (3 \cos (2 \pi x)+2)^2 (1263 \cos (6 \pi x)+19530)}{1638400}\nonumber\\
   &\quad+\frac{\sin ^4(\pi x) (5444642 \cos (2 \pi x)+2684168 \cos (4 \pi x))}{6553600 c}\nonumber\\
   &\quad+\frac{\sin ^4(\pi x) (913973 \cos (6 \pi x)+286934 \cos (8 \pi x)+59049 \cos (10 \pi x)+3718434)}{6553600 c}\nonumber\\
   &\quad+\frac{-48486154240 \cos (\pi x)+12614780944 \cos (2 \pi x)-4445962240 \cos (3 \pi x)}{53687091200}\nonumber\\
   &\quad+\frac{2269930440 \cos(4 \pi x)-754974720 \cos (5 \pi x)+634223408 \cos (6 \pi x)}{53687091200}\nonumber\\
   &\quad+\frac{-121298020 \cos (8 \pi x)-5870960 \cos (10 \pi x)+13794296 \cos (12 \pi x)}{53687091200}\nonumber\\
   &\quad+\frac{18977328
   \cos (14 \pi x)+1595169 \cos (16 \pi x)+38260958595}{53687091200}\nonumber
\end{align}

\subsubsection{Ising model}\label{app:TSDising}
Some results for $\varepsilon$ descendants in the Ising models:
\begin{align}
    T^{(2)}_{L_{-1}\ket{\varepsilon},\ket{\varepsilon}} &= \frac{ (\cos (2 \pi  x)+7)}{1024} \sin ^4\left(\frac{\pi  x}{2}\right) (11 + 92 \cos (\pi  x)+148 \cos (2 \pi  x) \nonumber\\
    & \phantom{= \frac{ (\cos (2 \pi  x)+7)}{1024}}+ 100 \cos (3 \pi  x)+ 25 \cos (4 \pi  x)) \\
    T^{(2)}_{L_{-2}\ket{\varepsilon},\ket{\varepsilon}} &= \frac{ (\cos (2 \pi  x)+7)}{4194304} \sin ^4\left(\frac{\pi  x}{2}\right) (2347723 + 2430412 \cos (\pi  x)+1872304 \cos (2 \pi  x) \nonumber\\
    & \phantom{= \frac{ (\cos (2 \pi  x)+7)}{4194304}} +1393796 \cos (3 \pi  x)+1144940 \cos (4 \pi  x)+751500
   \cos (5 \pi  x) \nonumber\\ 
   & \phantom{= \frac{ (\cos (2 \pi  x)+7)}{4194304}} +366480 \cos (6 \pi  x)+142884 \cos (7 \pi  x)+35721 \cos (8 \pi  x)) \\
   & = T^{(2)}_{L_{-1}^2\ket{\varepsilon},\ket{\varepsilon}} \nonumber\\
   T^{(2)}_{L_{-3}\ket{\varepsilon},\ket{\varepsilon}} &= \frac{ (\cos (2 \pi  x)+7)}{1073741824} \sin ^4\left(\frac{\pi  x}{2}\right) ( 912429118 + 1218353112 \cos (\pi  x) \nonumber\\
   & \phantom{= \frac{ (\cos (2 \pi  x)+7)}{1073741824}} +998414040 \cos (2 \pi  x)+780711528 \cos (3 \pi  x) \nonumber\\
   &\phantom{= \frac{ (\cos (2 \pi  x)+7)}{1073741824}} +629316847 \cos (4 \pi 
   x)+497152212 \cos (5 \pi  x) \nonumber\\
   &\phantom{= \frac{ (\cos (2 \pi  x)+7)}{1073741824}} +393829628 \cos (6 \pi  x)+276532300 \cos (7 \pi  x)\nonumber\\
   &\phantom{= \frac{ (\cos (2 \pi  x)+7)}{1073741824}} +168888850 \cos (8 \pi  x)+94527900 \cos (9 \pi  x) \nonumber\\
   &\phantom{= \frac{ (\cos (2 \pi  x)+7)}{1073741824}} +46637100 \cos (10 \pi 
   x)+18404100 \cos (11 \pi  x)\nonumber\\
   &\phantom{= \frac{ (\cos (2 \pi  x)+7)}{1073741824}} +4601025 \cos (12 \pi  x)) \\
   &= T^{(2)}_{L_{-1}^3\ket{\varepsilon},\ket{\varepsilon}} =T^{(2)}_{L_{-2}L_{-1}\ket{\varepsilon},\ket{\varepsilon}} \nonumber \nonumber
\end{align}
For $\sigma$ descendants:
\begin{align}
    T^{(2)}_{L_{-1}\ket{\sigma},\ket{\sigma}} &= -\frac{\cos ^2\left(\frac{\pi  x}{2}\right) \sqrt{\csc \left(\frac{\pi  x}{2}\right)}}{32768 \sin ^{\frac{7}{2}}\left(\frac{\pi  x}{2}\right) (\cos (\pi  x)+1)^{5/4}} \Bigg\{66752\ 2^{3/4} \sqrt{\sin ^7\left(\frac{\pi  x}{2}\right) \sin (\pi 
   x)} \nonumber\\
   &\phantom{=}+\sqrt[4]{\cos (\pi  x)+1} \big[ 46488 \cos (\pi  x)-16137 \cos (2 \pi  x)+3612 \cos (3 \pi  x)  \nonumber\\
   &\phantom{=} -1006 \cos (4 \pi  x)+76 \cos (5 \pi  x)-7 \cos (6 \pi  x)\big]-33026
   \sqrt[4]{\cos (\pi  x)+1}  \nonumber\\
   &\phantom{=} -16 \sqrt[4]{2} \sin ^4\left(\frac{\pi  x}{2}\right) \sqrt{\cos \left(\frac{\pi  x}{2}\right)} \big[-3888 \cos (\pi  x)-804 \cos (2 \pi 
   x) \nonumber\\
   &\phantom{=}  +48 \cos (3 \pi  x)+129 \cos (4 \pi  x)+7099\big]\Bigg\}\displaybreak[0] \\
   T^{(2)}_{L_{-2}\ket{\sigma},\ket{\sigma}} &= \left(33554432\ 2^{3/4} \sin
   ^{\frac{5}{4}}(\pi  x) \sqrt{\csc \left(\frac{\pi  x}{2}\right)}\right)^{-1}\Bigg\{ 2 \big[ 51995040 \cos (\pi  x) \nonumber\\
   &\phantom{=}  +6726368 \cos (2 \pi  x)-50083264 \cos (3 \pi  x)-27705396 \cos (4 \pi  x) \nonumber\\
   &\phantom{=} -23115584 \cos (5 \pi  x)+3517116 \cos (6 \pi  x)-1600624
   \cos (7 \pi  x) \nonumber\\
   &\phantom{=} +688029 \cos (8 \pi  x)-18480 \cos (9 \pi  x)+36 \cos (10 \pi  x)+93860567\big] \nonumber\\
   &\phantom{=} \times \sqrt[4]{\sin ^3(\pi  x) (\cos (\pi  x)+1)}+\big[-124036846 \sin (\pi 
   x)-131438432 \sin (2 \pi  x) \nonumber\\
   &\phantom{=}-31447124 \sin (3 \pi  x)+24870528 \sin (4 \pi  x)+31182128 \sin (5 \pi  x)\nonumber\\
   &\phantom{=}+21514960 \sin (6 \pi  x)-2930095 \sin (7 \pi  x)+1582144
   \sin (8 \pi  x)\nonumber\\
   &\phantom{=}-453993 \sin (9 \pi  x)+18480 \sin (10 \pi  x)-36 \sin (11 \pi  x)\big] \sqrt[4]{\cot \left(\frac{\pi  x}{2}\right)}\Bigg\} \\
   & = T^{(2)}_{L_{-1}^2\ket{\sigma},\ket{\sigma}} \nonumber\displaybreak[0]\\
   T^{(2)}_{L_{-3}\ket{\sigma},\ket{\sigma}} &= \left(64961380352\ 2^{3/4} \sin
   ^{\frac{5}{4}}(\pi  x) \sqrt{\csc \left(\frac{\pi  x}{2}\right)}\right)^{-1} \nonumber\\
   &\phantom{=} \times \Bigg\{ \big[-804063279604 \sin (\pi  x)-679262872576 \sin (2 \pi  x) \nonumber\\
   &\phantom{=} -676379752341 \sin (3 \pi  x)-151873435008 \sin (4 \pi  x)+78179526785 \sin (5 \pi  x)  \nonumber\\
   &\phantom{=}  +162043727616
   \sin (6 \pi  x)+203623756022 \sin (7 \pi  x)+116161263104 \sin (8 \pi  x)\nonumber\\
   &\phantom{=} +68995216314 \sin (9 \pi  x)+27584342272 \sin (10 \pi  x)-1884177607 \sin (11 \pi 
   x) \nonumber\\
   &\phantom{=} +2161088640 \sin (12 \pi  x)-1097984829 \sin (13 \pi  x)+30000 (1568 \sin (14 \pi  x) \nonumber\\
   &\phantom{=}+93 \sin (15 \pi  x))\big] \sqrt[4]{\cot \left(\frac{\pi  x}{2}\right)}-2
   \big[-463898496256 \cos (\pi  x) \nonumber\\
   &\phantom{=}-339771582456 \cos (2 \pi  x)+161014585088 \cos (3 \pi  x)\nonumber\\
   &\phantom{=}+347509883805 \cos (4 \pi  x)+308827933824 \cos (5 \pi  x)+269411694364 \cos
   (6 \pi  x) \nonumber\\
   &\phantom{=}+145953734016 \cos (7 \pi  x)+65832431142 \cos (8 \pi  x)+29792470912 \cos (9 \pi  x) \nonumber\\
   &\phantom{=}-3124129172 \cos (10 \pi  x)+2208128640 \cos (11 \pi  x)-1220635853
   \cos (12 \pi  x)\nonumber\\
   &\phantom{=}+47040000 \cos (13 \pi  x)+2790000 \cos (14 \pi  x)\nonumber\\
   &\phantom{=}-619071617270\big] \sqrt[4]{\sin ^3(\pi  x) (\cos (\pi  x)+1)}\Bigg\}
\end{align}
For degenerate states at level~3 the expressions are different, but we report here only one for simplicity.

\subsubsection{Three-states Potts model}\label{app:TSDpotts}
In the following an example of TSD between a descendant of $\varepsilon$ and the primary state itself in the Potts model:
\begin{align}
    T^{(2)}_{L_{-1}\ket{\varepsilon},\ket{\varepsilon}} &= \frac{1}{8} F\!\left(-\tfrac{8}{5},-\tfrac{1}{5};-\tfrac{2}{5};\eta \right){}^2 \sin^4\!\left(\frac{\pi  x}{2}\right) (4 \cos (\pi  x)+\cos (2 \pi  x)+19) \nonumber\\
    &\phantom{=}+\frac{1}{5200}\Bigg\{(F\!\left(-\tfrac{8}{5},-\tfrac{1}{5};-\tfrac{2}{5};\eta \right) \sin^{10}\!\left(\frac{\pi  x}{2}\right) \cos^8\!\left(\frac{\pi  x}{2}\right) \csc ^6(\pi  x)\nonumber\\
    &\phantom{=}  \times\bigg[(41600 F\!\left(-\tfrac{3}{5},\tfrac{4}{5};\tfrac{3}{5};\eta \right) (2 \cos (\pi  x)+\cos (2 \pi  x)-15)+32 \cos^2\!\left(\frac{\pi  x}{2}\right) \nonumber\\
    &\phantom{=}\times\bigg(9
   \sin (\pi  x) \left(49 F\!\left(\tfrac{12}{5},\tfrac{19}{5};\tfrac{18}{5};\eta \right) \sin ^3(\pi  x)+260 F\!\left(\tfrac{7}{5},\tfrac{14}{5};\tfrac{13}{5};\eta \right) \sin (2 \pi  x)\right)\nonumber\\
   &\phantom{=}  +520 F\!\left(\tfrac{2}{5},\tfrac{9}{5};\tfrac{8}{5};\eta \right) (31 \cos
   (2 \pi  x)+9)\bigg)\bigg]\Bigg\} \nonumber\\
   &\phantom{=}-\left(1782579200\ 2^{2/5} \Gamma
   \left(-\frac{8}{5}\right) \Gamma \left(\frac{17}{10}\right) \Gamma \left(\frac{12}{5}\right)\right)^{-1}\nonumber\\
   &\phantom{=}\times\Bigg\{\Gamma \left(-\frac{2}{5}\right) \Gamma \left(\frac{3}{10}\right) \Gamma \left(\frac{13}{5}\right) F\!\left(\tfrac{6}{5},\tfrac{13}{5};\tfrac{12}{5};\eta \right) \sin ^{\frac{28}{5}}(\pi  x) \nonumber\\
   &\phantom{=} \times\bigg[ 544 F\!\left(\tfrac{6}{5},\tfrac{13}{5};\tfrac{12}{5};\eta
   \right) (9600 \cos (\pi  x)-17164 \cos (2 \pi  x) +15129 \cos (4 \pi  x) \nonumber\\
   &\phantom{=} -1293) +208 \sin ^2(\pi  x) \bigg( 9016 F\!\left(\tfrac{26}{5},\tfrac{33}{5};\tfrac{32}{5};\eta \right) \sin ^6(\pi  x)+17 F\!\left(\tfrac{11}{5},\tfrac{18}{5};\tfrac{17}{5};\eta \right) \nonumber\\
   &\phantom{=}\times(3697 \cos
   (\pi  x)+6479 \cos (3 \pi  x)+800)+79488 F\!\left(\tfrac{21}{5},\tfrac{28}{5};\tfrac{27}{5};\eta \right) \sin ^4(\pi  x)  \nonumber\\
   &\phantom{=} \times \cos (\pi  x) +99 F\!\left(\tfrac{16}{5},\tfrac{23}{5};\tfrac{22}{5};\eta \right) \sin ^2(\pi  x) (1373 \cos (2 \pi  x)+843)\bigg)\bigg]\Bigg\}
\end{align}
where $F \equiv \; _2F_1 $ is the hypergeometric function and $\eta = \sin^2 \left( \frac{\pi x}{2} \right)$.
We computed also TSDs for higher level descendants, but the expressions are more complicated and we won't show them here.

\bibliography{lit}

\end{document}